\documentclass[fleqn,usenatbib]{mnras}

\usepackage{newtxtext,newtxmath}

\usepackage[T1]{fontenc}

\DeclareRobustCommand{\VAN}[3]{#2}
\let\VANthebibliography\thebibliography
\def\thebibliography{\DeclareRobustCommand{\VAN}[3]{##3}\VANthebibliography}

\usepackage{graphicx}	
\usepackage{amsmath}	
\usepackage{hyperref}
\usepackage{threeparttable}

\title[IMF in diverse stellar systems]{Initial mass function variability from the integrated light of diverse stellar systems}

\author[C. M. Cheng et al.]{
Chloe M. Cheng,$^{1,2,3}$\thanks{E-mail: cheng@strw.leidenuniv.nl (CMC)}
Alexa Villaume,$^{2,3}$
Michael L. Balogh,$^{2,3}$
Jean P. Brodie,$^{4,5}$
\newauthor
Ignacio Martín-Navarro,$^{6,7}$
Aaron J. Romanowsky,$^{8,9,10}$
and Pieter G. van Dokkum$^{11}$
\\
$^{1}$Leiden Observatory, Leiden University, P.O. Box 9513, 2300 RA Leiden, The Netherlands\\
$^{2}$Waterloo Centre for Astrophysics, University of Waterloo, Waterloo, Ontario, N2L 3G1, Canada\\
$^{3}$Department of Physics and Astronomy, University of Waterloo, Waterloo, Ontario N2L 3G1, Canada\\
$^{4}$Centre for Astrophysics and Supercomputing, Swinburne University of Technology, 1 John St, Hawthorn, VIC 3122, Australia\\
$^{5}$University of California Observatories, University of California, Santa Cruz, CA 95060, USA\\
$^{6}$Instituto de Astrofísica de Canarias, Vía Láctea s/n, E-38205 La Laguna, Tenerife, Spain\\
$^{7}$Departamento de Astrofísica, Universidad de La Laguna, E-38205 La Laguna, Tenerife, Spain\\
$^{8}$Department of Physics \& Astronomy, San José State University, One Washington Square, San Jose, CA 95192, USA\\
$^{9}$University of California Observatories, 1156 High Street, Santa Cruz, CA 95064, USA\\
$^{10}$Department of Astronomy \& Astrophysics, University of California Santa Cruz, 1156 High Street, Santa Cruz, CA 95604, USA\\
$^{11}$Astronomy Department, Yale University, 52 Hillhouse Ave, New Haven, CT 06511
}

\date{Accepted XXX. Received YYY; in original form ZZZ}

\pubyear{2023}

\begin{document}
\label{firstpage}
\pagerange{\pageref{firstpage}--\pageref{lastpage}}
\maketitle

\begin{abstract}
We present a uniform analysis of the stellar initial mass function (IMF) from integrated light spectroscopy of 15 compact stellar systems (11 globular clusters in M31 and 4 ultra compact dwarfs in the Virgo cluster, UCDs) and two brightest Coma cluster galaxies (BCGs), covering a wide range of metallicities ($-$1.7 $<$ [Fe/H] $<$ 0.01) and velocity dispersions (7.4 km~s$^{-1}$ $< \sigma <$ 275 km~s$^{-1}$).  The S/N $\sim 100$ \AA$^{-1}$ Keck LRIS spectra are fitted over the range $4000<\lambda/\mbox{\AA}<10,000$ with flexible, full-spectrum stellar population synthesis models.  
We use the models to fit simultaneously for ages, metallicities, and individual elemental abundances of the population, allowing us to decouple abundance variations from variations in IMF slope.    We show that compact stellar systems do not follow the same trends with physical parameters that have been found for early-type galaxies.  Most globular clusters in our sample have an IMF consistent with that of the Milky Way, over a wide range of [Fe/H] and [Mg/Fe].  There is more diversity among the UCDs, with some showing evidence for a bottom-heavy IMF, but with no clear correlation with metallicity, abundance, or velocity dispersion.  The two Coma BCGs have similar velocity dispersion and metallicity, but we find the IMF of NGC~4874 is consistent with that of the Milky Way while NGC~4889 presents evidence for a significantly bottom-heavy IMF.  For this sample, the IMF appears to vary between objects in a way that is not explained by a single metallicity-dependent prescription.
\end{abstract}




\section{Introduction}\label{sec:introduction}

A fundamental concept underpinning star formation (SF) and galaxy evolution is the distribution of a galaxy's birth stellar masses, called the initial mass function (IMF, e.g.\ \citealt{Bastian_2010, Conroy_2013, Hopkins_2018}).  This informs physical models of SF and is needed to translate observables (e.g.\ galaxy luminosities) into physically meaningful quantities (e.g.\ stellar masses and SF rates, \citealt{Hopkins_2018, Smith_2020}).

The IMF in the Milky Way (MW) is well characterized by a single power law with slope $\sim 2.35$ for masses $\gtrsim 0.5 ~\mathrm{ M}_\odot$ \citep{Salpeter}, with a shallower slope at lower masses \citep{Miller_Scalo_1979, Scalo_1986, Kroupa_2001, Chabrier_2003}.
Measurements of resolved star counts in nearby, diverse, SF regions are largely consistent with a single IMF of this shape  \citep{Sagar_2001, Bastian_2010, DaRio_2012, PenaRamirez_2012, Andersen_2017, Suarez_2019, Damian_2021}.  However, the limited number of galaxies close enough for their stars to be resolved is not representative of the full diversity of the population, and it is not yet possible to extend these techniques to more distant galaxies.  For this reason, the IMF shape is often assumed to be universal.  However, there is little theoretical reason for this to be the case (e.g.\ \citealt{Schwarzschild_1953, Larson_1986, Kroupa_2001, Hennebelle_2008, Krumholz_2011, Hopkins_2012, Hopkins_2013, Chabrier_2014}).  In particular, predictions that the IMF should depend on metallicity \citep[e.g.][]{2005ApJ...626..627O} and temperature of the cosmic microwave background \citep[e.g.][]{2010MNRAS.402..429S} may explain recent {\it JWST} observations of UV-bright galaxies at $z>10$ \citep{2023arXiv230504944T}.  Uncertainty over the variation in IMF shape is one of the most significant sources of systematic uncertainty in stellar population studies \citep{Hopkins_2018}.

To study the IMF in a wider diversity of stellar systems requires an inference based on measurements of their integrated light.  One approach is to compare this light with the total mass measured from stellar dynamics \citep[e.g.][]{Tinsley_1980, Conroy_2013, Hopkins_2018} or strong lensing \citep[e.g.][]{Treu_2010}.  The mass-to-light ratio measured in this way, (M/L)$_\mathrm{dyn}$,  is sensitive to the shape of the IMF, but also to any contribution from non-stellar dark matter. 

An alternative is to measure the equivalent widths of specific, strong absorption features  
(i.e.\ Lick indices, \citealt{Worthey_1994}), and compare these with stellar population synthesis (SPS) model predictions. Since dwarf and giant stars at the same effective temperature display differences discernible at a level of $\sim1-3\%$ in high S/N spectra \citep{Smith_2020}, dwarf- and giant-sensitive spectral features \citep[e.g.][]{Wing_Ford_1969} can be used to quantify the relative numbers of each type of star.  However, IMF-sensitive features are also sensitive to age, metallicity, and many elemental abundances, and IMF-driven fluctuations in individual indices are often too weak to exclude variation in these other parameters \citep{McConnell_2016, Zieleniewski_2017, Lonoce_2021}.  For this reason it is preferable to fit the full spectrum with SPS models, taking advantage of all available spectral information over a wide wavelength range.  Assuming the SPS models and their input stellar libraries are sufficiently detailed, one can then correctly measure ages, metallicities, and elemental abundances to separate them from IMF variations \citep{Lonoce_2021}.  This method requires much higher-quality, well-calibrated, and longer-wavelength-range data.  

Evidence for IMF variations has been found using all the above techniques.  For example, \cite{Cenarro_2003} found variations in the calcium triplet (CaT) index in massive, early-type galaxies (ETGs) that could not be explained by age, metallicity, and elemental abundances alone.  Their models thus require more flexibility, perhaps in the form of a variable IMF.  These results were reinforced by \cite{CvD_2012b}, who applied a full spectrum SPS model (\textsc{alf}\footnote{\url{https://github.com/cconroy20/alf}~.}, \citealt{CvD_2012a}) to the spectra of massive ETGs and  found that the IMF becomes more bottom-heavy with increasing [Mg/Fe] and velocity dispersion, $\sigma$.  Similar findings have since been confirmed by many others, using stellar population analyses and other methods \cite[e.g.][]{Spiniello_2012, Cappellari_2012, Cappellari_2013, La_Barbera_2013, Ferreras_2013, McDermid_2014, Posacki_2015}.

IMF variations are also found to correlate strongly with metallicity for some stellar systems (e.g.\ \citealt{Geha_2013, Martin-Navarro_2015, van_Dokkum_2017}), which suggests that metallicity may be a more fundamental determinant of IMF shape.  This idea has become widespread, with some recent theoretical studies assuming a metallicity-dependent IMF (e.g.\ \citealt{Clauwens_2016, Prgomet_2021, Sharda_2022}).  However, it is not clear if different metallicity-dependent relationships (e.g.\ \citealt{Geha_2013, Martin-Navarro_2015}) are consistent, and not all objects seem to follow these trends \citep{CvD_2012b, Newman_2017, Villaume_2017}.  Moreover, metallicity is mutually correlated with other parameters (e.g.\ velocity dispersion, $\alpha$-element abundances), and correlations between IMF variability and these parameters have also been found (e.g.\ \citealt{CvD_2012b, Geha_2013, Martin-Navarro_2015, Li_2017, Gennaro_2018, Martin-Navarro_2021}).  Comparing IMF inferences from different works, using different techniques, is very challenging as there are various systematic uncertainties and model assumptions that can lead to results that differ by more than their formal uncertainties, for the same objects \cite[e.g.][]{Smith_2014}.  Robust conclusions require a carefully controlled sample with self-consistent modeling assumptions \citep{Lyubenova_2016}.

Most of the work summarized above has focused on metal-rich, massive ETGs, restricted to a narrow parameter space in metallicity, density, and $\sigma$.  However, evidence for IMF variation has also been found from star counts in local, metal-poor stellar systems \citep[e.g.][]{Geha_2013, Gennaro_2018, Hallakoun_2021}.  Specifically, \citet{Geha_2013} found a trend with metallicity in metal-poor, ultra-faint dwarf galaxies that echoes the extragalactic result from \citet{Martin-Navarro_2015} (though note that they examine different mass ranges and metallicities).  To expand on this and test whether or not metallicity is the only factor that determines the IMF, we must study objects with similar metallicities but a wider range of $\sigma$.  

To this end, \citet[][V17 hereafter]{Villaume_2017} analysed long-slit spectra of a small sample of low-$\sigma$ compact stellar systems (CSSs): three globular clusters (GCs), two ultracompact dwarfs (UCDs) and a compact elliptical.  Their results indicated that the IMFs in these CSSs behave differently than in ETGs, with GCs having MW-to-bottom-light IMFs, and UCDs being less bottom-heavy than ETGs.  They also found IMF variations among systems with similar metallicities and abundance patterns, concluding metallicity cannot be the sole driver of variability.  These results are tantalizing, but with a sample of only five objects they are not conclusive.  

In this paper, we use the \textsc{Absorption Line Fitter} (\textsc{alf}) models to build on the results of V17 and measure the IMF in a larger sample of diverse stellar systems.  This paper is organized as follows: in Section~\ref{sec:sample_DR} we describe our sample and data reduction.  We describe the \textsc{alf} full spectrum SPS model in Section~\ref{sec:methods}.  We present our fit results in Section~\ref{sec:results} and discuss these results in Section~\ref{sec:discussion}.  We summarize and conclude in Section~\ref{sec:conclusion}.

\section{Sample and Data Reduction}\label{sec:sample_DR}
\subsection{Sample selection}\label{sec:sample_selection}
\begin{figure}
    \centering
    \includegraphics[width=\columnwidth]{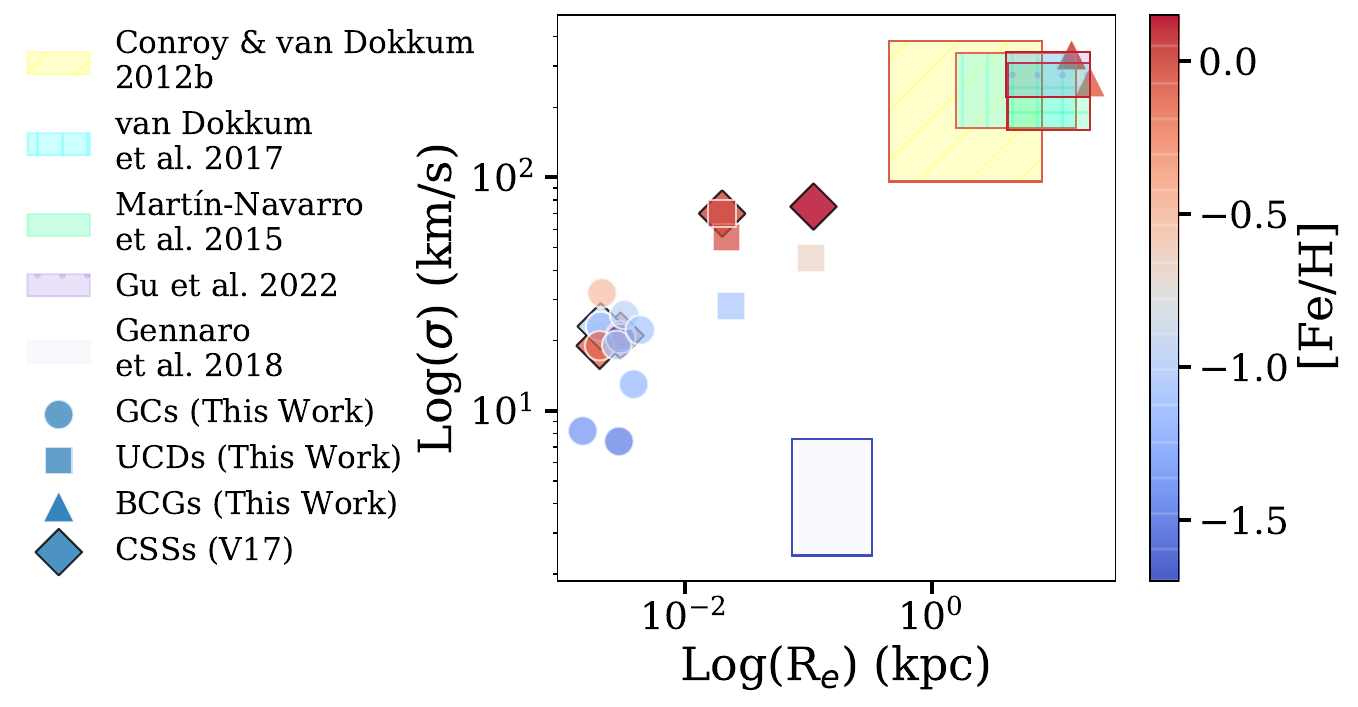}
    \caption{The parameter space spanned by our sample, compared to previous studies.  We show logarithmic effective radius ($R_\mathrm{e}$) versus logarithmic velocity dispersion ($\sigma$) for the GCs (circles), UCDs (squares), and BCGs (triangles).  These are colour-coded by metallicity.  We show the sample from V17 (diamonds), to demonstrate that we fill in their metallicity range.  The rectangles indicate parameter spaces covered by previous studies examining metal-rich ETGs \citep{CvD_2012b, Martin-Navarro_2015, van_Dokkum_2017, Gu_2022} and metal-poor ultra-faint dwarf galaxies (\citealt{Gennaro_2018}, an extension of \citet{Geha_2013}).  These are outlined by the colour representing the average sample metallicity.}
    \label{fig:paper_sample}
\end{figure}
We examine a diverse set of stellar systems, including eleven M31 GCs\footnote{Images of all M31 GCs can be found here: \url{https://lweb.cfa.harvard.edu/oir/eg/m31clusters/m31clusterso_frames.html}~.} \citep{Galleti_2004}, four Virgo cluster UCDs \citep{VUCD_img, M59UCD3_img}, and the centres of two Coma cluster brightest cluster galaxies (BCGs).  The full sample is reported in Table~\ref{tab:sample}.   We re-examine objects from V17, including new data for the globular cluster B058.  We do not combine the new B058 data with that examined in V17, but throughout we show two independent measurements of this object as an additional check on our systematics.  The data examined in V17 will be denoted as B058 2014 and the new data as B058 2016.  The remaining M31 GCs in our sample serve to expand the metallicity range of V17.  Among the UCD sample we include VUCD3 and M59-UCD3, which have evidence for central massive black holes \citep[][see Table~\ref{tab:literature_vals} for black hole masses]{Ahn_2017,Ahn_2018}.  The two Coma BCGs in our sample were previously studied by \citet[][Z17 hereafter]{Zieleniewski_2017}, who found near-MW IMFs in their cores, differing from the general trend for increasingly bottom-heavy IMFs with increasing $\sigma$ found in other studies.   We also re-fit the ETG data from \cite{CvD_2012b} with the newest version of the \textsc{alf} models for comparison.   

The sample is deliberately chosen to span a wide range of metallicities, stellar masses, $\sigma$, and densities.  This is illustrated in Figure~\ref{fig:paper_sample}, where we plot logarithmic effective radius ($R_\mathrm{e}$) versus logarithmic $\sigma$ for our sample, colour-coded by metallicity.  For comparison, we also show individual points from V17.  Rectangles indicate parameter spaces covered by previous studies, including those that examine metal-rich ETGs \citep{CvD_2012b, Martin-Navarro_2015, van_Dokkum_2017, Gu_2022} and metal-poor ultra-faint dwarf galaxies (UFDs, \citealt{Gennaro_2018}).  This shows that we cover a wider range of parameter space compared with previous work.  In particular, in regions of previously-studied metallicities, we also sample objects with diverse $\sigma$'s and densities.  

\begin{table*}
    \centering
    \caption{Observing parameters for the objects in our study.  In the $R_\mathrm{e}$ column, the fraction of $R_\mathrm{e}$ encompassed by our aperture is indicated in brackets.  References: (d) \citet{Federici_2012}; (e) From NED; (f) \citet{Holland_1998}; (g) \citet{Paturel_1992}; (h) \citet{Mei_2007}; (i) \citet{Mieske_2013}; (j) \citet{Tully_2016}; (k) \citet{Peacock_2010}; (l) \citet{Strader_2011}; (m) \citet{Barmby_2007}; (n) \citet{Barmby_2010}; (o) \citet{Sandoval_2015}; (p) \citet{Greene_2019}; (q) \citet{Galleti_2004}; (r) \citet{Paudel_2010}; (s) \citet{Kim_2014}; (t) \citet{Abazajian_2009}.}
    \label{tab:sample}
    \begin{threeparttable}
    \begin{tabular}{p{14mm}cccp{10mm}p{5mm}p{10mm}cccccc}
        \hline
         ID & RA & Dec & Type & Date(s) & Slit & Exposure & Blue S/N & Red S/N & Distance & $R_{\mathrm{e}}$ & r-Band & \\
          & & & & Observed & & Time (s) & (\AA$^{-1}$)\tnote{a} & (\AA$^{-1}$)\tnote{b} & (Mpc) & (pc) & Magnitude \\
          \hline 
         B012	&	00:40:32.54	&	+41:21:44.3	&	GC	& 	28/10/16	& 	1.0''	&	1300	&	99	& 	100	&	0.73\tnote{}{d}	&	3\tnote{k} ~(0.44)	&	$14.83\pm0.02$\tnote{k}	\\
        B058 (2014, 2016)	&	00:41:53.06	&	+40:47:09.9	&	GC	& 	20/12/14\tnote{c}, 28/10/16	& 	0.7'', 1.0''	&	4300, 900	&	206, 115	& 	233, 126	&	0.74\tnote{d}	&	2.07\tnote{k} ~(0.64)	&	$14.71\pm0.02$\tnote{k}	\\
        B067	&	00:42:03.14	&	+41:04:23.7	&	GC	& 	27/10/16	& 	1.0''	&	2700	&	121	& 	119	&	1.2\tnote{e}	&	1.48\tnote{k} ~(0.37)	&	$16.99\pm0.05$\tnote{k}	\\
B074	&	00:42:08.04	&	+41:43:21.7	&	GC	& 	29/10/16	& 	1.0''	&	1350	&	107	& 	106	&	0.783\tnote{f}	&	2.91\tnote{k} ~(0.45)	&	$16.37\pm0.03$\tnote{k}	\\
B107	&	00:42:31.22	&	+41:19:38.9	&	GC	& 	28/10/16	& 	1.0''	&	720	&	90	& 	110	&	0.783\tnote{f}	&	2.7\tnote{l} ~(0.48)	&	$15.48\pm0.02$\tnote{k}	\\
B163	&	00:43:18.10	&	+41:28:04.2	&	GC	& 	19/12/14\tnote{c}	& 	0.7''	&	4300	&	265	& 	277	&	1.02\tnote{g}	&	3\tnote{k} ~(0.44)	&	$14.65\pm0.02$\tnote{k}	\\
B193	&	00:43:45.42	&	+41:36:57.4	&	GC	& 	19/12/14\tnote{c}	& 	0.7''	&	4300	&	142	& 	138	&	0.61\tnote{g}	&	2.03\tnote{k} ~(0.69)	&	$14.95\pm0.03$\tnote{k}	\\
B225	&	00:44:29.82	&	+41:21:36.6	&	GC	& 	27/10/16	& 	1.0''	&	450	&	238	& 	299	&	0.81\tnote{d}	&	2.12\tnote{k} ~(0.63)	&	$13.82\pm0.02$\tnote{k}	\\
B338	&	00:40:58.83	&	+40:35:48.0	&	GC	& 	29/10/16	& 	1.0''	&	540	&	207	& 	227	&	0.73\tnote{d}	&	4.33\tnote{m} ~~(0.31)	&	$13.93\pm0.02$\tnote{k}	\\
B405	&	00:49:39.82	&	+41:35:29.7	&	GC	& 	29/10/16	& 	1.0''	&	600	&	140	& 	145	&	0.83\tnote{d}	&	3.83\tnote{n} ~(0.35)	&	$14.88\pm0.02$\tnote{k}	\\
G001	&	00:32:46.57	&	+39:34:40.6	&	GC	& 	28/10/16	& 	1.0''	&	540	&	245	& 	322	&	0.82\tnote{d}	&	3.23\tnote{m} ~~(0.41)	&	$13.208$\tnote{q}	\\
M59-UCD3	&	12:42:10.99	&	+11:38:41.6	&	UCD	& 	20/12/14\tnote{c}	& 	0.7''	&	4300	&	89	& 	108	&	14.9\tnote{h}	&	20\tnote{o} ~(4.75)	&	$16\pm0.05$\tnote{o}	\\
VUCD3	&	12:30:58.02	&	+12:25:51.5	&	UCD	& 	19/04/17	& 	0.7''	&	1800	&	49	& 	67	&	16.5\tnote{i}	&	21.6\tnote{i} ~(3.99)	&	$18.1$\tnote{r}	\\
VUCD4	&	12:31:04.93	&	+11:56:38.2	&	UCD	& 	19/04/17	& 	0.7''	&	3600	&	67	& 	76	&	16.5\tnote{i}	&	23.6\tnote{i} ~(3.21)	&	$18.6$\tnote{r}	\\
VUCD7	&	12:31:53.42	&	+12:15:54.3	&	UCD	& 	19/04/17	& 	1.0''	&	2400	&	101	& 	113	&	16.5\tnote{i}	&	105\tnote{i} ~(1.03)	&	$17.34$\tnote{s}	\\
NGC 4874	&	12:59:35.68	&	+27:57:43.8	&	BCG	& 	19/04/17	& 	0.7''	&	270	&	70	& 	87	&	104.71\tnote{j}	&	13600\tnote{p} ~(0.07)	&	$12.101\pm0.002$\tnote{t}	\\
NGC 4889	&	13:00:08.36	&	+27:58:47.4	&	BCG	& 	19/04/17	& 	0.7''	&	270	&	111	& 	142	&	92.04\tnote{j}	&	19200\tnote{p} ~(0.11)	&	$11.976\pm0.002$\tnote{t}	\\
         \hline
    \end{tabular}
    \begin{tablenotes}
        \item[a] Median S/N over the blue wavelength regions that we fit.
        \item[b] Median S/N over the red wavelength regions that we fit.
        \item[c] Data analyzed in V17.
    \end{tablenotes}
    \end{threeparttable}
\end{table*}

\subsection{Observing strategy}\label{sec:observing_strategy}
We observed these objects over several nights in April 2016, October 2016, and April 2017, obtaining optical and near-infrared spectroscopy with the Low Resolution Imaging Spectrometer (LRIS, \citealt{LRISr, LRIsb, LRIS_new}) on the Keck I telescope at the W. M. Keck Observatory.  Observation dates and exposure times are indicated in Table~\ref{tab:sample}.  

LRIS is an instrument for visible-wavelength imaging and spectroscopy \citep{LRISr}.  It consists of CCD detectors covering blue and red wavelengths.  Both cameras have a pixel scale of 0.135'' pixel$^{-1}$.  The 680 nm dichroic was used to split the light between the blue and red arms.  On the blue side, the 300 lines mm$^{-1}$ grism blazed at 5000 \AA\ was used to give a spectral wavelength coverage of $\sim 3500 - 7500$ \AA.  On the red side, the 600 lines mm$^{-1}$ grating blazed at 10 000 \AA\ was used to give coverage of $\sim 7300 - 10~ 600$ \AA.  The effective wavelength range that we consider here is $4000 - 10~ 150$ \AA.  We used the 0.7\arcsec- and 1.0\arcsec-width, 3\arcmin-long, long slits, as indicated in Table~\ref{tab:sample}.  Slit widths were chosen based on object apparent sizes.  The spectral full-width-at-half-maximum resolution is 8.4--9.2 \AA\ and $\sim$ 4.7 \AA\ for the blue and red arms, respectively, for a 1\arcsec\ slit width.  The fraction of $R_\mathrm{e}$ that our apertures encompass is indicated in Table~\ref{tab:sample}.

\subsection{Data reduction}\label{sec:data_reduction}
We reduce\footnote{A detailed tutorial of how the data reduction was done for this paper can be found here: \url{https://github.com/chloe-mt-cheng/imf_css/tree/main/DR_tutorial}~.} the data using the semi-automated reduction package \textsc{PypeIt}\footnote{\url{https://pypeit.readthedocs.io/en/latest/}~.} version 1.5 \citep{pypeit:zenodo, pypeit:joss_pub}.  This follows standard procedures, including overscan bias subtraction and slit tracing via flat field images.  We do not perform flat fielding, as the extra random noise this introduces is larger than any small-scale, systematic sensitivity variations.

In the following sections, we detail aspects of the data reduction that we treat with particular care due to the sensitivity of our method to data systematics. For example, near-perfect (Poisson-limited) sky subtraction and telluric correction are necessary.  This is because the IMF only impacts spectral features at a level of $\sim1-3\%$ (see, e.g.\ \citealt{Smith_2020}).  Additionally, several of the most important IMF-sensitive features are near or overlap with sky lines (e.g.\ NaD near $\sim5900$ \AA) and/or regions heavily affected by telluric absorption (e.g.\ NaI near $\sim8200$ \AA, Wing--Ford band).  

\begin{figure}
    \centering
    \includegraphics[width=\columnwidth]{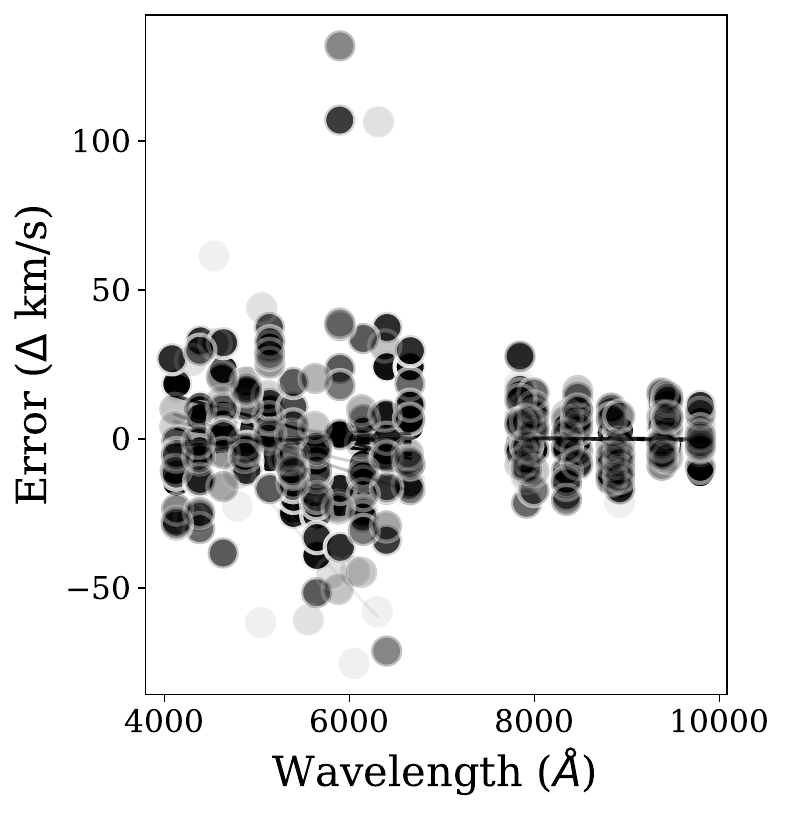}
    \caption{Remaining error in the wavelength solution as a function of wavelength, after flexure correction is applied.  Each object is shown as a series of grey points, to illustrate the range of wavelength solution errors.  The absolute errors are relatively small ($\lesssim 150$ km~s$^{-1}$) and randomly scattered around zero.  This indicates that there is no longer a systematic pattern to the errors.}
    \label{fig:paper_centroids}
\end{figure}

\subsubsection{Wavelength solution}\label{sec:wavelength_soln}
\textsc{PypeIt} produces a master arc frame using individual frames taken using all of the Keck arc lamps (Hg, Cd, and Zn on the blue side and Ne, Ar, Kr, and Xe on the red side).  These were taken for each night and instrument set up.  We use this master arc frame for the wavelength calibration and to correct for spectral tilt \citep{pypeit:joss_pub}.  

We make several modifications to the default \textsc{PypeIt} wavelength solution. 
On the blue side, we use the \texttt{full\_template} method with a polynomial of order $n = 5$.  Arc spectra are cross-correlated against an archived template spectrum with an algorithm to reduce non-linearities \citep{pypeit:joss_pub}.  The resulting root mean square (RMS) of the residuals is $\sim 0.14 - 0.31$ pixels ($\sim 12.6 - 28.0$ km~s$^{-1}$).    

We alter the wavelength solution for the red frames as well, from the default \texttt{holy\_grail} method (where detected lines are matched with those expected from observed arc lamps) to the \texttt{full\_template} method.  Since \textsc{PypeIt} does not have an archived template spectrum for the 600/10 000 grating, we use \textsc{PypeIt}'s interactive routine to manually produce a wavelength solution, which we then apply to all of our red-side calibrations.  This has now been included in the \textsc{PypeIt} code base.  We find that the RMS of the residuals is similarly small, $\sim 0.095 - 0.424$ pixels ($\sim 4.5 - 20.3$ km~s$^{-1}$).

\subsubsection{Flexure correction}\label{sec:flexure}
The wavelength solution is further corrected for spectral flexure.  We do not use \textsc{PypeIt}'s default, single-pixel shift correction, as we find that a more complex, wavelength-dependent solution is required.  This is based on examination of sky emission lines. 

On the red side, we compare the extracted sky spectra to a Paranal sky model \citep{paranal1, paranal2}, provided in \textsc{PypeIt}.  We select ten strong sky emission lines across the wavelength range and measure the centroids of the corresponding lines in the data and the model, using a Gaussian fitting routine.  We find the differences between each set of centroids and fit a straight line to these deviations.  This linear function is interpolated over the entire red wavelength range and subtracted from the wavelengths to produce a final, corrected wavelength array on the red side.

\begin{figure*}
    \centering
    \includegraphics[width=\textwidth]{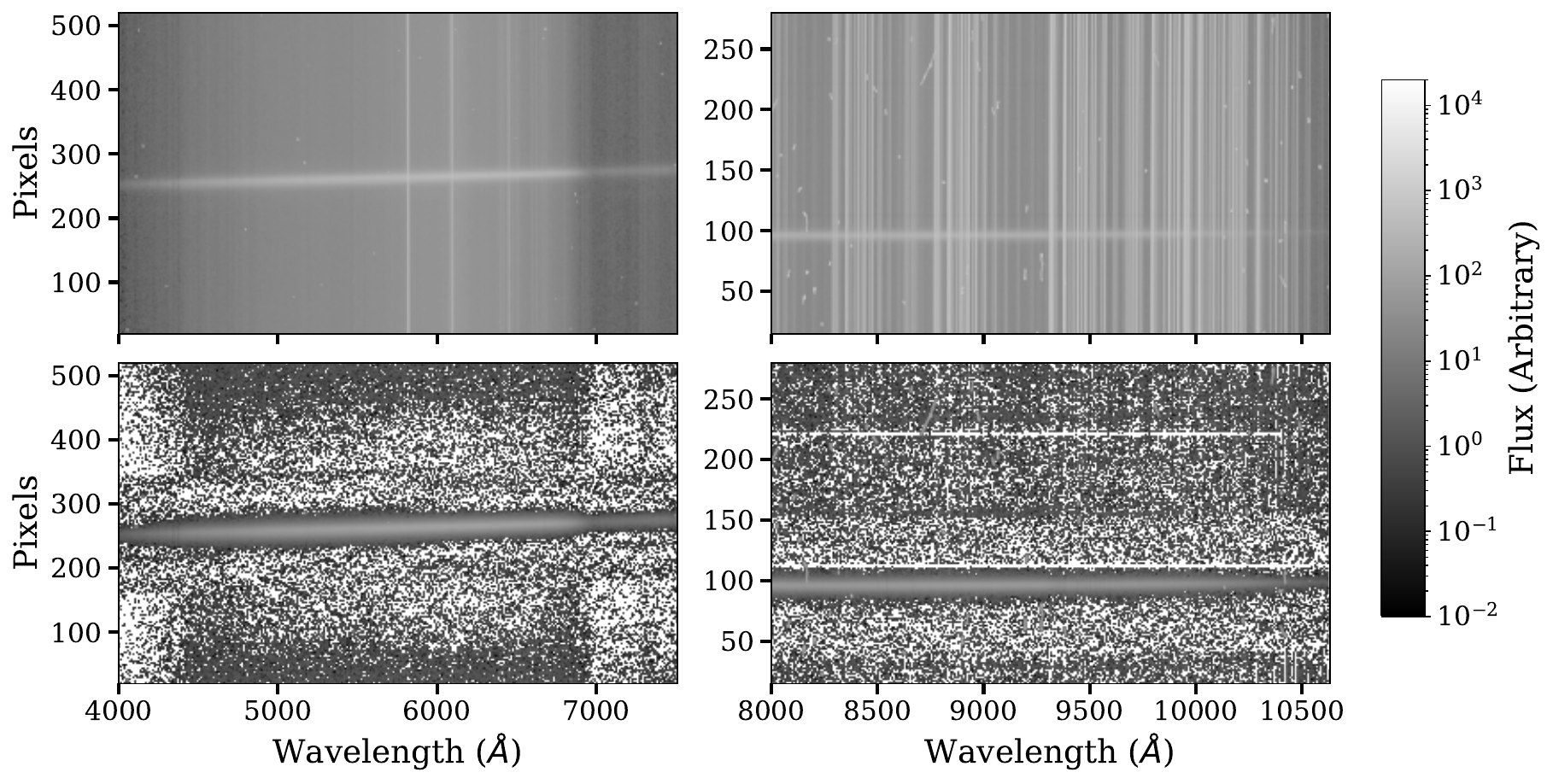}
    \caption{Examples of two-dimensional spectra from the B058 2016 data.  The left and right panels show the blue and red spectra, respectively, while the top and bottom panels show the spectra before and after sky subtraction.  The horizontal strips in each panel represent the object traces.  The bright vertical lines represent sky emission lines.  In the bottom panels, the sky lines are well-subtracted and largely not visible across the object traces.}
    \label{fig:paper_skysub}
\end{figure*}

We cannot effectively perform this procedure on the blue side due to a lack of strong sky emission lines, so we use a similar method as in \cite{van_Dokkum_2012}.  We create a template spectrum of the data with literature age and metallicity (see references in Table~\ref{tab:literature_vals}), using the \texttt{write\_a\_model} simple stellar population (SSP) framework in \textsc{alf}.  We split the template and our spectra into regions of $\sim 250$ \AA\ and compare these to measure the redshift in each region.  Assuming the flexure is a linear function of the redshifted template wavelength, we fit a straight line to these redshifts and use this to solve for coefficients in a flexure function.  We shift our blue wavelengths by this flexure function to produce a final, flexure-corrected wavelength solution.  For details, see Appendix~\ref{sec:flexure_appendix}.  

In Figure~\ref{fig:paper_centroids}, we show the remaining error in the wavelength solution after we apply the flexure correction.  To produce this plot, we redo the redshift and centroid measurements on the blue and red sides, respectively, after we apply the flexure correction.  The "errors" on the y-axes are the redshifts for the blue procedure and the centroid differences for the red procedure, respectively, converted to km~s$^{-1}$.  It is evident that the wavelength-dependence has been largely removed, as any remaining trend with wavelength is small compared with the random scatter around zero.  We also compare the flexure-corrected sky emission line wavelengths to the expected values and find that they are now consistent within measurement uncertainties.

\subsubsection{Sky subtraction}\label{sec:sky_sub}
We sky subtract the two-dimensional (2D) spectra prior to extracting the one-dimensional (1D) spectra.  \textsc{PypeIt} first executes a global sky subtraction, using a 2D b-spline algorithm \citep{Kelson_2003} .  Here, the background is well-sampled and accurately modeled by making use of information about camera distortions and spectral curvature.  This method is insensitive to bad pixels (e.g.\ cosmic rays, hot pixels, etc.), making the cleaning of such pixels simple after the sky subtraction is complete \citep{Kelson_2003}.  The sky model is then locally refined around the science target during object extraction \citep{pypeit:joss_pub}.  This is done by interpolating sky regions on either side of the object trace and fitting a weighted least-squares polynomial to the sky background at each wavelength, with weights inversely proportional to the variances of individual sky pixels \citep{Horne_1986}.

In Figure~\ref{fig:paper_skysub}, we show examples of 2D spectra from the 2016 B058 data.  The left and right panels show the blue and red spectra and the top and bottom panels show the spectra before and after sky subtraction, respectively.  Bright, horizontal strips represent the object traces while bright vertical lines represent sky emission lines.  In the top panels, many strong sky lines cross the object traces, particularly on the red side.  However, these are well-subtracted in the bottom panels, with a uniform background near the traces and very few lines visible across the traces themselves.

\subsubsection{Object extraction}\label{sec:object_extract}
To extract objects, we use the \cite{Horne_1986} algorithm.  In summary, the sky-subtracted image is summed over the pixels which include the object, with nonuniform pixel weights applied to minimize statistical noise and retain photometric accuracy.  The object profile is fitted with a polynomial.  This accurate characterization of the spatial profile across the wavelength range allows the algorithm to correct for cosmic rays (CRs), as a CR event results in a distortion of the object profile that can be easily recognized and masked.  For some of our bright sources, we find that this can lead to excessive masking of key spectral features ($\gtrsim 10\%$ of pixels, \citealt{pypeit:joss_pub}).  To avoid this, we increase the threshold for the CR rejection where necessary, until $\lesssim 10\%$ of pixels are masked and masked pixels are randomly scattered across the object trace.  For especially bad cases (i.e.\ where increasing the CR rejection threshold indefinitely does not reduce the masking), we suppress the masking entirely.

\begin{figure*}
    \centering
    \includegraphics[width=\textwidth]{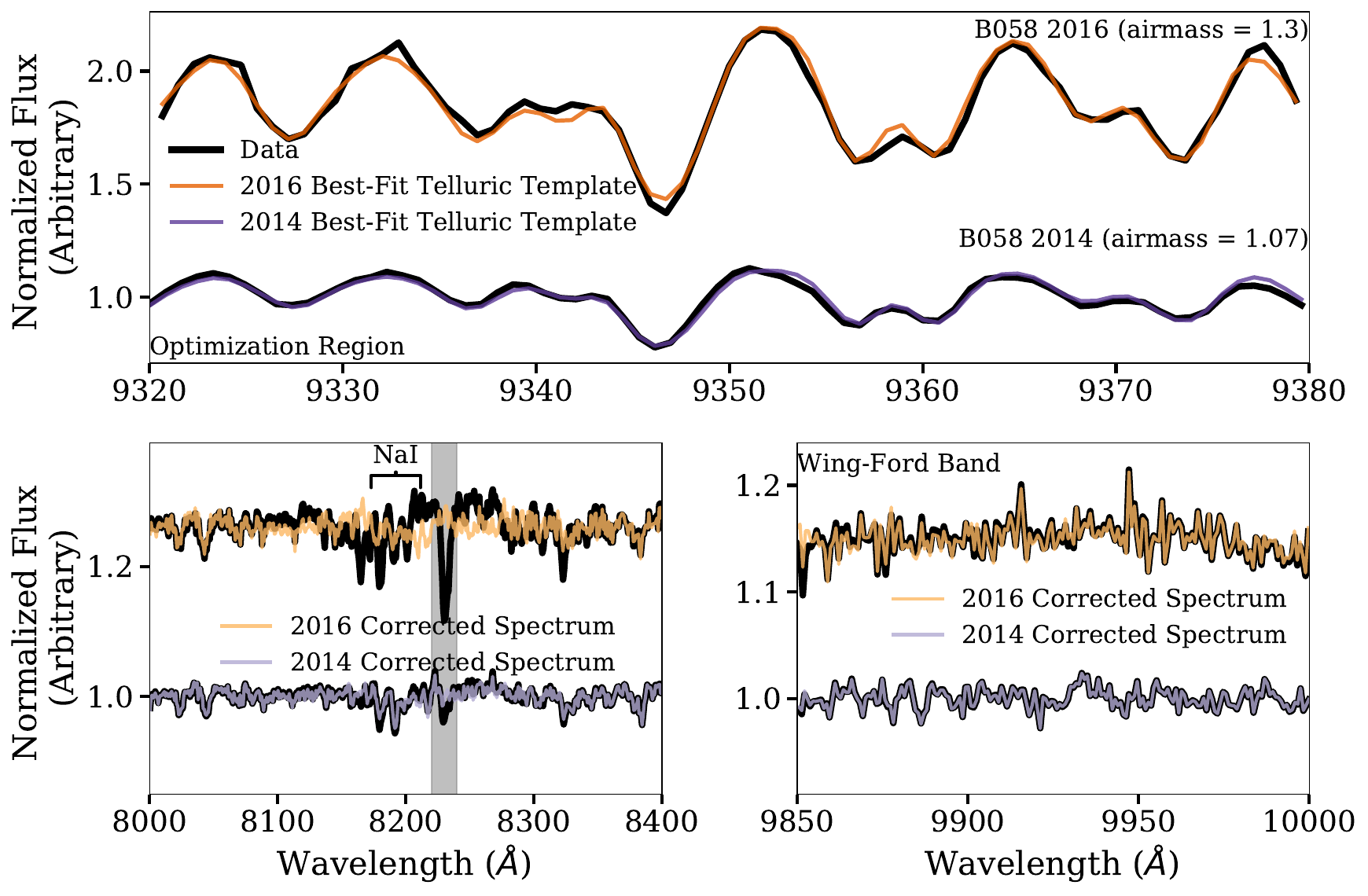}
    \caption{A demonstration of our telluric correction.  In the top panel, we show the normalized B058 2016 and B058 2014 spectra, prior to telluric correction (black lines), over the narrow wavelength range containing many strong telluric lines that we use to select the atmospheric template.  We compare each to the best-fitting template (dark orange for B058 2016/dark purple for B058 2014).  In the bottom panels, we show the spectra prior to telluric correction compared to the final, corrected spectra (light orange for B058 2016/light purple for B058 2014).  The left panel shows the spectra near the NaI spectral feature and the right panel shows them over the Wing--Ford band.  Each set of spectra is arbitrarily shifted vertically, for clarity.  In the left panel, the grey region indicates the unreliably-corrected H$_2$O feature, around which we inflate the spectral uncertainties and weight the fits to zero (see Section~\ref{sec:methods}).}
    \label{fig:paper_telluric}
\end{figure*}

\subsubsection{Flux calibration and co-addition}\label{sec:coadd}
We do not flux calibrate the data.  \textsc{alf} does not require spectra to be fluxed, as we subtract the continuum prior to fitting (see Section~\ref{sec:methods}).  

After the main reduction process, we use \textsc{PypeIt}'s external co-adding routine to combine all of the 1D spectra for each object.  This is done by optimally weighting each one-dimensional spectrum by its S/N at each pixel.  The spectra are also cleaned of CRs again at this stage, by scaling the fluxes using the root mean square of the squared S/N \citep{pypeit:joss_pub}.  This combination step leaves us with one spectrum for each of the blue and red sides of each object.

\begin{figure*}
    \centering
    \includegraphics[width=\textwidth]{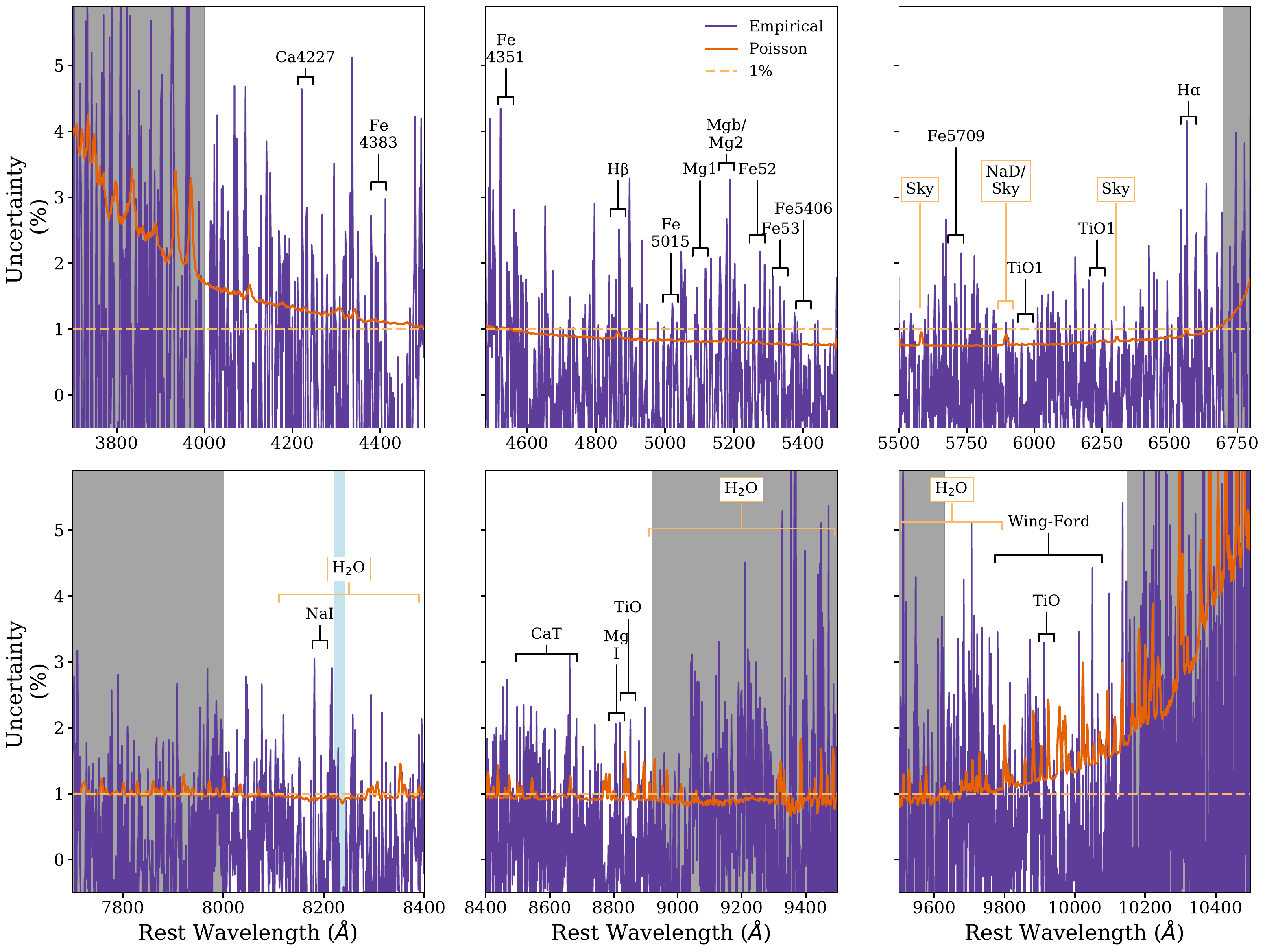}
    \caption{Quadrature sum of the Poisson uncertainties in the B058 2014 and B058 2016 spectra (dark orange) compared to the empirical uncertainties between the two dates (purple).  The dashed light orange line indicates 1\% uncertainty.  Prominent telluric and sky features are labelled in light orange and spectral features of interest are labelled in black.  The grey regions indicate areas of the spectra that we do not fit or use in our analysis.  The blue region indicates the area where we inflate the uncertainties and weight the fits to zero (see Section~\ref{sec:methods}).  In the areas of the spectra relevant for fitting and analysis, the empirical uncertainties are largely comparable to the Poisson uncertainties.  In areas where the empirical uncertainties are larger, we are still able to constrain the majority of them to $\lesssim 5$\%.}
    \label{fig:paper_MAD_B058}
\end{figure*}

\subsubsection{Telluric correction}\label{sec:telluric}
We perform telluric correction using a modified version of the method in \cite{van_Dokkum_2012}.  We scale a template spectrum to the observed atmospheric absorption, where the scaling is parameterized by Equation (2) in \cite{van_Dokkum_2012}.  We smooth each template from the Mauna Kea telluric grids included with \textsc{PypeIt}, originally produced via the Line-By-Line Radiative Transfer Model\footnote{\url{http://rtweb.aer.com/lblrtm.html}~.} \citep{Clough_2005, Gullikson_2014}, to the LRIS instrumental resolution and scale each of them.  We continuum-normalize the scaled template and target spectra by dividing them by a polynomial of order 4 over the region $9250 - 9650$ \AA.  We choose the best-fitting template by minimizing the $\chi^2$ of the difference between the scaled template and target spectra, over the optimization region $9320 - 9380$ \AA\ (which contains many strong atmospheric absorption lines and no strong galaxy absorption features).  This best-fitting template is used to divide the target spectrum to produce the final, corrected spectrum.  See \cite{van_Dokkum_2012} for more details.  We find that we are unable to reliably correct the telluric H$_2$O feature near $\sim 8200$ \AA\ for all objects.  This is problematic because it is close to a crucial NaI feature.  To account for this, we weight the fits around the H$_2$O feature to zero and inflate its uncertainties, so that it is effectively not considered in the fits.  

We demonstrate this procedure in Figure~\ref{fig:paper_telluric}.  At the top of each panel, we show the arbitrarily shifted, continuum-normalized B058 2016 spectrum prior to telluric correction, and at the bottom we show the same for the B058 2014 spectrum (black).  In the top panel, we compare each spectrum to the best-fitting template.  In the bottom panel, we compare each spectrum prior to telluric correction to the final, corrected spectrum, over regions with strong telluric and stellar absorption features (e.g.\ near NaI and the Wing--Ford band).  The grey region indicates the unreliably-corrected H$_2$O line.  Both sets of data are fitted well by the scaled template over all regions, with most atmospheric features removed.  

\begin{figure*}
    \centering
    \includegraphics[width=\textwidth]{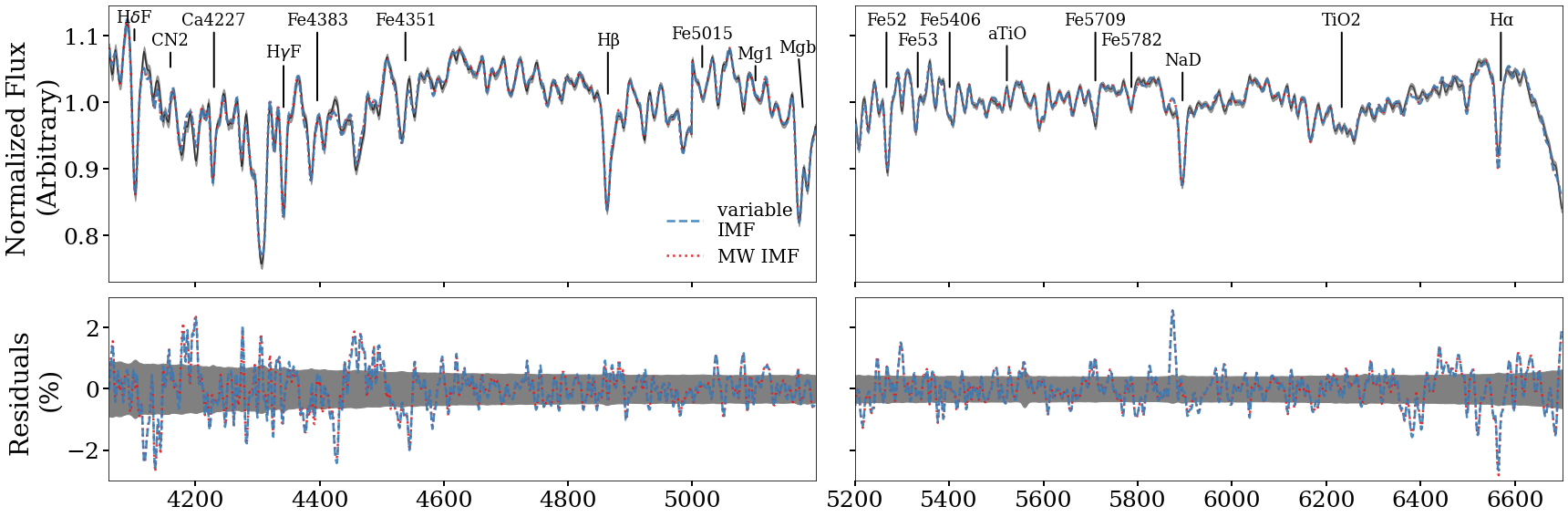}
    \includegraphics[width=\textwidth]{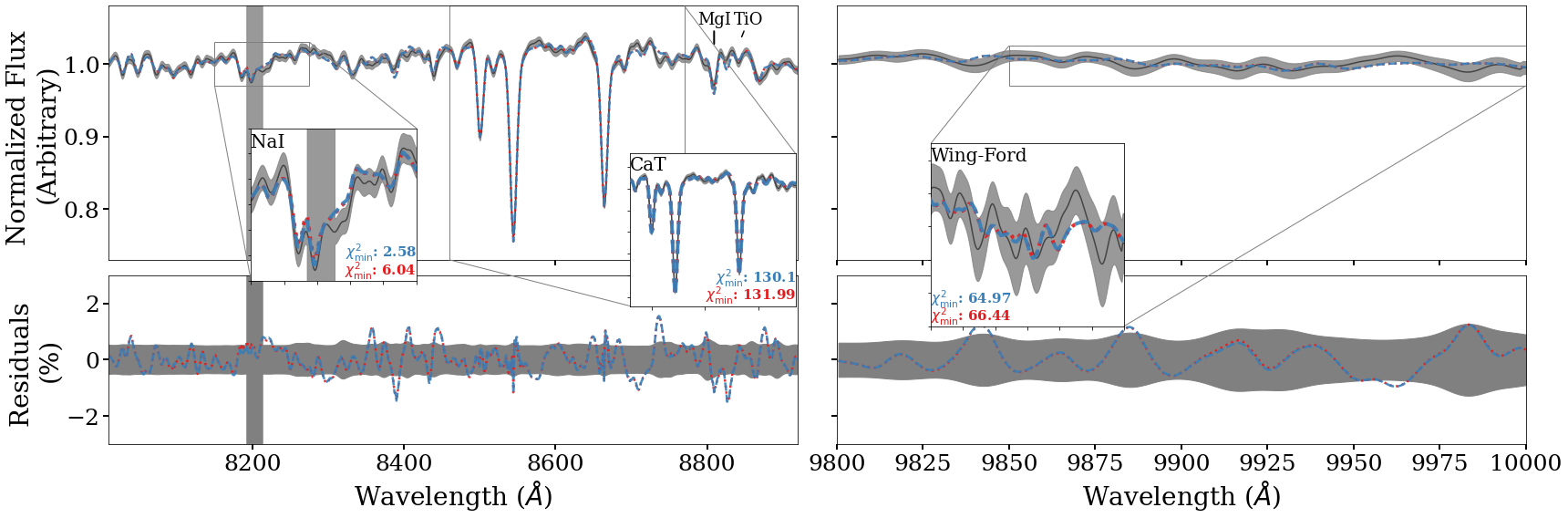}
    \caption{\textit{Upper panels}: The best-fit \textsc{alf} models for VUCD7 over the fitted wavelength regions.  The data are shown in black and the grey bands indicate the spectral uncertainties.  A fit allowing for a variable IMF is shown in blue and a fit where we fix a Kroupa IMF is shown in red.  Important IMF-sensitive features and the $\chi^2_{\mathrm{min}}$ over these features for each respective fit are shown in the inset panels.  This is similar to Figure (3) in \citet{Gu_2022}.  \textit{Lower panels}: The fit residuals.  They grey band indicates the uncertainty in the residuals for the variable IMF fit.}
    \label{fig:paper_fits_vucd7}
\end{figure*}

\begin{figure*}
    \centering
    \includegraphics[width=\textwidth]{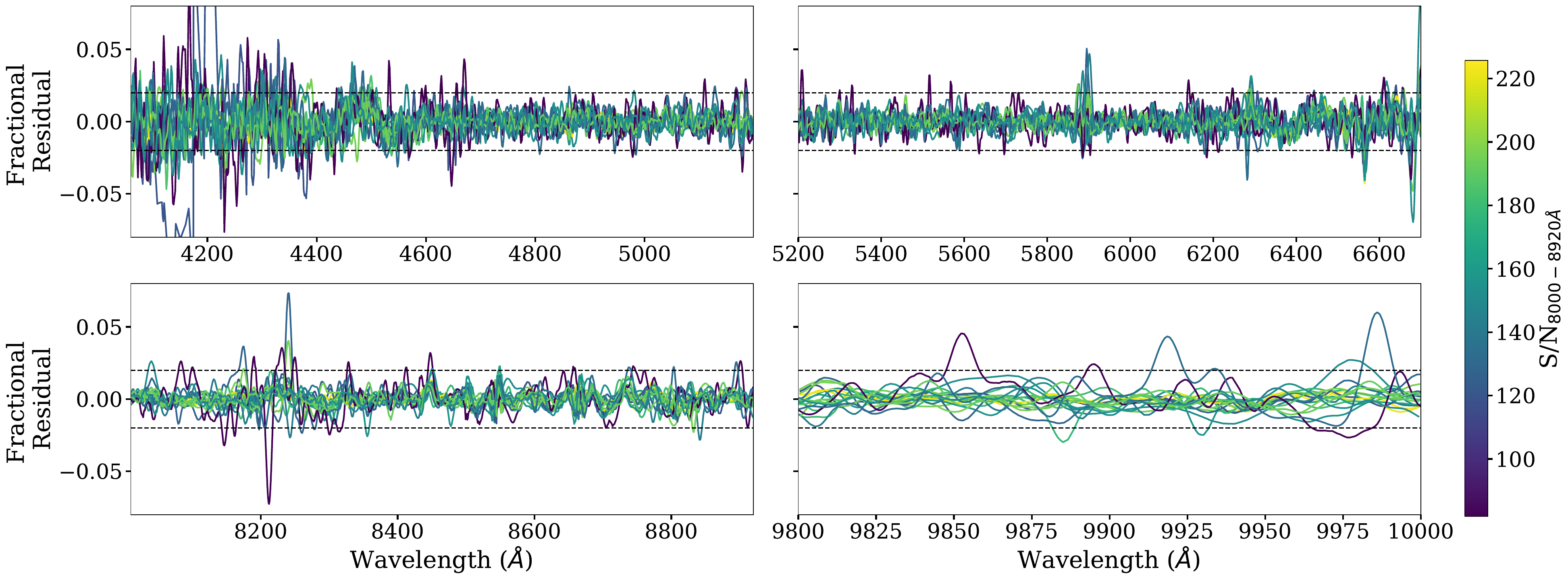}
    \caption{The fractional residuals for all of our objects for the fits where we allow for a variable IMF.  These are colour-coded by the median S/N across $8000-8920$ \AA.  This is similar to the top panel of Figure (4) in \citet{Gu_2022}.  The fractional residuals are $\lesssim 2$\% on average (horizontal dotted lines), with higher S/N objects having smaller fractional residuals.}  
    \label{fig:paper_residuals}
\end{figure*} 

\subsubsection{Data reduction quality}\label{sec:DR_QA}
We demonstrate the quality of our data reduction in Figure~\ref{fig:paper_MAD_B058}.  Here, we compare the quadrature sum of the Poisson statistical uncertainties of observations of B058 2014 and B058 2016 in orange, and the empirical uncertainty between the two objects (purple).  The Poisson statistical uncertainty represents the minimal uncertainties (from photon counting and data calibrations), while the empirical uncertainty includes systematic errors (e.g.\ noise from sky background and telluric absorption features remaining after correction).  The grey regions indicate areas of the spectra that we do not fit.  The blue region indicates the specially-treated H$_2$O line (see Section~\ref{sec:telluric}).  In most areas of the spectra that are important to our analysis, the empirical uncertainties are comparable to Poisson, and are close to $\sim 1\%$ (dashed light orange line) at most wavelengths.   For example, in the case of the NaD feature near $\sim 5900$ \AA, where a sky line contributes significant noise, the empirical uncertainty is comparable to the Poisson uncertainty at $\lesssim 2\%$.  Even at wavelengths where our empirical uncertainties are larger than the Poisson expectation, they are still $\lesssim 5\%$ across the majority of the wavelength range. 

There are certain areas to note.  For example, as discussed above, a sky line lies directly on top of the NaD feature near $\sim 5900$ \AA.  Even though this sky line contributes noise to the NaD feature, we are able to constrain the empirical uncertainty to be at a level of $\lesssim 2\%$.  Similarly, we are able to characterize the uncertainty across the entire spectrum.  Thus, Figure~\ref{fig:paper_MAD_B058} demonstrates that, given our data reduction process, it is possible to minimize the data systematics introduced by different observing conditions such that the empirical uncertainty is on par with the Poisson uncertainty.

\begin{table*}
	\centering
	\caption{Values of fitted parameters from \textsc{alf}.}
	\label{tab:fitparams}   
	\begin{tabular}{cccccc}
		\hline
		ID & Age (Gyr) & [Fe/H] & [Mg/Fe] & M/L$_V$ (2PL) & M/L$_V$ (MW)\\
		\hline
		B012	&	$11.67_{-0.57}^{+0.72}$	&	$-1.7_{-0.03}^{+0.03}$	&	$0.34_{-0.03}^{+0.03}$	&	$1.9_{-0.2}^{+0.21}$	&	$2.26_{-0.13}^{+0.12}$	\\
        B058 (2014, 2016)	&	$13.49_{-0.73}^{+0.37}$, $8.64_{-0.33}^{+0.35}$	&	$-1.02_{-0.01}^{+0.01}$, $-0.95_{-0.09}^{+0.02}$	&	$0.33_{-0.02}^{+0.02}$, $0.33_{-0.03}^{+0.04}$	&	$2.42_{-0.16}^{+0.17}$, $2.03_{-0.17}^{+0.27}$	&	$2.79_{-0.09}^{+0.04}$, $2.01_{-0.04}^{+0.05}$	\\
        B067	&	$13.45_{-0.57}^{+0.38}$	&	$-1.56_{-0.02}^{+0.02}$	&	$0.34_{-0.02}^{+0.03}$	&	$2.32_{-0.17}^{+0.2}$	&	$2.63_{-0.08}^{+0.06}$	\\
        B074	&	$12.07_{-0.58}^{+0.68}$	&	$-1.51_{-0.02}^{+0.02}$	&	$0.46_{-0.05}^{+0.05}$	&	$2.14_{-0.2}^{+0.24}$	&	$2.36_{-0.08}^{+0.09}$	\\
        B107	&	$8.71_{-0.86}^{+1.28}$	&	$-0.86_{-0.04}^{+0.03}$	&	$0.16_{-0.03}^{+0.03}$	&	$2.79_{-0.26}^{+0.41}$	&	$2.27_{-0.11}^{+0.21}$	\\
        B163	&	$11.84_{-0.72}^{+0.67}$	&	$-0.2_{-0.01}^{+0.01}$	&	$0.23_{-0.01}^{+0.01}$	&	$2.95_{-0.18}^{+0.25}$	&	$4.11_{-0.13}^{+0.12}$	\\
        B193	&	$13.46_{-0.38}^{+0.35}$	&	$-0.19_{-0.01}^{+0.01}$	&	$0.24_{-0.01}^{+0.01}$	&	$3.3_{-0.18}^{+0.25}$	&	$4.62_{-0.08}^{+0.07}$	\\
        B225	&	$10.66_{-0.39}^{+0.39}$	&	$-0.42_{-0.01}^{+0.01}$	&	$0.23_{-0.01}^{+0.01}$	&	$2.7_{-0.17}^{+0.27}$	&	$3.24_{-0.08}^{+0.07}$	\\
        B338	&	$13.88_{-0.33}^{+0.09}$	&	$-1.07_{-0.1}^{+0.02}$	&	$0.34_{-0.03}^{+0.02}$	&	$2.58_{-0.29}^{+0.36}$	&	$2.66_{-0.06}^{+0.15}$	\\
        B405	&	$11.41_{-0.34}^{+0.53}$	&	$-1.22_{-0.02}^{+0.02}$	&	$0.43_{-0.03}^{+0.03}$	&	$2.25_{-0.18}^{+0.19}$	&	$2.37_{-0.05}^{+0.07}$	\\
        G001	&	$8.9_{-0.28}^{+0.28}$	&	$-0.74_{-0.01}^{+0.01}$	&	$0.42_{-0.02}^{+0.02}$	&	$5.05_{-0.38}^{+0.34}$	&	$2.28_{-0.04}^{+0.06}$	\\
        M59-UCD3	&	$9.67_{-0.51}^{+0.33}$	&	$-0.02_{-0.01}^{+0.01}$	&	$0.19_{-0.01}^{+0.01}$	&	$4.54_{-0.91}^{+0.81}$	&	$4.11_{-0.14}^{+0.11}$	\\
        VUCD3	&	$13.66_{-0.51}^{+0.25}$	&	$-0.08_{-0.02}^{+0.02}$	&	$0.4_{-0.02}^{+0.02}$	&	$8.04_{-1.97}^{+1.59}$	&	$5.01_{-0.11}^{+0.07}$	\\
        VUCD4	&	$12.27_{-1.12}^{+0.9}$	&	$-1.01_{-0.02}^{+0.02}$	&	$0.46_{-0.04}^{+0.04}$	&	$2.07_{-0.17}^{+0.2}$	&	$2.52_{-0.12}^{+0.11}$	\\
        VUCD7	&	$13.86_{-0.23}^{+0.1}$	&	$-0.75_{-0.01}^{+0.01}$	&	$0.43_{-0.02}^{+0.01}$	&	$4.43_{-0.38}^{+0.36}$	&	$3.21_{-0.03}^{+0.02}$	\\
        NGC 4874	&	$11.6_{-1.4}^{+1.28}$	&	$-0.01_{-0.04}^{+0.04}$	&	$0.32_{-0.04}^{+0.04}$	&	$5.9_{-1.07}^{+1.31}$	&	$5.08_{-0.41}^{+0.36}$	\\
        NGC 4889	&	$13.77_{-0.39}^{+0.17}$	&	$0.08_{-0.02}^{+0.02}$	&	$0.25_{-0.02}^{+0.02}$	&	$15.89_{-2.0}^{+2.43}$	&	$5.9_{-0.12}^{+0.12}$	\\
		\hline
	\end{tabular}
\end{table*}

\section{Methods}\label{sec:methods}
To model our data and derive stellar parameters, we fit the spectrum of each object with the \textsc{Absorption Line Fitter} (\textsc{alf}), a full spectrum SPS model developed in \cite{CvD_2012a} and updated in \cite{Conroy_2018} with expanded stellar parameter coverage.  The empirical SSPs underpinning the \textsc{alf} models were created with the MIST isochrones \citep{MIST} and the Spectral Polynomial Interpolator (SPI, \citealt{IRTF})\footnote{\url{https://github.com/AlexaVillaume/SPI_Utils}~.}.  With SPI, we use the MILES optical stellar library \citep{Sanchez_Blazquez_2006}, the Extended IRTF stellar library (E-IRTF, \citealt{IRTF}), and a large sample of M-dwarf spectra \citep{Mann_2015} to create a data-driven model from which we can generate stellar spectra as a function of $T_{\mathrm{eff}}$, surface gravity, and metallicity.  

The empirical parameter space is set by the E-IRTF and \citet{Mann_2015} samples, which together span $-2.0\lesssim$ [Fe/H] $\lesssim 0.5$ and $3.5\lesssim\log{(T_{\mathrm{eff}})}\lesssim 3.9$.  To preserve the quality of interpolation at the edges of the empirical parameter space, we augment the training set with a theoretical stellar library (C3K, \citealt{MIST}).  The \textsc{alf} models allow for variable abundance patterns by differentially including theoretical element response functions.  We use the measured Mg abundances for the MILES stellar library stars from \citet{Milone_2011} to derive the [Mg/Fe] versus [Fe/H] relation in our model. 

In all cases we assume a stellar population comprised of two SSPs of different ages.  While an improvement over a single SSP, this may still not be sufficient to capture the extended star formation histories of UCDs and BCGs.  As shown by \citet{CvD_2012b}, degeneracies between age and IMF are small for old stellar populations ($>3$ Gyr).  Nonetheless, extension of \textsc{alf} to include a star formation history parameterization would be worthwhile for future work.

We parameterize the IMF as a double power-law (PL) with a break point at $m = 0.5 \mathrm{M}_\odot$, similar to the \cite{Kroupa_2001} IMF, and a fixed low-mass cutoff ($m_c$) at 0.08 $\mathrm{M}_\odot$.  Above 1.0 $\mathrm{M}_\odot$, the IMF slope is assumed to have the \cite{Salpeter} value of 2.35:
\begin{equation}\label{eq:2PL}
    \frac{dN}{dm} = \begin{cases}
    k_1m^{-\alpha_1}, &0.08 < m < 0.5 \\
    k_2m^{-\alpha_2}, &0.5 < m < 1.0 \\
    k_3m^{-2.3}, &m\geq1.0
    \end{cases}
\end{equation}
For a MW IMF, we use the \cite{Kroupa_2001} values of $\alpha_1 = 1.3$ and $\alpha_2 = 2.3$.  We note that this approach to full-spectrum fitting of old stellar populations is effective at identifying a bottom-heavy IMF ($\alpha_1>1.3$ and/or $\alpha_2>2.3$) due to the distinctive signature of low-mass dwarfs, but not necessarily a bottom-light IMF \citep{CvD_2012a} in which such dwarfs are underrepresented.  As these are old populations, there are also no strong constraints on the high-mass end of the IMF.

\textsc{alf} fits the spectra over pre-defined wavelength intervals.  It continuum-normalizes the target spectrum by multiplying it by a high order polynomial of order $n\equiv(\lambda_{\mathrm{max}} - \lambda_{\mathrm{min}})/100$\AA\ within each wavelength interval.  It uses a Fortran implementation of the Markov chain Monte Carlo algorithm \texttt{emcee} \citep{emcee} to 
sample the posteriors of 46 stellar parameters including redshift, $\sigma$, total metallicity, a two-component star formation history comprised of two bursts of star formation with free ages and a relative mass fraction, 18 elemental abundances, the contribution to the fraction of light at 1 $\mu$m by a hot star component, and two higher order terms of the line-of-sight velocity distribution.  It also fits for several systematic parameters to characterize observed errors.  For further details, see \cite{CvD_2012a} and \cite{Conroy_2018}.  We fit our data as in V17, using $512$ walkers, $25,000$ burn-in steps, and a $1,000$ step production run.  

The velocity dispersions of the CSSs are small compared with the intrinsic resolution of the \textsc{alf} models of $\sim 100$ km~s$^{-1}$.  To account for this, we broaden the spectra of the CSSs minimally, by 110 km~s$^{-1}$.  We use a modified version of the \textsc{Prospector} \texttt{smoothspec}\footnote{\url{https://prospect.readthedocs.io/en/latest/api/utils_api.html}~.} function, which uses Fast Fourier Transform convolutions to smooth the spectrum by a wavelength-dependent line-spread function.  Using a model spectral library, the function generates a model spectrum and then smoothes at each model generation step by the difference between the observed-frame instrumental resolution and the rest-frame library resolution.  This is calculated for the model systemic velocity, velocity dispersion, and the instrumental line-spread function parameters.  In this routine, it is assumed that the instrumental and library resolutions are approximated by a Gaussian at each wavelength, so the difference kernel can also be represented by a wavelength-dependent Gaussian.  Then, the model spectrum is resampled onto a space in which the kernel is not varying with wavelength.  See \cite{Prospector} for more details.  We broaden the blue and red sides individually so the discontinuity in the centre of the spectrum from the two LRIS detectors is not broadened into adjacent spectral features.  Prior to smoothing, we also mask unphysical artifacts for this reason.  We discuss some smoothing tests in Appendix~\ref{sec:smoothing}.

We also smooth the BCGs minimally, to 360 km~s$^{-1}$.  This is not necessary for the full spectrum fitting, as the velocity dispersion of the BCGs is much larger than that corresponding to the resolution of the \textsc{alf} models.  However, we also make use of \textsc{alf}'s index fitting feature (see Appendix~\ref{sec:indices}), which require the spectra to be smoothed.  Thus, for consistency, we smooth for the full spectrum fitting as well. 

In Figure~\ref{fig:paper_fits_vucd7}, we show the spectra and best-fit \textsc{alf} models for VUCD7 as an example.  The remaining objects are shown in Appendix~\ref{sec:other_fits} (available online).  These are similar to Figure (3) in \cite{Gu_2022}.  In the upper panels, we show the fully-reduced spectrum, along with fits where we allow for a variable IMF (blue) and where we fix a Kroupa IMF (red).  In the lower panels, we show the fit residuals.  The residual uncertainties for the variable IMF fit are indicated by grey bands.  We highlight key, IMF-sensitive absorption features in the inset panels.

\begin{figure*}
    \centering
    \includegraphics[width=\linewidth
    ]{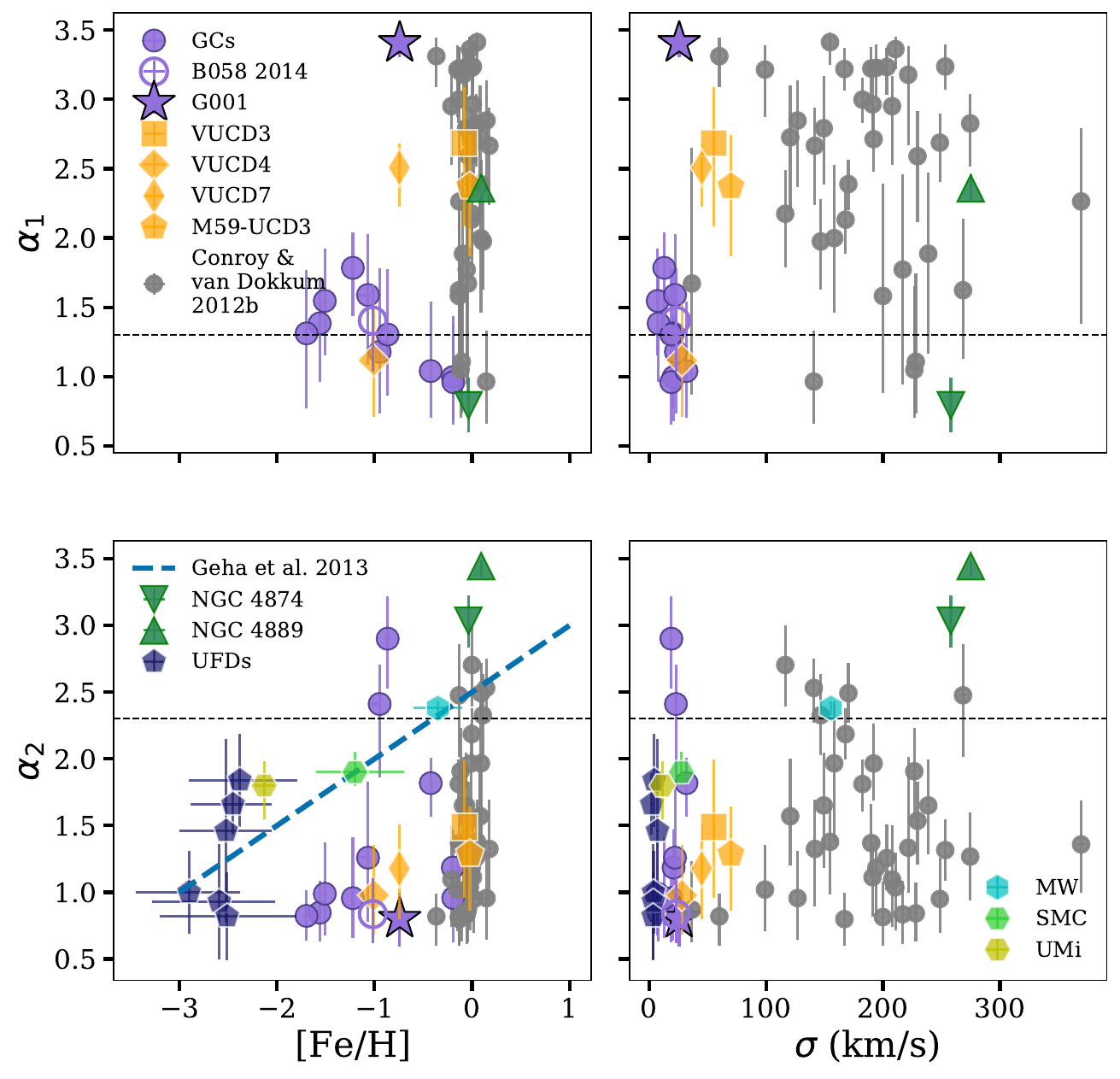}
    \caption{Comparison of IMF slope trends with metallicity and $\sigma$ from G13.  Our fitted low-mass slopes ($\alpha_1$) are in the top panel, which cannot be directly compared to G13 as their mass range differs.  We compare our intermediate-mass slopes ($\alpha_2$), which cover similar mass ranges, in the bottom panels.  We include individual points from Figure (5) in G13, including those representing the MW \citep{Bochanski_2010}, the SMC \citep{Kalirai_2013}, and UMi \citep{Wyse_2002}.  We also add the UFDs from the extended sample in \citet{Gennaro_2018}.  The re-fitted ETG data from \citet{CvD_2012b} are shown in grey.  In the left panels we plot our fitted [Fe/H] and on the right we plot literature $\sigma$ (see Table~\ref{tab:literature_vals})  In the bottom-left panel, we include the linear fit to metallicity defined in G13 (the \citet{Kroupa_2001} empirical relation plus a zero-point shift, dashed blue).  In each panel, the horizontal dashed line represents the corresponding slope for a Kroupa IMF.}
    \label{fig:paper_metallicity_trends}
\end{figure*}

We show the fractional fit residuals for all objects in Figure~\ref{fig:paper_residuals}, colour-coded by the median S/N across $8000-8920$ \AA.  This is similar to the top panel of Figure (4) in \cite{Gu_2022}.  In general, our objects are well-fitted by \textsc{alf}.  Most residuals are small ($\lesssim 2\%$).  

\section{Results}\label{sec:results}
We report our fitted [Fe/H], age of the dominant (older) stellar population, [Mg/Fe], and $V$-band (M/L)$_*$ with a variable IMF  in Table~\ref{tab:fitparams}.  For the same fit parameters, the code also provides the $V$-band (M/L)$_*$ that would be obtained with a fixed, Kroupa IMF, and we list this as M/L$_V$ (MW).  We also fit several other stellar parameters, which are listed in Tables~\ref{tab:appendix_fitparams1}, \ref{tab:appendix_fitparams2}, and \ref{tab:appendix_fitparams3} (available online). 

\begin{figure*}
    \centering
    \includegraphics[width=\textwidth]{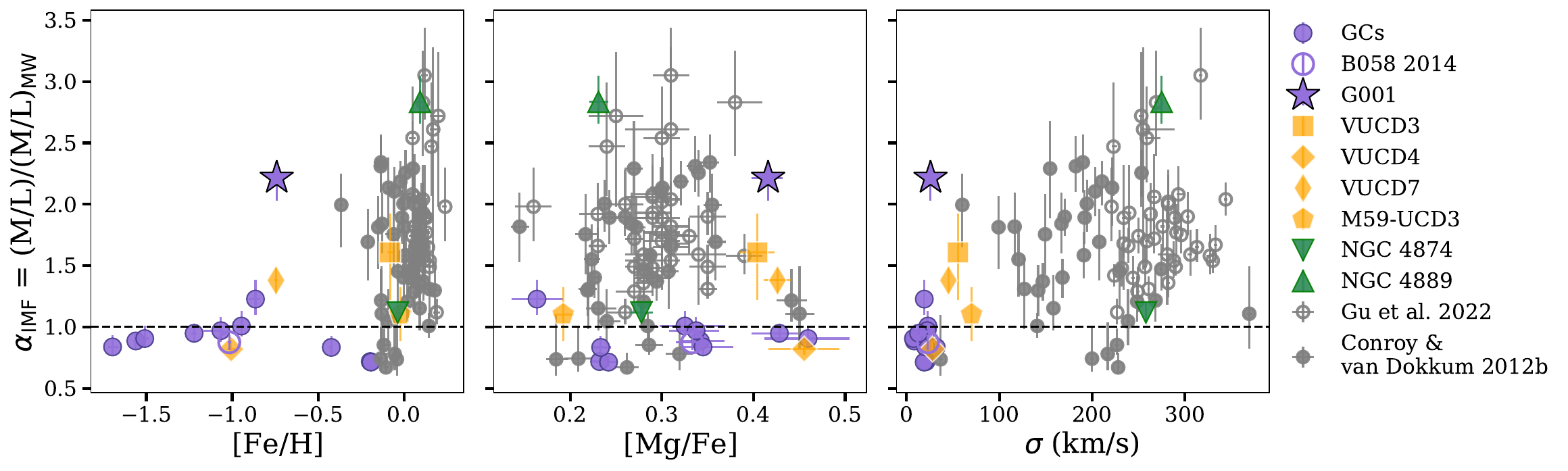}
    \caption{The IMF mismatch parameter, $\alpha_{\mathrm{IMF}}$, as a function of stellar parameters, including fitted [Fe/H] (left), fitted [Mg/Fe] (middle), and literature velocity dispersion ($\sigma$, right).  This is similar to Figure 3 in V17.  The GCs are shown as purple points.  G001 (star) is highlighted to show its deviation from the main body of GC results.  The newly reduced B058 data from V17 (open symbol) is highlighted to demonstrate the consistency between data taken on different dates and reduced differently.  The UCDs are shown in orange and the BCGs are shown in green.  We also include the sample of ETG values from \citet{CvD_2012b} (data re-fitted here) as filled grey circles and from \citet{Gu_2022} as open grey circles.  The dashed line represents the value of $\alpha_{\mathrm{IMF}}$ for a Kroupa IMF.}
    \label{fig:paper_fig3}
\end{figure*}

In Figure~\ref{fig:paper_metallicity_trends}, we show the slopes of the fitted 2PL IMF (Equation~\ref{eq:2PL}) as a function of our fitted [Fe/H] (top panels) and literature $\sigma$ (bottom panels, see Table~\ref{tab:literature_vals}).  
In each panel, the horizontal dashed line represents the respective slope for a \cite{Kroupa_2001} IMF.  We also show the re-fitted ETGs from \cite{CvD_2012b} as grey circles.  It is difficult to determine a clear picture from the IMF slopes, but we measure non-Kroupa IMF shapes for many objects.  There is also a diversity between and within different stellar systems that does not appear to be simply explained as a metallicity- or $\sigma$-dependence. 

In Figure~\ref{fig:paper_fig3} we present our results in terms of the IMF mismatch parameter ($\alpha_{\mathrm{IMF}}$, \citealt{Treu_2010}), which is the ratio between the best-fit (M/L)$_*$ with a variable IMF to (M/L)$_*$ for the same physical parameters but assuming a \cite{Kroupa_2001} IMF.  We compare our sample, the re-fitted \cite{CvD_2012b} ETGs, and the \cite{Gu_2022} ETGs to the fitted [Fe/H] (left), [Mg/Fe] (middle), and literature $\sigma$ (right).  Figure~\ref{fig:paper_fig3} is comparable to Figure (3) in V17.

In general, we confirm the results of V17, that CSSs do not follow the [Fe/H] and [Mg/Fe] trends established for ETGs (e.g.\ \citealt{CvD_2012b, Martin-Navarro_2015, van_Dokkum_2017}).  With the exception of G001, GCs have IMFs consistent with Kroupa or even slightly bottom-light (within $2\sigma$) over a large range of [Fe/H] and [Mg/Fe].  This is also apparent in the Kroupa-like $\alpha_1$ for these objects in Figure~\ref{fig:paper_metallicity_trends}.  UCDs are more bottom-heavy (VUCD3, VUCD7) or Kroupa-like (M59-UCD3, VUCD4).  Regarding the BCGs, while they are both consistent with the main body of ETG results from the literature, NGC 4874 is more similar to a Kroupa IMF (similar to the result found in Z17), while NGC 4889 is bottom-heavy.  This latter result may be inconsistent with Z17; we revisit this in Section~\ref{sec-BCGs}.  

\begin{figure*}
    \centering
    \includegraphics[width=\textwidth]{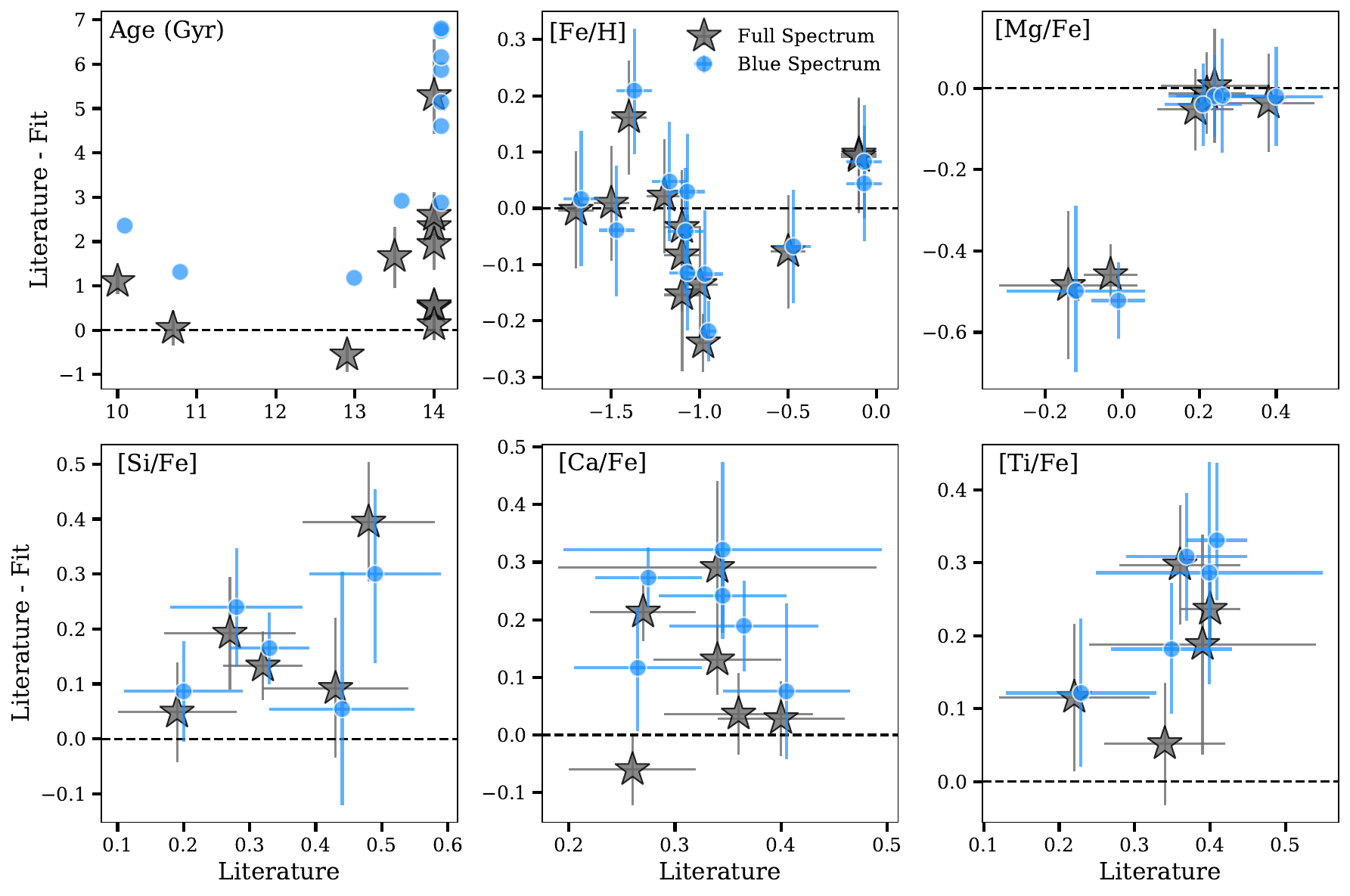}
    \caption{A comparison between parameters derived from the full spectrum fits of the integrated light and literature parameters for the GCs (see Table~\ref{tab:literature_vals}).  We compare fits over the blue wavelength range only (blue points) versus fits over the entire spectrum (grey stars), as described in Section~\ref{sec:systematics}.  The blue points have been given a small $x$-offset so that the error bars do not completely overlap with the grey ones.  The consistency that we see between the blue and full wavelength range fits indicates that our sky subtraction and telluric correction are robust.}
    \label{fig:paper_lit_fit_compare}
\end{figure*}

\begin{table*}
	\centering
	\caption{Literature parameters.  A dashed entry indicates no literature value available.  References: (b) \citet{Caldwell_2011}; (c) \citet{Sakari_2021}; (d) \citet{Janz_2016}; (e) From NED; (f) \citet{Paudel_2010}; (g) \citet{Strader_2011}; (h) \citet{Janz_2016}; (i) \citet{Forbes_2014}; (j) \citet{Veale_2018}; (k) \citet{Sandoval_2015}; (l) \citet{Mieske_2013}; (m) \citet{Greene_2019}; (n) \citet{Sakari_2016}; (o) \citet{Colucci_2014}; (p) \citet{Ahn_2018}; (q) \citet{Ahn_2017}; (r) $\sigma$-based mass from \citet{Dullo_2019}; (s) \citet{McConnell_2012}.}
	\label{tab:literature_vals}
  \begin{threeparttable}    
	\begin{tabular}{cccccccccccc}
		\hline
		ID & Age (Gyr) & $\sigma$ (km~s$^{-1}$) & [Fe/H] & [Mg/Fe] & [Si/Fe] & [Ca/Fe] & [Ti/Fe] & Black Hole \\
		\hline
		B012	&	14\tnote{b}	&	20.4\tnote{g}	&	$-1.7\pm0.1$\tnote{b}	&	$-0.14\pm0.18$\tnote{n}	&	$0.43\pm0.11$\tnote{n}	&	$0.4\pm0.06$\tnote{o}	&	-	&	unconfirmed	\\
        B058\tnote{a}	&	14\tnote{b}	&	23\tnote{g}	&	$-1.1\pm0.1$\tnote{b}	&	-	&	-	&	-	&	-	&	unconfirmed	\\
        B067	&	14\tnote{b}	&	8.2\tnote{g}	&	$-1.4\pm0.1$\tnote{b}	&	-	&	-	&	-	&	-	&	unconfirmed	\\
        B074	&	14\tnote{b}	&	7.4\tnote{g}	&	$-1.5\pm0.1$\tnote{b}	&	-	&	-	&	-	&	-	&	unconfirmed	\\
        B107	&	14\tnote{b}	&	19.1\tnote{g}	&	$-1\pm0.1$\tnote{b}	&	-	&	-	&	-	&	-	&	unconfirmed	\\
        B163\tnote{a}	&	13.5\tnote{b}	&	21\tnote{g}	&	$-0.1\pm0.1$\tnote{b}	&	$0.22\pm0.1$\tnote{n}	&	$0.19\pm0.09$\tnote{n}	&	$0.27\pm0.05$\tnote{n}	&	$0.22\pm0.1$\tnote{n}	&	unconfirmed	\\
        B193\tnote{a}	&	12.9\tnote{b}	&	19\tnote{g}	&	$-0.1\pm0.1$\tnote{b}	&	$0.19\pm0.1$\tnote{n}	&	$0.27\pm0.1$\tnote{n}	&	$0.34\pm0.15$\tnote{n}	&	$0.36\pm0.08$\tnote{n}	&	unconfirmed	\\
        B225	&	10.7\tnote{b}	&	32\tnote{g}	&	$-0.5\pm0.1$\tnote{b}	&	$0.24\pm0.14$\tnote{n}	&	$0.32\pm0.06$\tnote{n}	&	$0.34\pm0.06$\tnote{n}	&	$0.39\pm0.15$\tnote{n}	&	unconfirmed	\\
        B338	&	14\tnote{b}	&	22.2\tnote{g}	&	$-1.1\pm0.1$\tnote{b}	&	-	&	-	&	-	&	-	&	unconfirmed	\\
        B405	&	14\tnote{b}	&	13\tnote{g}	&	$-1.2\pm0.1$\tnote{b}	&	$-0.03\pm0.07$\tnote{o}	&	$0.48\pm0.1$\tnote{o}	&	$0.26\pm0.06$\tnote{o}	&	$0.4\pm0.04$\tnote{o}	&	unconfirmed	\\
        G001	&	10\tnote{c}	&	25.8\tnote{g}	&	$-0.98\pm0.05$\tnote{c}	&	$0.38\pm0.12$\tnote{o}	&	-	&	$0.36\pm0.07$\tnote{c}	&	$0.34\pm0.08$\tnote{o}	&	debated	\\
        M59-UCD3\tnote{a}	&	11.7\tnote{d}	&	70\tnote{h}	&	$-0.01\pm0.04$\tnote{k}	&	-	&	-	&	-	&	-	&	$4.2^{+2.1}_{-1.7}\times10^6$ M$_\odot$\tnote{p}	\\
        VUCD3	&	13.754\tnote{e}	&	55.2\tnote{i}	&	-0.011\tnote{l}	&	-	&	-	&	-	&	-	&	$4.4^{+2.5}_{-3.0}\times10^6$ M$_\odot$\tnote{q}	\\
        VUCD4	&	11.9\tnote{f}	&	28.1\tnote{i}	&	-1.1\tnote{l}	&	-	&	-	&	-	&	-	&	unconfirmed	\\
        VUCD7	&	10.7\tnote{f}	&	45.1\tnote{i}	&	-0.66\tnote{l}	&	-	&	-	&	-	&	-	&	unconfirmed	\\
        NGC 4874	&	13.453\tnote{e}	&	258\tnote{j}	&	$-0.13\pm0.06$\tnote{m}	&	-	&	-	&	-	&	-	&	$8.1^{+7.9}_{-7.9}\times10^8\mathrm{ M}_\odot$\tnote{r}	\\
        NGC 4889	&	13.486\tnote{e}	&	275\tnote{j}	&	$0.01\pm0.02$\tnote{m}	&	-	&	-	&	-	&	-	&	$2.1^{+1.6}_{-1.6}\times10^{10}\mathrm{ M}_\odot$\tnote{s}	\\
		\hline
	\end{tabular}
  \begin{tablenotes}
      \item[a] Analyzed in \cite{Villaume_2017}.
  \end{tablenotes}
  \end{threeparttable}
\end{table*}

In Figure~\ref{fig:paper_lit_fit_compare} we compare our fitted values for age, [Fe/H], [Mg/Fe], [Si/Fe], [Ca/Fe], and [Ti/Fe] to literature values (see Table~\ref{tab:literature_vals}).  Here we consider only the GCs, to compare our results with the wide body of literature ages and abundances, which are not as robust for the UCDs.  Note that the literature abundances are derived using a variety of methods, including Lick indices, spectral line synthesis, and full spectrum SPS fitting.  Our fitted [Fe/H] are largely consistent with  literature results.  The discrepancies between our fitted ages and the literature can primarily be attributed to the fact that the H$\beta$ feature, which is a strong indicator of age, is highly degenerate with the blue horizontal branch in GCs \citep{Rich_2005, Ocvirk_2010}.  As such, it is difficult to derive accurate ages from \textsc{alf}.  We see some significant differences for the other abundances ([Mg/Fe], [Si/Fe], [Ca/Fe], [Ti/Fe]), which may stem from aperture effects or different abundance measurement methods.  The agreement between our abundances and those from \cite{Colucci_2014} and \cite{Sakari_2016} is particularly poor, despite the fact that the abundances for objects common between these two studies are largely in agreement, albeit with some minor offsets \citep{Sakari_2016}.  However, we note several key differences that may affect this comparison.  In particular, our S/N is significantly higher.  The wavelength ranges are also different.  \cite{Colucci_2014} examined $3800 - 8300$ \AA, excluding important features like CaT and MgI and TiO near $8800$ \AA.  Meanwhile, \cite{Sakari_2016} studied $15, 100 - 16, 900$ \AA, which is significantly redder than our spectra and excludes optical features that are also crucial for our measurements.  Finally, we note that all of these studies use different abundance determination methods (\citealt{Colucci_2014} measured the equivalent widths of spectral features of interest and \citealt{Sakari_2016} synthesized spectra of regions around each line of interest), and that ours is the only one to take advantage of the entire spectrum.  In light of these differences, we emphasize that this comparison is not intended to demonstrate the precision of our results, but rather to indicate the similarities and differences between different methods.

\begin{figure}
    \centering
    \includegraphics[width=\columnwidth]{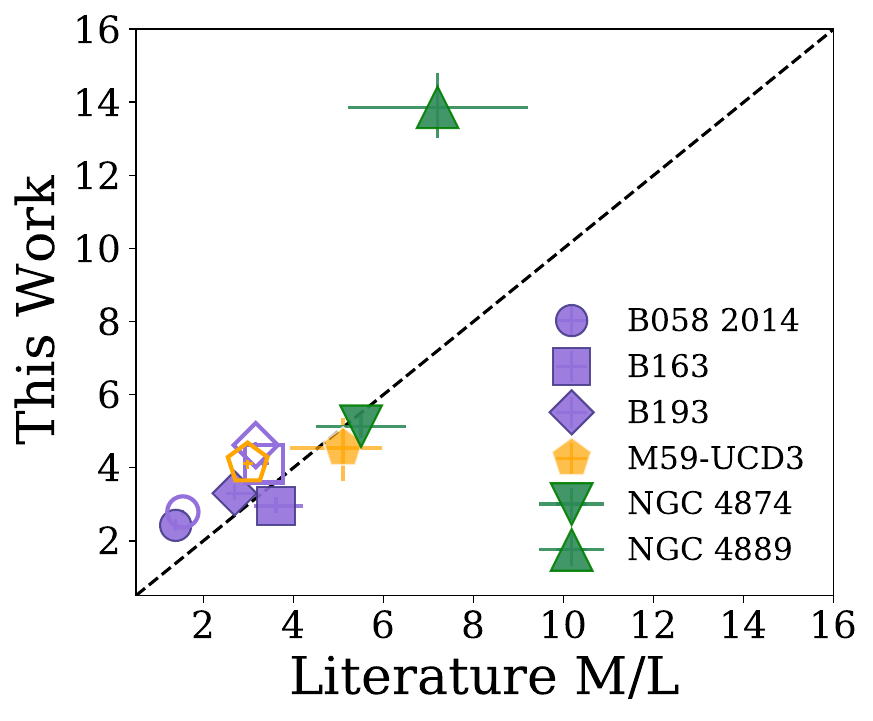}
    \caption{Comparison between M/L derived in this study vs. V17 and Z17.  Filled symbols represent M/L measured allowing for a variable IMF, while open symbols fix the IMF at the Kroupa value.  The GCs (B058, B163, and B193) are shown as purple points, M59-UCD3 is shown as an orange pentagon, and the BCGs are shown as green triangles.  For the V17 points, we compare M/L in the $V$-band (used throughout this study).  To be consistent with Z17, we re-derive M/L in the $r$-band for the BCGs and plot them here.}
    \label{fig:paper_V17_compare}
\end{figure}

Finally, we compare the $V$-band (M/L)$_*$ ratio for objects common between this study and V17 (i.e.\ B058, B163, B193, and M59-UCD3), as well as those from Z17 (NGC 4874 and NGC 4889) in Figure~\ref{fig:paper_V17_compare}.  For the V17 objects, we show our $V$-band (M/L)$_*$ from \textsc{alf} for the data taken in 2014 that we have re-reduced here on the $y$-axis (reported in Table~\ref{tab:fitparams}), compared to the results in V17 for the same data on the $x$-axis (reported in their Table 1).  For the BCGs, we re-derive (M/L)$_*$ in the $r$-band and compare to those that Z17 compute using SSP models.  The filled symbols allow for a variable IMF while the open symbols have a fixed Kroupa IMF.  In general, our results are consistent with V17, with minor differences that do not change their conclusions or affect ours.  The most significant difference is for M59-UCD3, where we find $\alpha_{\mathrm{IMF}}=1.1\pm 0.2$, compared with $\alpha_{\mathrm{IMF}}=1.7\pm 0.4$ in V17 (however the variable-solution (M/L)$_*$ is very similar).  Although we also find evidence that the IMF is bottom-heavy ($\alpha_1>1.3$ in Figure~\ref{fig:paper_metallicity_trends}), our best-fit solution has a larger age and lower metallicity than V17, which leads to a larger (M/L)$_{V, \mathrm{MW}}$ and hence lower $\alpha_{\mathrm{IMF}}$.  The reason for this difference is not clear; we have confirmed that differences in the data are consistent at the $<2$\% level except for a different treatment of the continuum.  In any case, both measurements are consistent with a modestly bottom-heavy IMF.  For NGC 4889 we find a (M/L)$_*$ that is significantly larger than that of Z17.  We discuss this further in Section~\ref{sec-BCGs}.

\section{Discussion}\label{sec:discussion}
\subsection{The relationship between IMF and metallicity}\label{sec:metallicity_trend}
Our results shown in Figures~\ref{fig:paper_metallicity_trends} and ~\ref{fig:paper_fig3} confirm the conclusion in V17, that GCs and UCDs show less variation in IMF shape compared to ETGs at similar metallicities.  This is primarily driven by the GCs: all of the GCs apart from G001 have an IMF that is Kroupa-like at low stellar masses and, thus, an $\alpha_{\mathrm{IMF}}\approx 1$ over $\sim1.5$ dex in [Fe/H].

G001 is intriguing because the mechanism of its formation remains highly debated, due to its unusual abundance patterns and unique location in M31's outer halo \citep{Meylan_2001, Sakari_2021}.  It has been argued to host an intermediate-mass black hole \citep{Gebhardt_2002, Gebhardt_2005}. It could have originated as a nuclear star cluster \citep{Baumgardt_2003}, or it may be a GC with a large spread in Fe \citep{Meylan_2001, Nardiello_2019}. Our evidence for a bottom-heavy IMF further supports a different origin from most massive GCs.  In combination with the UCD measurements, this also suggests that the IMF may be probing differences in the SF mechanisms between stellar systems with a stripped nucleus versus massive GC origin.  In particular, high-mass UCDs are considered to be stripped nuclei while the lower-mass population may be comprised of a combination of stripped nuclei and massive star clusters \citep{Mieske_2006, Brodie_2011, Chilingarian_2011, Norris_2011, DaRocha_2011, Mieske_2013, Pfeffer_2014, Pfeffer_2016, Voggel_2019}.  Differentiating UCD formation pathways has been challenging, but perhaps it is possible to link the IMF to their origins.  Further study is needed to explore this idea.

Recently, \citet{B+23} measured the IMF for 120 MW GCs using resolved star counts, and concluded that they are bottom-light compared with Kroupa.  In the mass range $0.4\lesssim M_*/M_\odot\lesssim 1$ they find a slope of $\sim -1.65\pm 0.2$, somewhat steeper than the $\alpha_2$ values we measure in Figure~\ref{fig:paper_metallicity_trends}.  At lower masses they find an even flatter slope of $-0.3$, significantly flatter than our results.   We note that the integrated-light approach is effective at identifying bottom-heavy IMFs, due to the presence of strong dwarf-sensitive features  (e.g.\ \citealt{CvD_2012b, Villaume_2017, Newman_2017, Conroy_2017}).  However, it is more challenging to distinguish bottom-light IMFs when these features are weak.    

We compare these results with the comparable, commonly-used, metallicity-dependent IMF relationship from \cite{Geha_2013} (G13).  
G13 use resolved star counts, and are sensitive to IMF variations in the range $0.5~\mathrm{ M}_\odot-0.8~\mathrm{ M}_\odot$, thus most closely related to our intermediate-mass slopes $\alpha_2$ from Equation~(\ref{eq:2PL}). In the bottom panels of Figure~\ref{fig:paper_metallicity_trends} we compare our measurements directly with those for the MW, Small Magellanic Cloud, and Ursa Minor from Figure (5) in G13 (hexagons), and with those from Figure 5 in \cite{Gennaro_2018} (an extension of G13, blue pentagons).  We also plot the empirical relationship from G13 with [Fe/H] (dashed blue, relationship is from \cite{Kroupa_2001} with a zero-point shift applied to fit their data).  We find that the G13 relation does not represent well the low-metallicity compact stellar systems in our sample, which show substantial variation, with many of them lying well below the relation.     We also note that the recent results of \citet{B+23} for MW GCs range from $0\lesssim\alpha_2<2.0$ with little dependence on metallicity.  Furthermore, even within galaxies, several IFU-based studies found that metallicity alone is not able to explain observed two-dimensional IMF variations in massive ETGs (e.g. \citealt{Parikh_2018, Sarzi_2018, Martin-Navarro_2019, Zhou_2019, Barbosa_2019, Martin-Navarro_2021}).   It may therefore be an important oversimplification to incorporate a purely metallicity-dependent IMF prescription in models of SF (e.g.\ \citealt{Dopcke_2013, Hopkins_2013, Clauwens_2016, Gutcke_2019, Chon_2021, Prgomet_2021, Sharda_2022}).  For example, \cite{Tanvir_2023} performed radiation-magnetohydrodynamic simulations and found that surface density is the more important driver of IMF variations compared to metallicity.

\subsection{Comparisons with dynamical mass measurements}\label{sec:dynamics}
\begin{figure*}
    \centering
    \includegraphics[width=\textwidth]{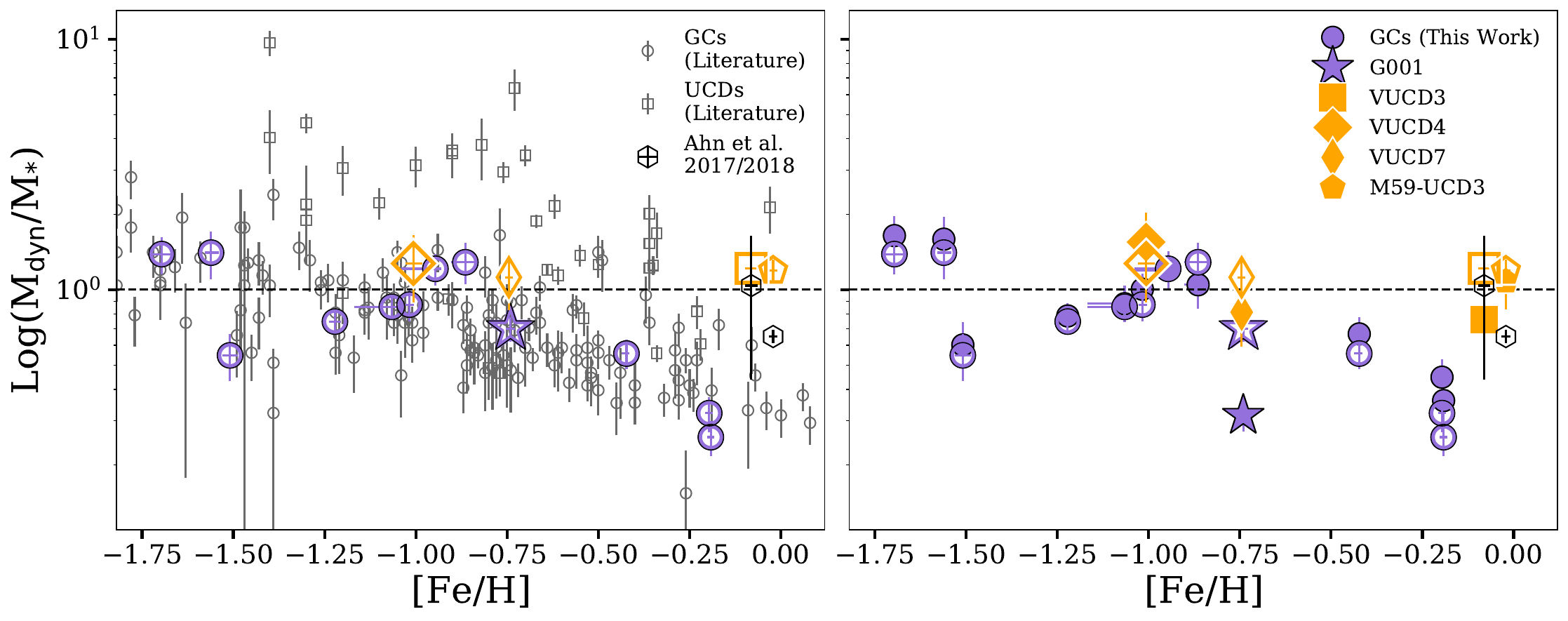}
    \caption{Logarithmic M$_{\mathrm{dyn}}/$M$_*$ vs. [Fe/H] for CSSs in the metallicity range spanned by our sample.  Our measurements are highlighted with symbols and colours similar to Figure~\ref{fig:paper_fig3}.  The dashed line in each panel indicates where M$_{\mathrm{dyn}} =$ M$_*$.  \textit{Left panel}: Stellar masses computed assuming a fixed, Kroupa IMF.  Dynamical masses and [Fe/H] are taken from \citet[][VUCDs]{Mieske_2013}, \citet[][M59-UCD3]{Liu_2015}, \citet[][GCs]{Strader_2011} and  \citet[][the remainder]{Norris_2014}. 
    Open black symbols include a correction for a SMBH contribution to the dynamical mass, based on the SMBH masses from \citet[][VUCD3]{Ahn_2017} and  \citet[][M59-UCD3]{Ahn_2018}. 
    \textit{Right panel}: Solid points show the results for objects in our sample when fitted with a variable IMF.  Open symbols are repeated from the left panel for comparison, and assume a Kroupa IMF.
    }
    \label{fig:paper_dynamical_compare}
\end{figure*}
Assumptions about the IMF impact inferences on the total stellar mass of a stellar system measured from photometry, which can be compared directly with dynamical estimates ($M_{\rm dyn}$).  This is illustrated in Figure~\ref{fig:paper_dynamical_compare}, where we compare $M_{\mathrm{dyn}}/M_*$ for our sample with other GCs and UCDs from the literature.  In the left panel we show our results when $M_*$ is derived assuming a Kroupa IMF, which can be compared directly with the published results for other systems.  Two of the UCDs in our sample have a central supermassive black hole (SMBH) which contributes to their dynamical mass.  We show the effect of correcting for the SMBH mass in both panels, by reducing the dynamical mass by the amount determined in \cite{Ahn_2017} and \cite{Ahn_2018}, shown as open black symbols. 

We confirm a general trend for a decreasing ratio of $M_{\mathrm{dyn}}/M_*$ with increasing metallicity for the GCs in our sample, as also seen by \citet{Strader_2011}.  Numerous effects have been suggested to explain this trend for the M31 GCs, including dynamical effects which remove low-mass stars (e.g.\ a metallicity-dependent spread in mass segregation, \citealt{Shanahan_2015}; metallicity- or density-dependent relaxation times, \citealt{Bianchini_2017}; or dynamical evolution,  \citealt{Bianchini_2017, Dalgleish_2020}) and differences in GC stellar populations (e.g.\ the age--metallicity relation, \citealt{Haghi_2017}; or a metallicity-dependent $\alpha$-enhancement, \citealt{Baumgardt_2020}). The four UCDs in our sample all have $M_{\mathrm{dyn}}/M_*\approx 1$, in contrast with the large values measured for many other UCDs \citep[e.g.][]{Hasegan_2005, Mieske_2008, Dabringhausen_2008, Frank_2011, Strader_2013}.

Allowing the IMF parameters to vary yields the results shown as the filled points in the right panel of Figure~\ref{fig:paper_dynamical_compare}.  In general, the changes are small.  For the two UCDs with bottom-heavy IMFs, the resulting reduction in $M_{\mathrm{dyn}}/M_*$ is of comparable magnitude to the contribution from SMBHs.

For the three most metal-rich GCs in our sample, the non-Kroupa IMF leads to a smaller stellar mass, bringing them closer to the dynamical mass, though they are still significantly larger.  This is as expected from the fact that we find the GCs have a near-constant $\alpha_{\mathrm{IMF}}$ across all metallicities (Figure~\ref{fig:paper_fig3}).  As noted earlier, however, integrated light measurements are not sensitive to the bottom-light IMFs that have been found from integrated star counts for many GCs \citep{B+23}.  Systematic differences in important stellar population parameters (e.g.\ age, $\alpha$-element abundances) are not likely to be responsible for this trend, since all of our GCs are old and IMF shape is independent of $\alpha$-element abundance (middle panel of Figure~\ref{fig:paper_fig3}).  Furthermore, it may be difficult to explain the trend as due to some dynamical process that removes low-mass stars in a metallicity-dependent manner, as these should be captured by our low-mass IMF measurements.  A possible solution is a metallicity-dependent remnant retention prescription \citep[e.g.][]{Belczynski_2010, Sippel_2012, Morscher_2015, Zonoozi_2016,Mahani_2021}.  \textsc{alf} currently uses the metallicity-independent prescription in \cite{Renzini_1993}.  Removing all remnants from our fits indeed results in $M_*$ that is in much better agreement with $M_{\mathrm{dyn}}$ for the metal-rich GCs; however, this is an unphysical extreme.  Finally, it is interesting that the lowest metallicity GCs in our sample show evidence for $M_*<M_{\rm dyn}$, which could be indirect evidence for a dark matter component (e.g.\ \citealt{Trenti_2015, Penarrubia_2017, Wirth_2020, Carlberg_2022, Errani_2022}).  

\subsection{The IMF in the cores of BCGs}\label{sec-BCGs}
We include the two Coma BCGs in our sample, NGC 4874 and NGC 4889, previously studied by Z17.  For NGC 4874 we confirm the result of Z17, that the IMF in the core is consistent with a Kroupa parameterization.  However,  for NGC 4889 we find evidence for a bottom-heavy IMF, in contrast to Z17.  The difference is unlikely to be due to aperture effects.  As the spectral extraction is weighted toward pixels with the highest S/N, our effective aperture within the longslit is $\sim 0.7''\times4.12''$  for NGC 4889 and $\sim 0.7''\times3.51''$  for NGC 4874, corresponding to $\sim <0.07R_\mathrm{e}$ for NGC 4874 and $\sim <0.11R_\mathrm{e}$ for NGC 4889 (Table~\ref{tab:sample}).  This is somewhat smaller than the aperture of Z17 ($0.2R_\mathrm{e}$, see their Table 1), but both are within the region where bottom-heavy IMF signatures are typically observed \citep{van_Dokkum_2017}.  Moreover, our data for NGC 4874 are {\it more} centrally concentrated than for NGC 4889, so this is unlikely to explain the discrepancy with Z17 for the latter object.  We note that our spectra have comparable or larger S/N per \AA\ compared with Z17, but cover a much wider wavelength range.  Our full-spectrum fitting approach  extends over 4000--10000 \AA, while Z17 examined line indices over $\sim 8140-9970$\AA.  In Appendix~\ref{sec:indices} we demonstrate that the use of the full optical spectrum results in both a systematic shift in $\alpha_{\rm IMF}$ as well as increased precision, relative to line index fitting.  In particular,
our bluer wavelength coverage allows better constraints on key element abundances that can be degenerate with the IMF parameters, as shown in \citet{CvD_2012a}.  

Our analysis therefore suggests that the cores of NGC 4874 and NGC 4889 indeed have different IMFs.  This difference may be due to different assembly or merger histories, since the BCGs likely originate from separate clusters that have since merged  \citep{Gu_2018}.  For example, \cite{Nipoti_2020} found that dry mergers tend to make the $\alpha_{\mathrm{IMF}}$ profile in an ETG shallower due to mixing with stellar populations with more bottom-light IMFs and the destruction of IMF gradients. 

\begin{figure*}
    \centering
    \includegraphics[width=\textwidth]{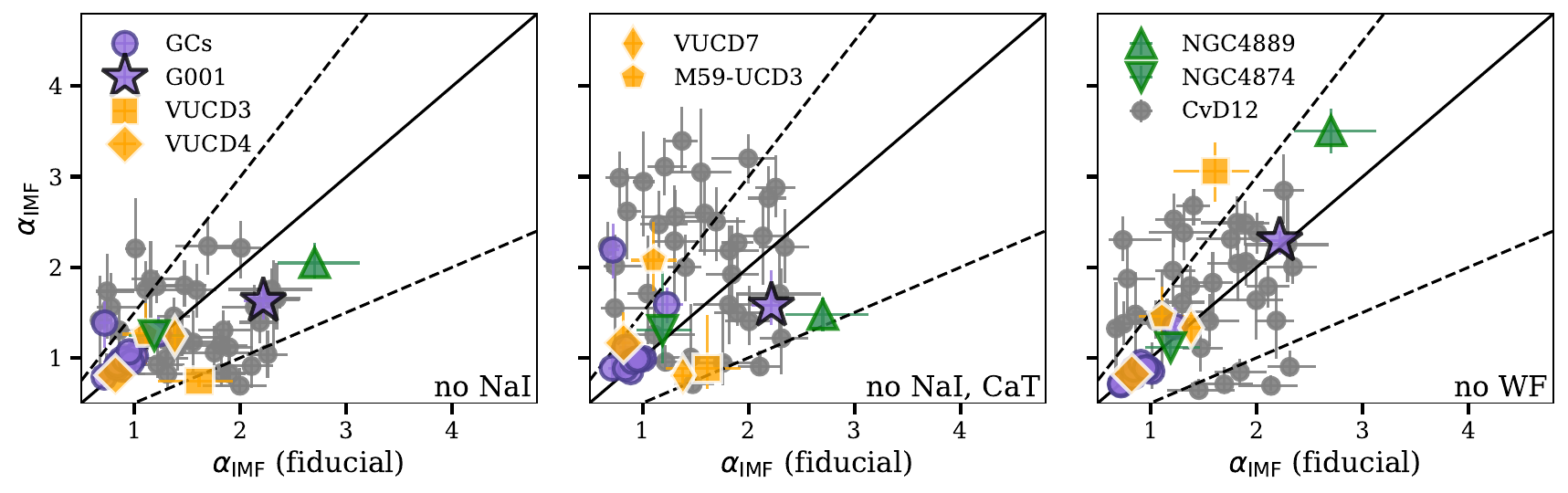}
    \caption{$\alpha_{\mathrm{IMF}}$ measured from fits where we exclude different IMF-sensitive features compared to $\alpha_{\mathrm{IMF}}$ from our fiducial, full-spectrum fits.  In the left panel, we exclude NaI, in the middle we exclude NaI and CaT, and on the right we exclude the Wing--Ford band.  We refit the \citet{CvD_2012b} data, shown as grey points.  The colours and symbols for the objects in our sample are similar to previous plots.  The solid black line shows the one-to-one relationship and the dashed lines show slope changes of $\pm50$\%.  This is similar to the top panel of Figure (12) in \citet{CvD_2012b}.}
    \label{fig:paper_CvD_fig12}
\end{figure*}

\subsection{Systematic effects and model assumptions}\label{sec:systematics}
Precise and accurate measurements of the IMF from integrated spectroscopy require a careful understanding of systematic uncertainties, as illustrated in part by observed discrepancies between different studies of the same objects  \citep[e.g.][]{Smith_2014, Lyubenova_2016}.  We have taken care to minimize object-to-object systematics by fitting a range of objects observed with the same instrument set-up and fitted with the same spectral fitting procedure.  Here we explore the impact of several choices that potentially influence the results.

In Figure~\ref{fig:paper_lit_fit_compare} we compare ages and abundances fitted in two different ways.  The first is the standard fit to the whole spectrum, using the "full fit" mode of \textsc{alf} that was used throughout the paper.  Alternatively, we fit only the blue wavelength range with \textsc{alf}'s "simple fit" mode, which fits for a single age, metallicity, velocity, $\sigma$, and the abundances of C, N, O, Mg, Si, Ca, Ti, and Na.  For the abundance parameters, the more restricted fits are generally consistent with the full spectrum fits.  Significant differences are seen in the ages, as expected since the the full mode fits for a two-component age.  The general consistency demonstrates that the differences in data reduction for the blue and red spectra, and differences in intrinsic resolution, do not have a significant impact on the results. 

A recent burst of SF or excess AGB/RGB stars could bias our measurements, since the light would be dominated by these populations \citep{Schiavon_2002_47Tuc, Girardi_2010, Strader_2011}.  We include a two-component stellar population in our fits and find that the younger component is always negligible ($<$0.01\%).  We also do not see trends between [Ba/Fe] and [Mg/Fe] that might be expected if there were an excess AGB/RGB population \citep{Conroy_2013_basr}.

Finally, we consider the sensitivity of our IMF measurements to specific spectral features that are known to be IMF- or abundance-sensitive.  In Figure~\ref{fig:paper_CvD_fig12} we show how $\alpha_{\mathrm{IMF}}$ varies as different IMF-sensitive features are excluded, similar to the test shown in Figure (12) in \cite{CvD_2012b}.  Here, we re-fit both\footnote{We note some minor differences between the fits to the different data sets.  For our data, to exclude the IMF-sensitive features, we mask and interpolate over them prior to smoothing, and exclude these regions from the wavelengths over which we fit.  We cannot do this for the \cite{CvD_2012b} data, so we simply exclude the regions from the fitted wavelength regions.  We note that the \cite{CvD_2012b} data also covers a slightly different wavelength range, namely that they do not have the NaD feature.} our data and the data from \cite{CvD_2012b}.  
In general there is a correlation, though with considerable scatter.  However, our overall conclusions about IMF trends do not depend on the scale of this scatter. Notably, our conclusion that $\alpha_{\rm IMF}>1$ for NGC 4889 still holds, even when important surface gravity-sensitive features are excluded.  Secondly, our result that $\alpha_{\rm IMF}\approx 1$ is independent of metallicity for most GCs in the sample is also robust to the exclusion of these features.  Qualitatively, and similar to findings in \cite{CvD_2012b}, an increasing trend is always present, no matter which set of features is considered.  This demonstrates that our fundamental results are not influenced by any one feature.  As found in \cite{CvD_2012b}, there are offsets in the slopes of the correlations between $\alpha_{\mathrm{IMF}}$ derived using different sets of features.  As they discussed, this indicates that different features prefer different M/L values, suggesting some model systematics remain.  For example, these features are sensitive to different masses, hinting that a more flexible IMF shape may be needed to fully explain the data.

It is also informative to consider the fits over these features, as shown for NGC 4874 and NGC 4889 in the insets of Figures~\ref{fig:paper_fits_ngc4874} and \ref{fig:paper_fits_ngc4889}, respectively (and for other objects in Section~\ref{sec:methods} and Appendix~\ref{sec:other_fits}).  For each spectrum, we show a fit allowing for a variable IMF (blue line) and a fit where we fix the IMF to the \cite{Kroupa_2001} MW value (red line).  We also compute the $\chi^2_{\mathrm{min}}$ for the fits in the region around each feature.  For NGC 4889, differences between the (preferred) bottom-heavy IMF and the Kroupa IMF fits are most evident in the NaI line.  Though the differences are small, the fact that the fixed-Kroupa model is unable to reproduce the depth of this line is an example of why the additional model flexibility is required. 

\section{Summary and Conclusions}\label{sec:conclusion}
There is now strong evidence that the IMF is not universal in the local Universe, but varies among different stellar systems.  While there is evidence that the IMF depends on either stellar metallicity, and/or galaxy velocity dispersion $\sigma$, this is based on observations of primarily metal-rich, early-type galaxies, representing a narrow region of mass-metallicity-density parameter space.  To improve on this, we expand the parameter space of variable IMF measurements by examining a diverse sample of extragalactic objects, including compact stellar systems (M31 GCs and Virgo cluster UCDs) and brightest cluster galaxies from the Coma cluster (Figure~\ref{fig:paper_sample}).  

Our main results, reflected in Figures~\ref{fig:paper_fig3} and \ref{fig:paper_dynamical_compare}, are summarized here:
\begin{itemize}
    \item GCs (excluding G001) have IMFs consistent with that of the MW or slightly bottom-light over a wide range of [Fe/H] and [Mg/Fe].  
    \item While some of the UCDs in our sample are also consistent with a Kroupa IMF, others show evidence for a bottom-heavy form. 
    \item For the two Coma BCGs, NGC 4874 has an IMF similar to that of the MW, while NGC 4889 has evidence for a bottom-heavy IMF.
    \item As we find most of the GCs in our sample to be consistent with a Kroupa IMF, there is little impact on the derived stellar mass by allowing the IMF to vary in the fit.  Although the changes go in the direction of reconciling stellar masses with dynamical masses for metal rich systems, it does not remove the trend noted by \citet{Strader_2011}.
\end{itemize}

In general, we find that the IMF shape varies among the stellar systems in our sample, in a way that is not directly linked to either metallicity or $\sigma$.  In particular, we confirm the conclusion in V17 that the IMFs of CSSs show less variation (generally consistent with Kroupa), compared with ETGs at similar metallicity.  This indicates that the form of the metallicity-dependence of the IMF may have to be reconsidered over this expanded parameter space.  The lack of IMF variation in CSSs is a potential way forward to better understand GCs and UCDs, and star formation more broadly.

\section*{Acknowledgements}

We acknowledge and are grateful to have the opportunity to work on the land on which the University of Waterloo operates.  The Waterloo, Kitchener, and Cambridge campuses of the University of Waterloo are situated on the Haldimand Tract, land granted to the Haudenosaunee of the Six Nations of the Grand River, and are within the territory of the Neutral, Anishinaabe, and Haudenosaunee peoples.  We also wish to recognize and acknowledge the very significant cultural role and reverence that the summit of Maunakea has always had within the Indigenous Hawaiian community.  We are most fortunate to have the opportunity to conduct observations from this mountain.  Finally, CMC is grateful to have had the opportunity to conduct the majority of this research on the land which we now call Toronto.  For thousands of years it has been the traditional land of the Huron-Wendat, the Seneca, and the Mississaugas of the Credit River, and is home to many Indigenous peoples from across Turtle Island.

We thank the anonymous referee for a useful report that improved this manuscript.  We would like to thank J. Xavier Prochaska, Joseph Hennawi, and the rest of the \textsc{PypeIt} development team for helpful insights into troubleshooting and improving our data reduction.  Thank you also to Vincent Hénault-Brunet, Michael Hudson, and James Taylor for useful conversations.  This research was enabled in part by support provided by the facilities of the Shared Hierarchical Academic Research Computing Network (SHARCNET: \hyperlink{www.sharcnet.ca}{www.sharcnet.ca}), Compute/Calcul Canada and the Digital Research Alliance of Canada (\hyperlink{https://alliancecan.ca/}{https://alliancecan.ca/}), and supported by an NSERC Discovery Grant (MLB).  The data presented herein were obtained at the W. M. Keck Observatory, which is operated as a scientific partnership among the California Institute of Technology, the University of California, and the National Aeronautics and Space Administration. The Observatory was made possible by the generous financial support of the W. M. Keck Foundation.  This research has made use of the NASA/IPAC Extragalactic Database (NED), which is operated by the Jet Propulsion Laboratory, California Institute of Technology, under contract with the National Aeronautics and Space.  This research has also made use of the SIMBAD database, operated at CDS, Strasbourg, France Administration.

\section*{Data Availability}
The Keck LRIS spectra used here are publicly available via the Keck Observatory Archive.  Other data products will be made available upon reasonable request.



\bibliographystyle{mnras}
\bibliography{main} 




\appendix

\section{Flexure Correction Methodology}\label{sec:flexure_appendix}
Here we describe our blue-side flexure correction in detail.  As discussed in Section~\ref{sec:flexure}, we create a template spectrum of the data with literature age and metallicity, using the \texttt{write\_a\_model} simple stellar population framework in \textsc{alf}.  Letting $\lambda_t$ be the template wavelength and $\lambda_o$ be the observed wavelength of the data, we see that the observed wavelength will be equal to the redshifted template wavelength, plus some wavelength-dependent function of flexure, $\delta(\lambda)$:
\begin{equation}\label{eq:lambda_o}
    \lambda_o = \lambda_t(1 + z) + \delta(\lambda).
\end{equation}
Here, we assume $\delta(\lambda)$ is a linear function of the redshifted template wavelength:
\begin{equation}\label{eq:delta_lambda}
    \delta(\lambda) = b\lambda_t(1 + z) + a
\end{equation}
Thus, our goal is to recover $\delta(\lambda)$ by disentangling both $z$ and $\delta(\lambda)$ from $\lambda_o$, by completing the following:
\begin{enumerate}
    \item Divide the data into regions of $\sim 250$ \AA\ (similar to \citealt{van_Dokkum_2012}) and continuum-normalize the observed regions.
    \item Divide the template into similar regions, by taking the central wavelength, $\lambda_c$, in each of the bins in step (i) and use this to define regions of the same length, plus an overhang of $\sim 20$ \AA\ on either side to account for wavelength shift from redshift and spectral flexure.  Continuum-normalize the template regions.
    \item Compare the corresponding observed and template regions and measure the redshift of each region using a routine based on the \textsc{astropy specutils} \texttt{template\_redshift}\footnote{\url{https://specutils.readthedocs.io/en/stable/api/specutils.analysis.template_redshift.html}~.} function.  This measured redshift will be different from the assumed literature redshift as a result of the spectral flexure. 
    \item Multiply the measured redshift in each chunk by the de-redshifted $\lambda_c$ (where we assume that the literature value of $z$ in SIMBAD is correct), such that the measured redshifts are in the rest-frame.  Fit a straight line to these measured redshifts:
    \begin{equation}\label{eq:lambda_error}
        \lambda_{\mathrm{error}} = c\lambda_c + d
    \end{equation}
    \item Use Equation~\ref{eq:lambda_error} to solve for the coefficients $a$ and $b$ in  Equation~\ref{eq:delta_lambda}.  We start from an expression for the difference between the observed and template wavelength,as 
    \begin{equation}\label{eq:Delta_lambda}
        \lambda_o - \lambda_t = \Delta\lambda = \lambda_t[(1 + z)(1 + b) - 1] + a.
    \end{equation}
    This is equivalent to $\lambda_{\mathrm{error}}$ in Equation~\ref{eq:lambda_error}, with $c = (1 + z)(1 + b) - 1$ and $d = a$.  Thus  
    \begin{equation}\label{eq:ab_coeffs}
        a = d \hspace{0.5cm} \& \hspace{0.5cm} b = \frac{c - z}{1 + z}.
    \end{equation}
    \item Create a "correction factor", which represents the flexure $\delta(\lambda)$, by rewriting Equation~\ref{eq:delta_lambda} in terms of $\lambda_o$:
    \begin{equation}\label{eq:flexure_function}
        \delta(\lambda) = \frac{b\lambda_0 + a}{1 + b}
    \end{equation}
    with $a$ and $b$ given by Equation~\ref{eq:ab_coeffs}.  We subtract this from the wavelength array to produce a final, flexure-corrected wavelength solution.
\end{enumerate}

\section{Smoothing}\label{sec:smoothing}
\begin{figure*}
    \centering
    \includegraphics[width=\textwidth]{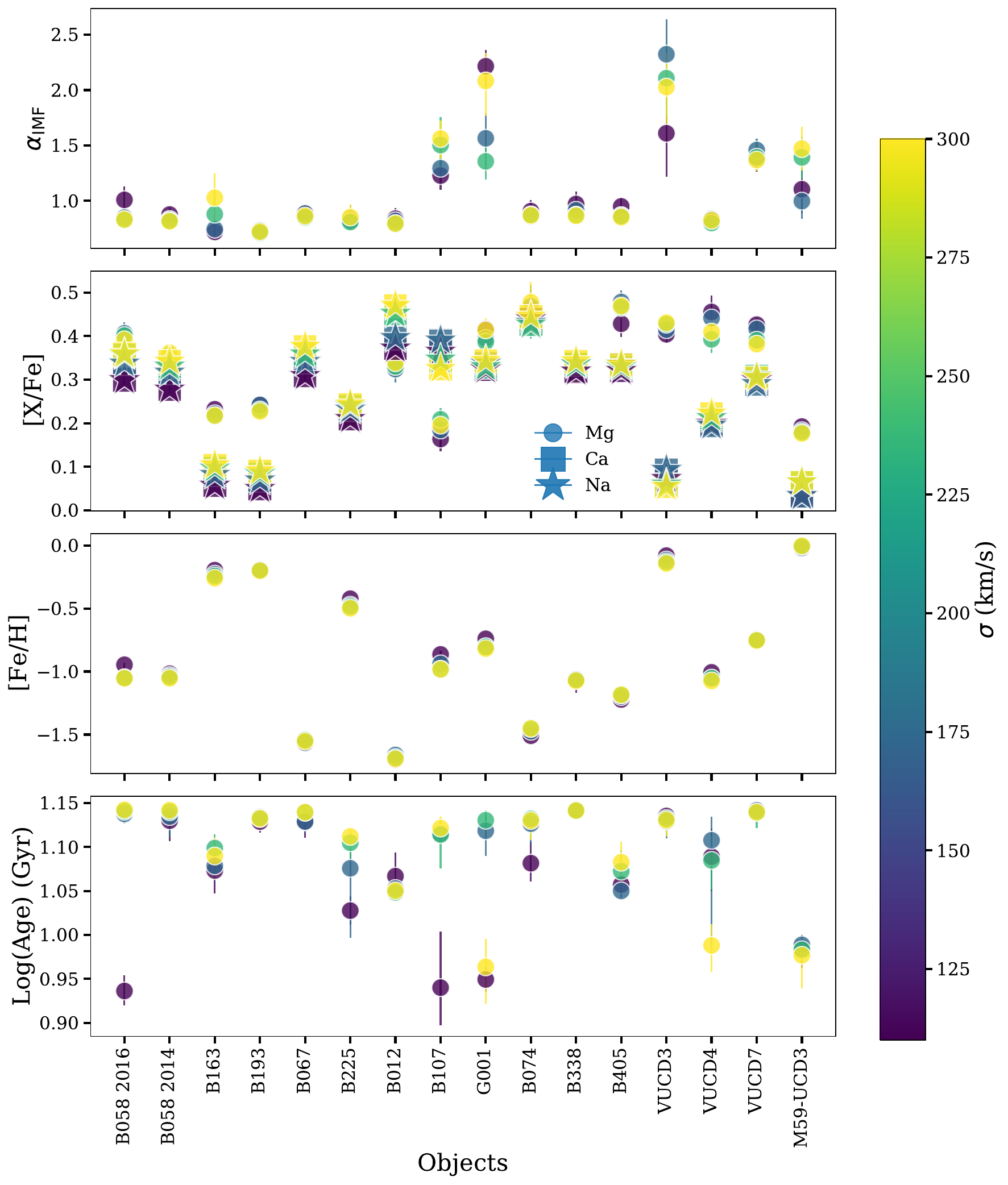}
    \caption{Comparison of the amount of spread in fitted parameters for different smoothing routines.  From top to bottom, we show $\alpha_{\mathrm{IMF}}$, [X/Fe] (where X is Mg, Ca, or Na), [Fe/H], and $
    \log(\text{age})$ on the $y$-axis for each object ($x$-axis).  These points are colour-coded by the amount by which the spectra were smoothed prior to fitting.}
    \label{fig:paper_smoothgrad}
\end{figure*}
As discussed in Section~\ref{sec:methods}, we smooth the spectra prior to fitting.  To better understand how our results depend on the amount of smoothing, we smooth the spectra of the CSSs by 110 km~s$^{-1}$, 200 km~s$^{-1}$, 250 km~s$^{-1}$, and 300 km~s$^{-1}$.  We note that extensive testing of the robustness of the \textsc{alf} models to smoothing has also been done in \cite{Choi_2014}.  The impact on fitted parameters $\alpha_{\mathrm{IMF}}$, [X/Fe], [Fe/H], and $\log(\text{age})$ are shown in Figure~\ref{fig:paper_smoothgrad}, colour-coded by the amount of smoothing.  We also tested a variety of different smoothing routines (not shown here), including smoothing the spectra with no modifications, masking and interpolating over unphysical artifacts, masking and interpolating over areas with poor telluric correction or sky subtraction, and weighting areas with poor telluric correction or sky subtraction to zero in the fits and inflating the uncertainties.   In general, these lead to comparable differences in fit parameters to shows shown in Figure~\ref{fig:paper_smoothgrad}. 

Although there is some dependence of fit parameters on the smoothing, this is not at a level that affects our main conclusions about variation of $\alpha_{\mathrm{IMF}}$ between stellar systems.   

\section{Indices versus Full Spectrum Fitting}\label{sec:indices}
\begin{table*}
	\centering
	\caption{Equivalent widths of the indices used in Z17, measured in this work using the index-fitting mode in \textsc{alf} (compare with Table (4) in Z17).}
	\label{tab:all_EWs}
  \begin{threeparttable}    
	\begin{tabular}{cccccccccc}
		\hline
		ID & H$\beta$ (\AA) & Fe52 (\AA) & Fe53 (\AA) & Mgb (\AA) & NaI (\AA) & CaT (\AA) & MgI (\AA) & TiO & FeH (\AA)\tnote{a} \\
		\hline
		NGC 4874	&	1.30$\pm$0.10	&	2.55$\pm$0.11	&	1.82$\pm$0.12	&	4.02$\pm$0.11	&	0.44$\pm$0.07	&	5.48$\pm$0.20	&	0.19$\pm$0.05	&	0.93$\pm$2.80$\times10^{-3}$	&	0.06$\pm$0.15	\\
            NGC 4889	&	1.44$\pm$0.07	&	2.33$\pm$0.07	&	1.58$\pm$0.07	&	4.02$\pm$0.07	&	0.81$\pm$0.04	&	4.69$\pm$0.12	&	-0.04$\pm$0.03	&	0.95$\pm$1.70$\times10^{-3}$	&	0.37$\pm$0.08	\\
		\hline
	\end{tabular}
  \begin{tablenotes}
      \item[a] The Wing--Ford band.
  \end{tablenotes}
  \end{threeparttable}
\end{table*}
Here we compare full-spectrum fitting to spectral index analysis in the context of the two BCGs, by refitting the data using the index fitter mode in \textsc{alf}.  In addition to fitting the models, this mode calculates the equivalent widths (EWs) of spectral indices using a method similar to that in \cite{IRTF}.  In summary, we create 1000 bootstrapped realizations of each spectrum based on the spectral uncertainty, measure the EWs in each using Equation (11) from \cite{IRTF}, and plot the median and $1\sigma$ errors of the distribution.  We fit all 25 available indices and add the red TiO index measured in Z17.  We also perform a third fit to the data, this time only fitting the five red indices examined in Z17.  We report the EWs in Table~\ref{tab:all_EWs}.

\begin{figure}
    \centering
    \includegraphics[width=\columnwidth]{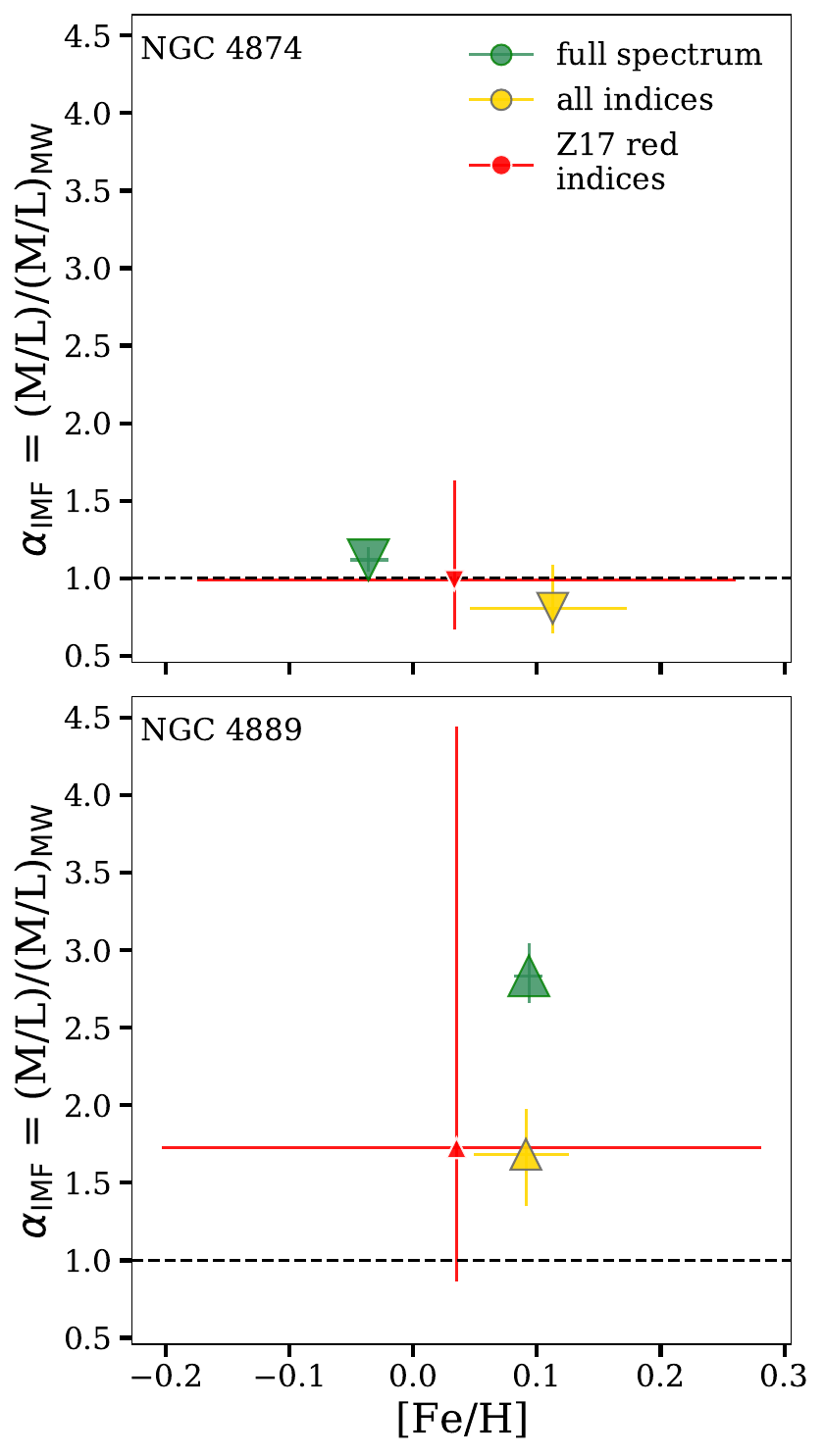}
    \caption{The IMF-mismatch parameter $\alpha_{\mathrm{IMF}}$ as a function of [Fe/H] for different types of fits to the BCG spectra.  In the left-hand panel we show NGC 4874, and on the right we show NGC 4889.  This is analogous to the left-most panel in Figure~\ref{fig:paper_fig3}.  Green symbols represent the results from our full spectrum fit (the same points as in Figure~\ref{fig:paper_fig3}), yellow symbols represent the results from our index fits using all of the available indices in \textsc{alf}, and red symbols represent the results from our index fits using only the red indices from Z17.  The dashed line at 1 represents a Kroupa IMF.}
    \label{fig:paper_fig3_index}
\end{figure}

In Figure~\ref{fig:paper_fig3_index}, we show $\alpha_{\mathrm{IMF}}$ as a function of [Fe/H] for NGC 4874 (left) and NGC 4889 (right).  We re-plot our full-spectrum results in green (i.e.\ the points from Figure~\ref{fig:paper_fig3}), the index-fitter results in yellow, and the red-index fits in red.  For both objects and all fits, we note that the $\alpha_{\mathrm{IMF}}$ values are consistent with each other regardless of whether we perform full- or index-fitting.  However, the error bars are much larger on the index fits, in particular the fits where only the red indices are taken into account.  Thus, our full spectrum method represents a dramatic increase in precision compared to spectral index analysis.  The [Fe/H] values are not as consistent for NGC 4874, but this may be because the full-spectrum fits are able to take into account many more small Fe-sensitive lines across the spectrum, compared to the handful of Fe-sensitive indices measured in the index fits.  For NGC 4874, all fits are consistent with having a Kroupa IMF, which is furthermore consistent with the result found in Z17.  For NGC 4889, all fits are consistent with a bottom-heavy IMF, but when only the red indices are fit, the result is consistent with a Kroupa IMF within uncertainties.  This demonstrates clearly that, while index- and full-fitting results are entirely consistent with each other, the additional information obtained from fitting the full spectrum, and in particular spectral information from blue wavelengths, is necessary to fully and precisely characterize the IMF (see also \citealt{CvD_2012a}).

\bsp	
\label{lastpage}

\clearpage
\pagenumbering{gobble}
\section{Other Fits}\label{sec:other_fits}
Here we show the fits and fit residuals for all objects in our sample not shown in Section~\ref{sec:methods}.  These Figures are available online.

\begin{figure*}
    \centering
    \renewcommand\thefigure{D1} 
    \includegraphics[width=\textwidth]{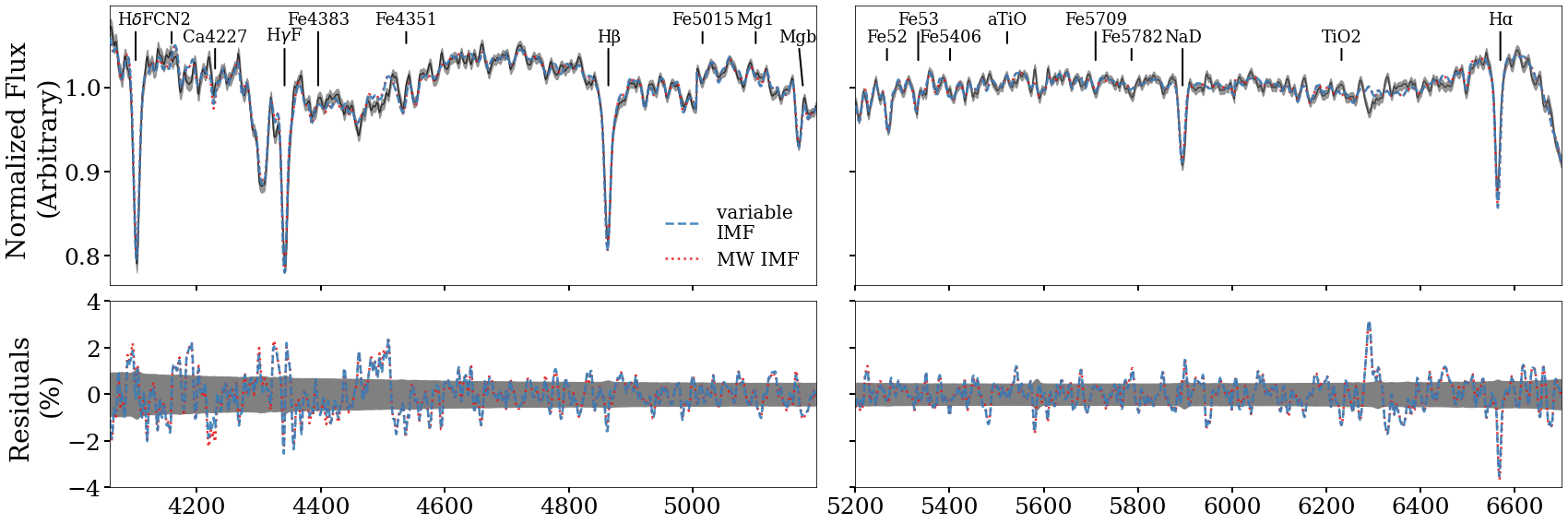}
    \includegraphics[width=\textwidth]{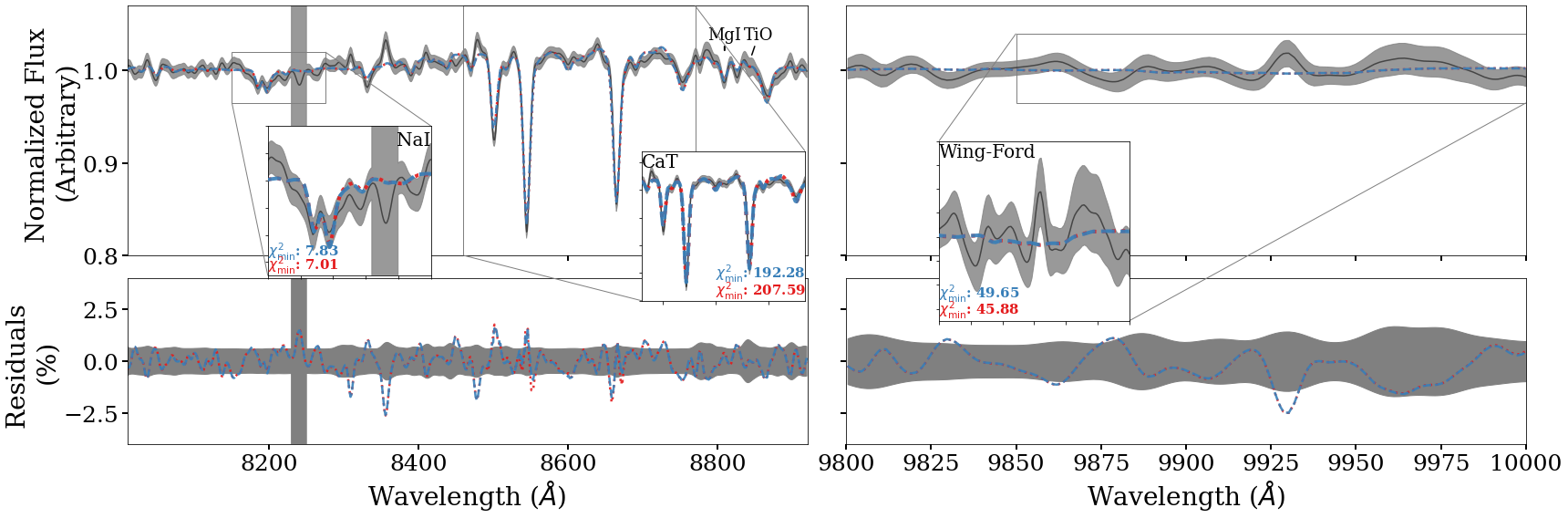}
    \caption{The same as Figure~\ref{fig:paper_fits_vucd7} but for B012 (GC).}
    \label{fig:paper_fits_B012}
\end{figure*}

\begin{figure*}
    \centering
    \renewcommand\thefigure{D2} 
    \includegraphics[width=\textwidth]{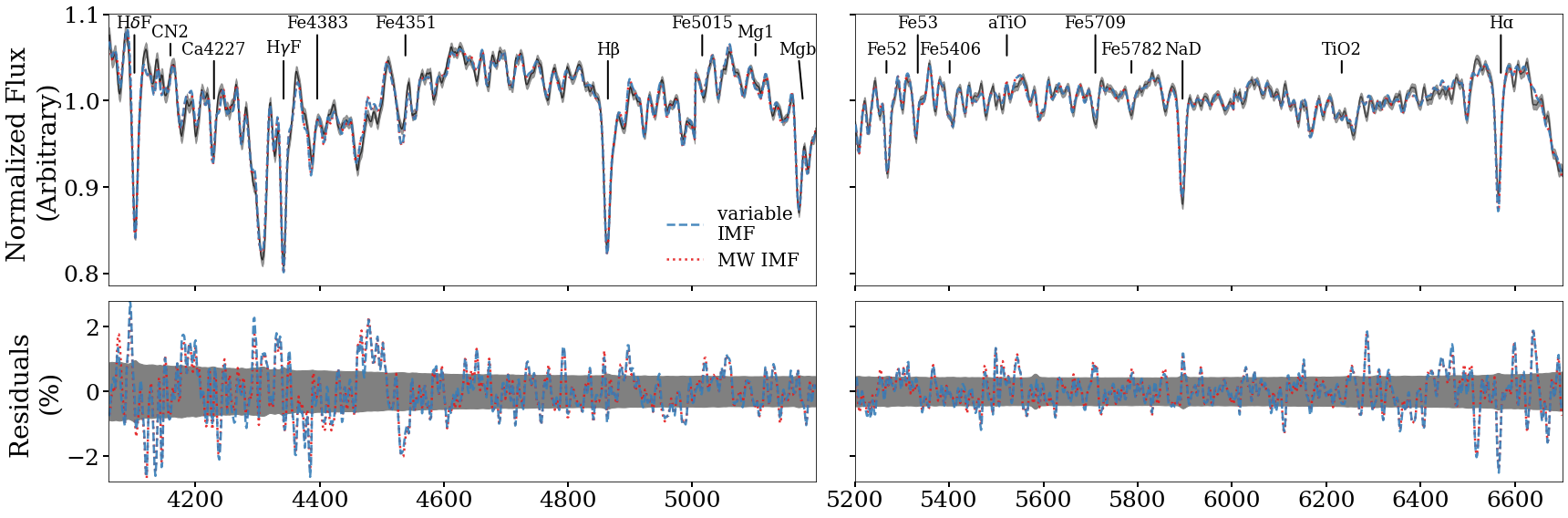}
    \includegraphics[width=\textwidth]{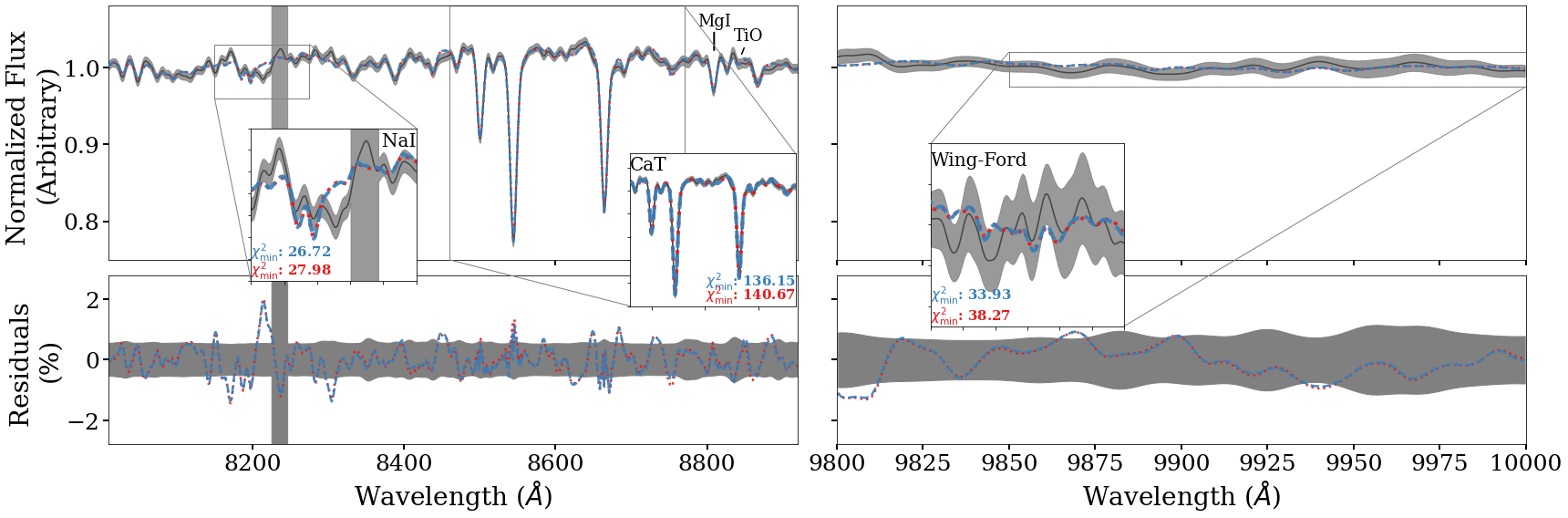}
    \caption{The same as Figure~\ref{fig:paper_fits_vucd7} but for B058 2016 (GC).}
    \label{fig:paper_fits_B058_new}
\end{figure*}

\begin{figure*}
    \centering
    \renewcommand\thefigure{D3} 
    \includegraphics[width=\textwidth]{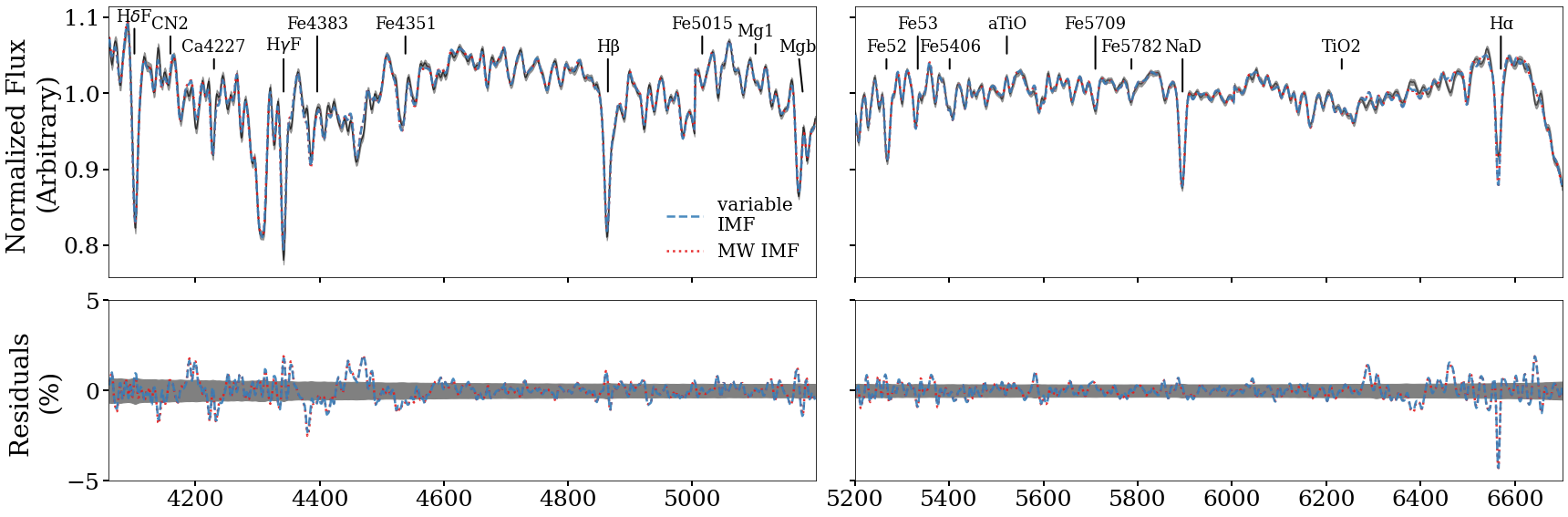}
    \includegraphics[width=\textwidth]{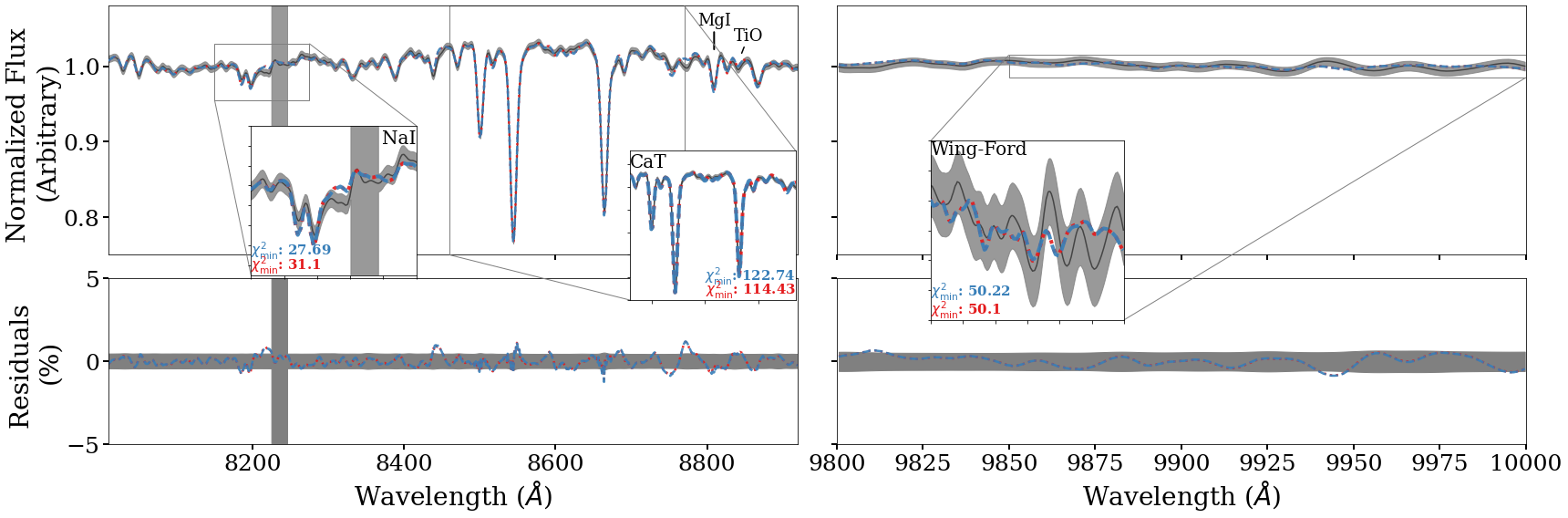}
    \caption{The same as Figure~\ref{fig:paper_fits_vucd7} but for B058 2014 (GC).}
    \label{fig:paper_fits_B058_old}
\end{figure*}

\begin{figure*}
    \centering
    \renewcommand\thefigure{D4} 
    \includegraphics[width=\textwidth]{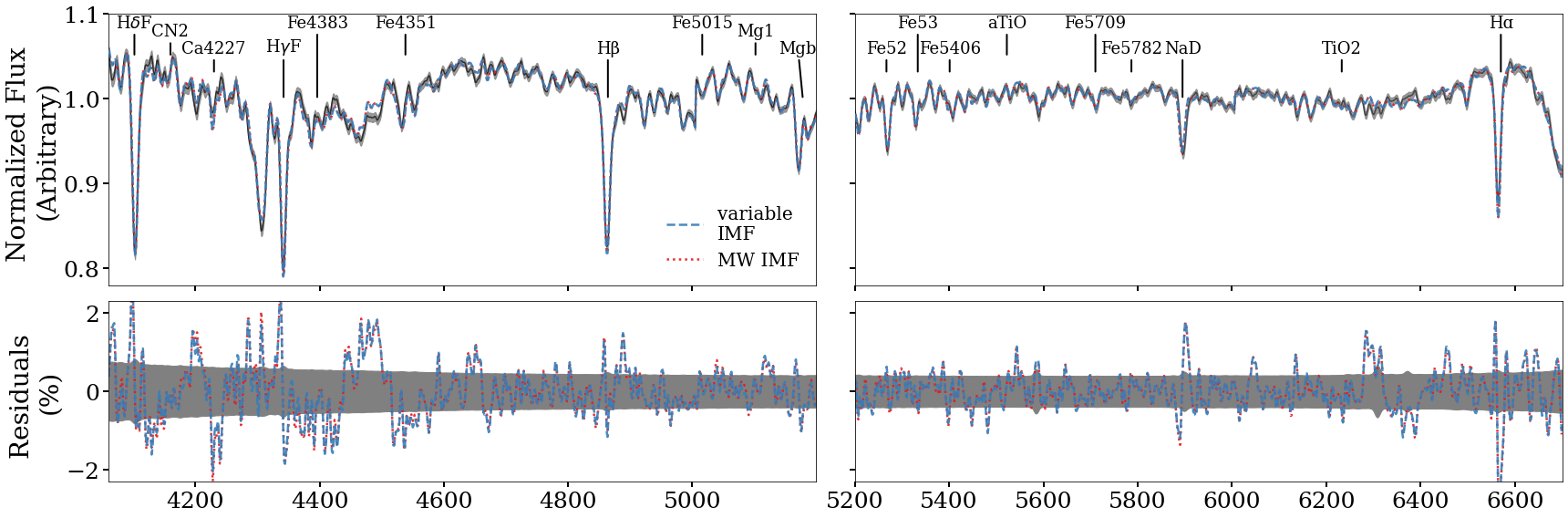}
    \includegraphics[width=\textwidth]{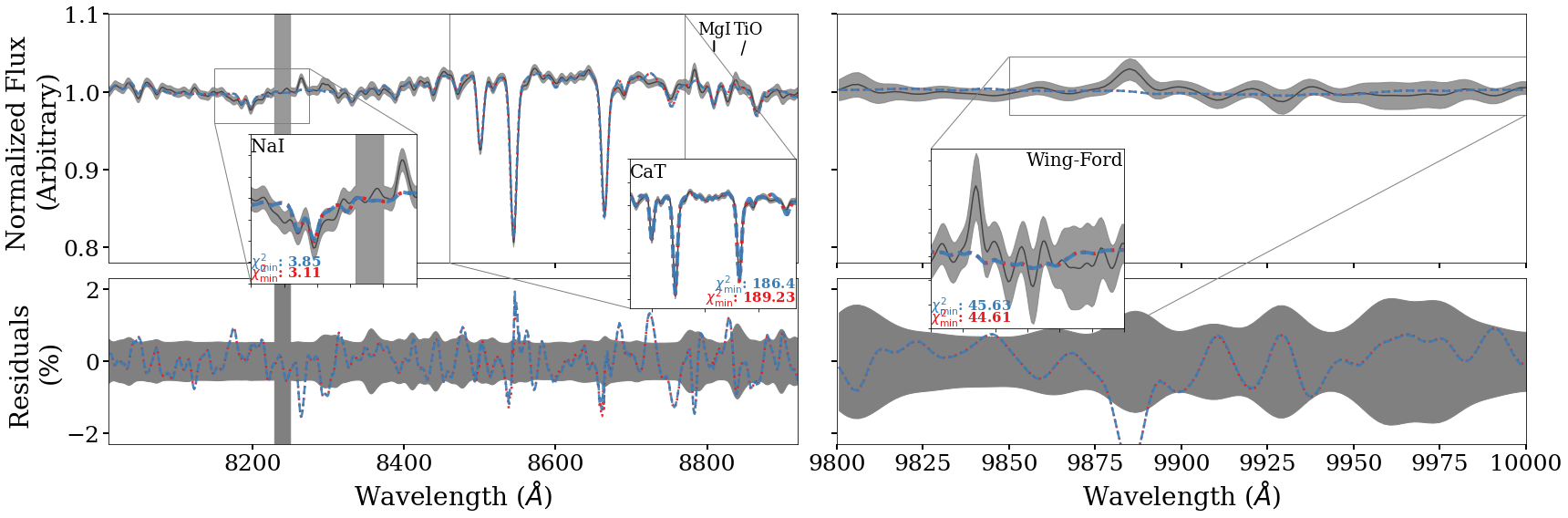}
    \caption{The same as Figure~\ref{fig:paper_fits_vucd7} but for B067 (GC).}
    \label{fig:paper_fits_B067}
\end{figure*}

\begin{figure*}
    \centering
    \renewcommand\thefigure{D5} 
    \includegraphics[width=\textwidth]{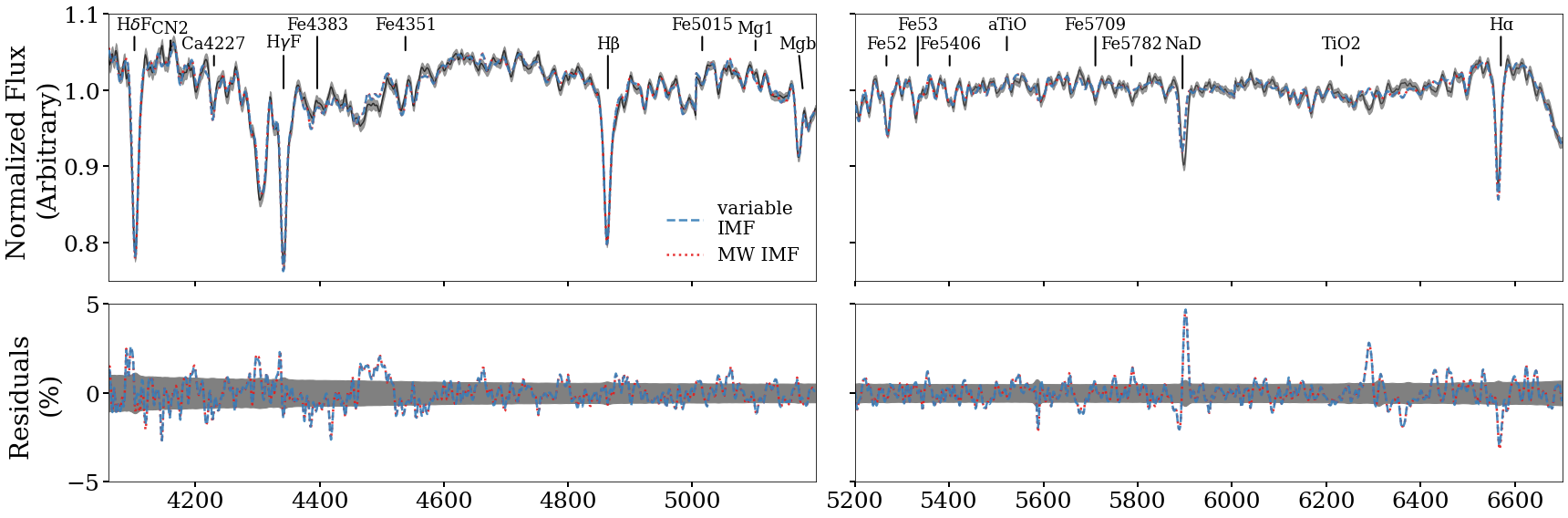}
    \includegraphics[width=\textwidth]{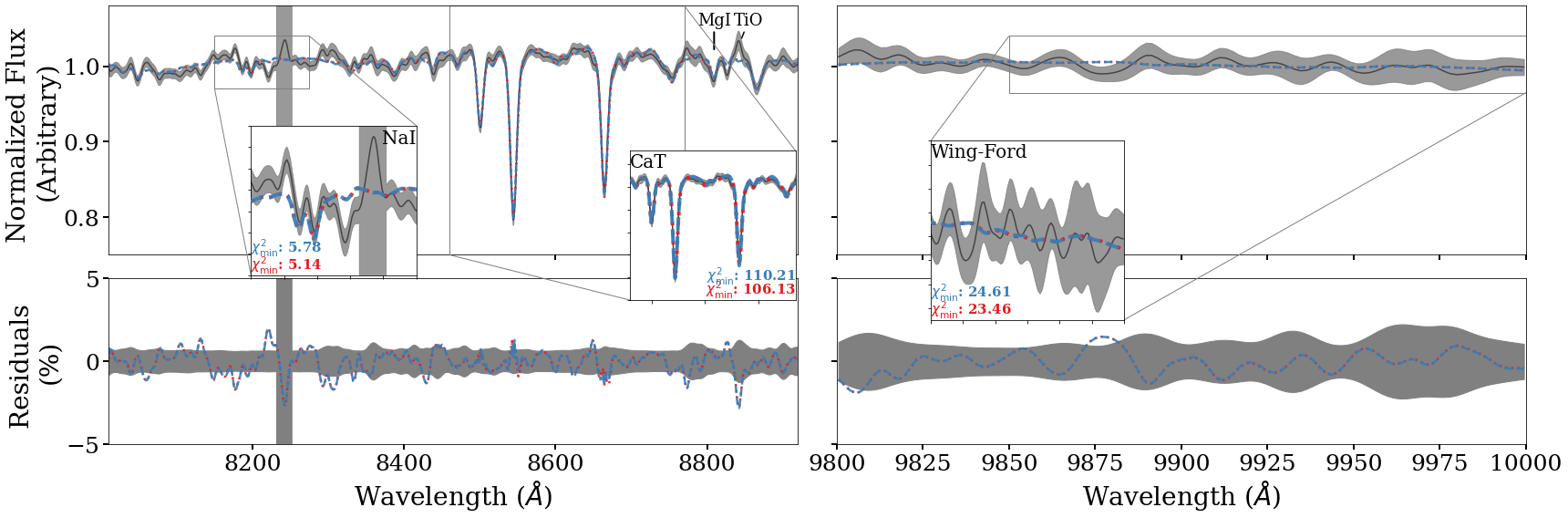}
    \caption{The same as Figure~\ref{fig:paper_fits_vucd7} but for B074 (GC).}
    \label{fig:paper_fits_B074}
\end{figure*}

\begin{figure*}
    \centering
    \renewcommand\thefigure{D6} 
    \includegraphics[width=\textwidth]{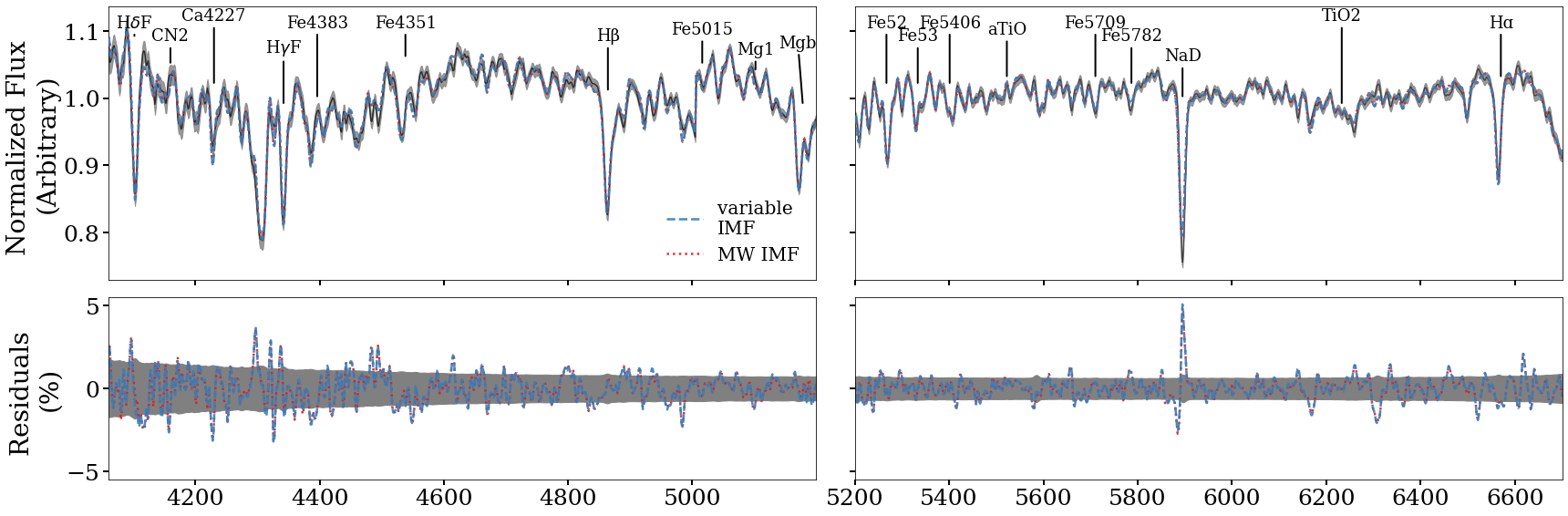}
    \includegraphics[width=\textwidth]{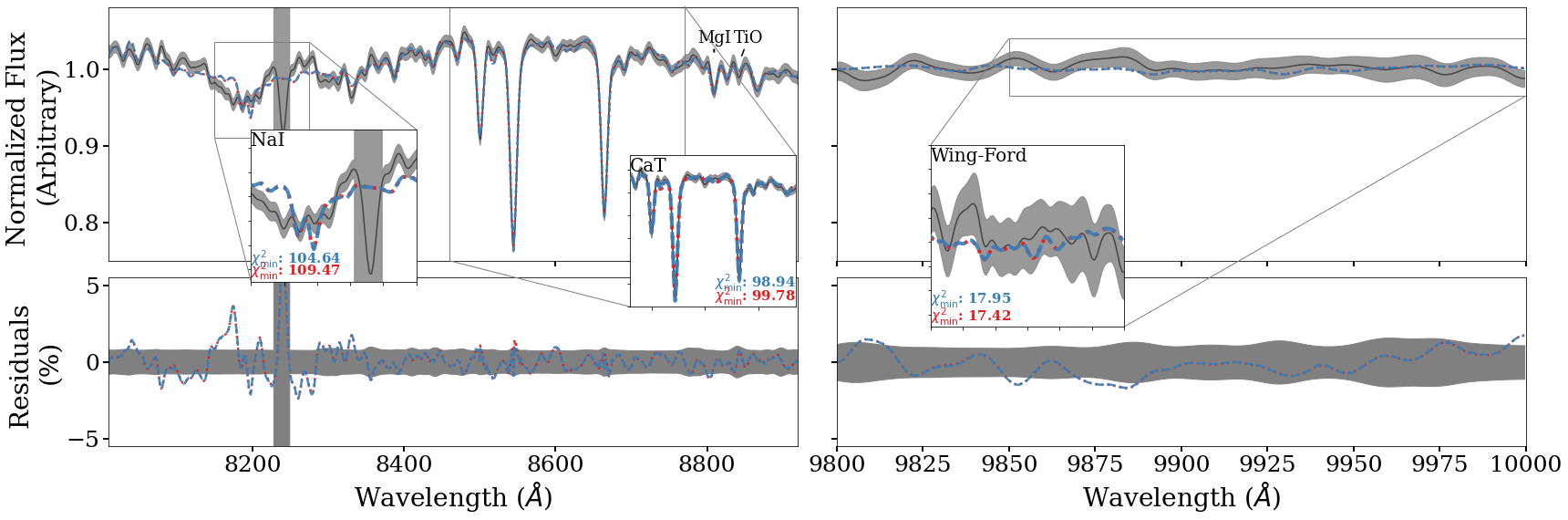}
    \caption{The same as Figure~\ref{fig:paper_fits_vucd7} but for B107 (GC).}
    \label{fig:paper_fits_B107}
\end{figure*}

\begin{figure*}
    \centering
    \renewcommand\thefigure{D7} 
    \includegraphics[width=\textwidth]{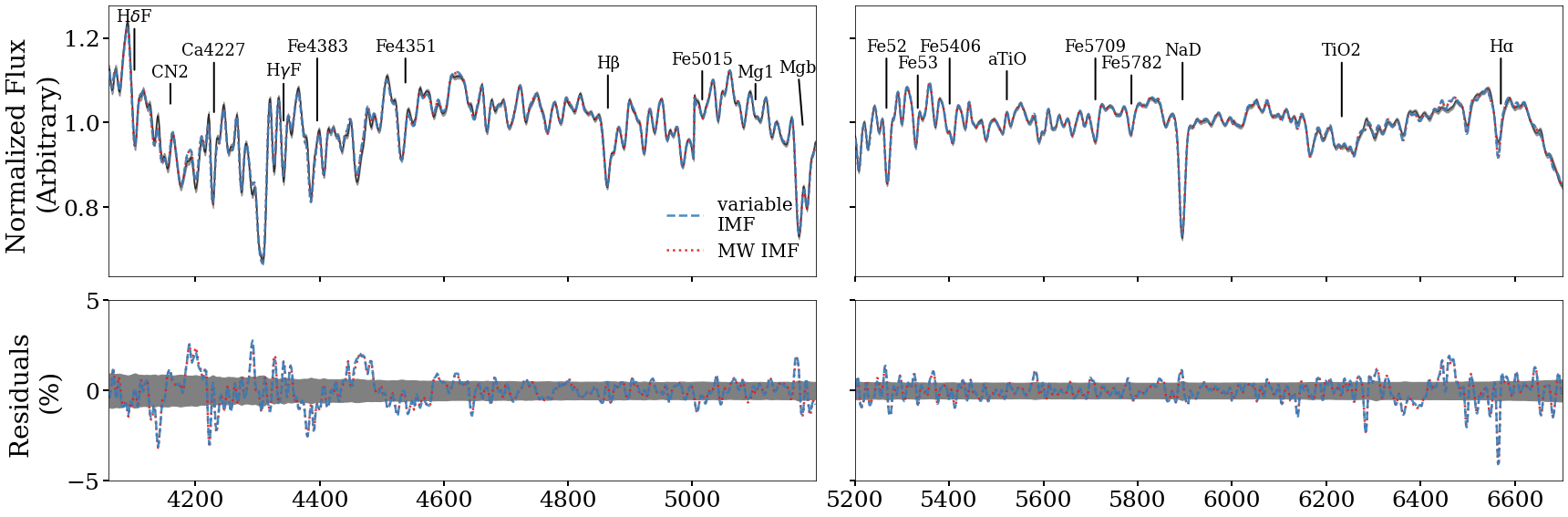}
    \includegraphics[width=\textwidth]{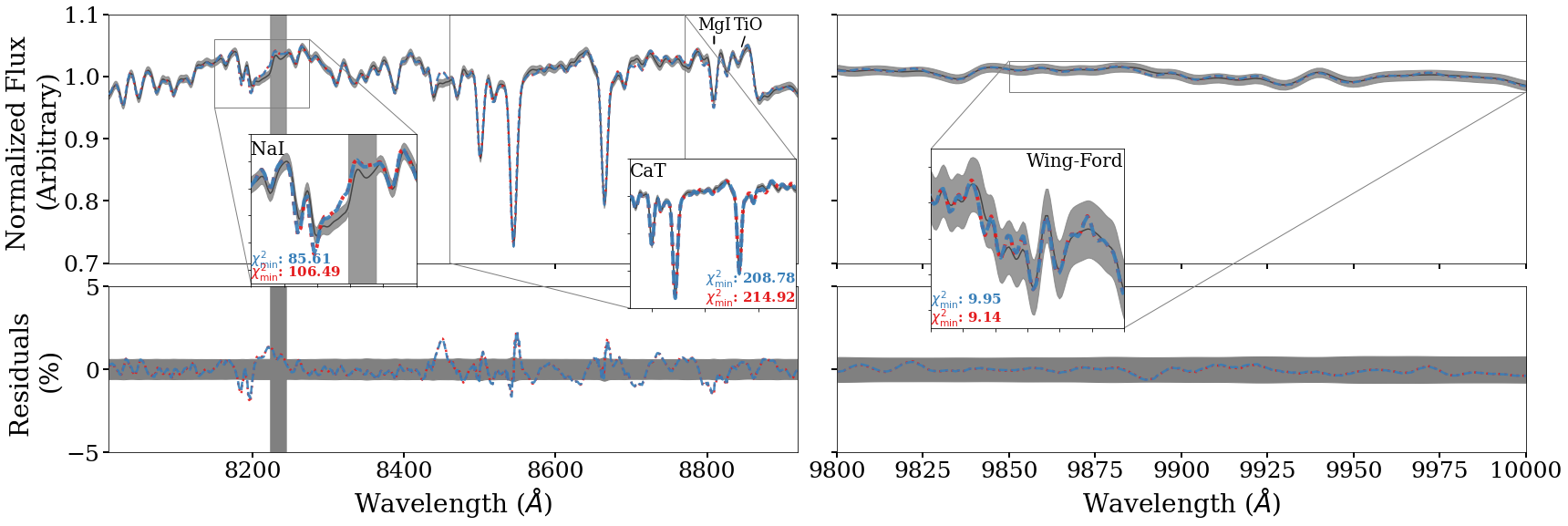}
    \caption{The same as Figure~\ref{fig:paper_fits_vucd7} but for B163 (GC).}
    \label{fig:paper_fits_B163}
\end{figure*}

\begin{figure*}
    \centering
    \renewcommand\thefigure{D8} 
    \includegraphics[width=\textwidth]{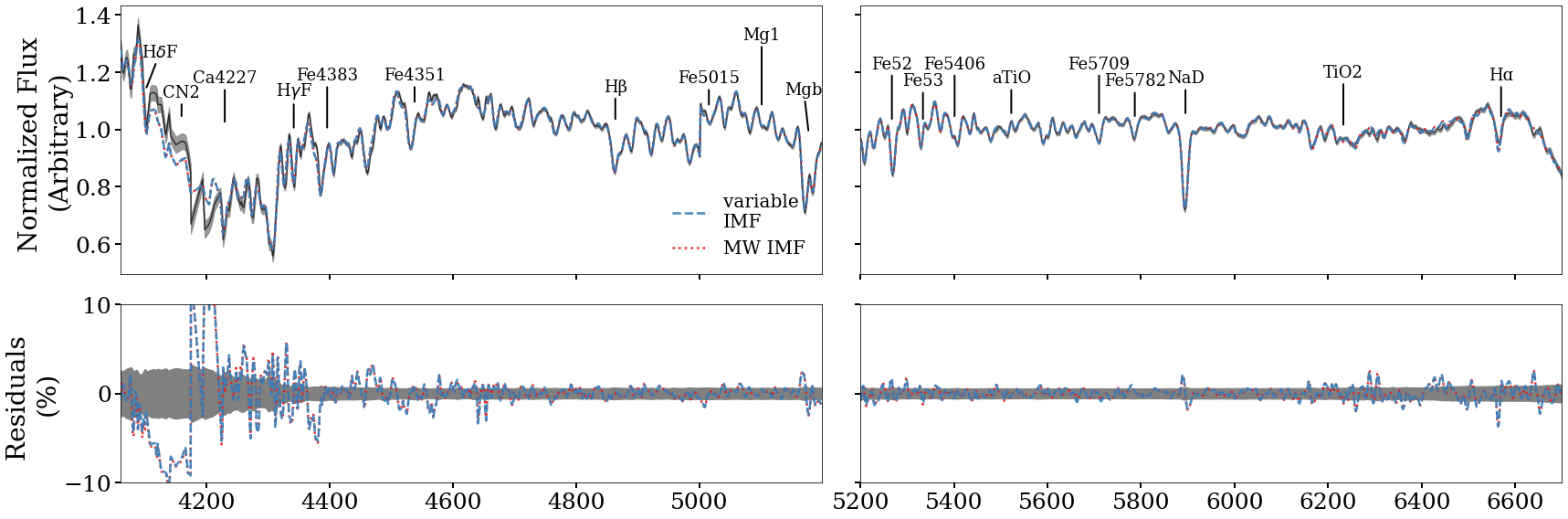}
    \includegraphics[width=\textwidth]{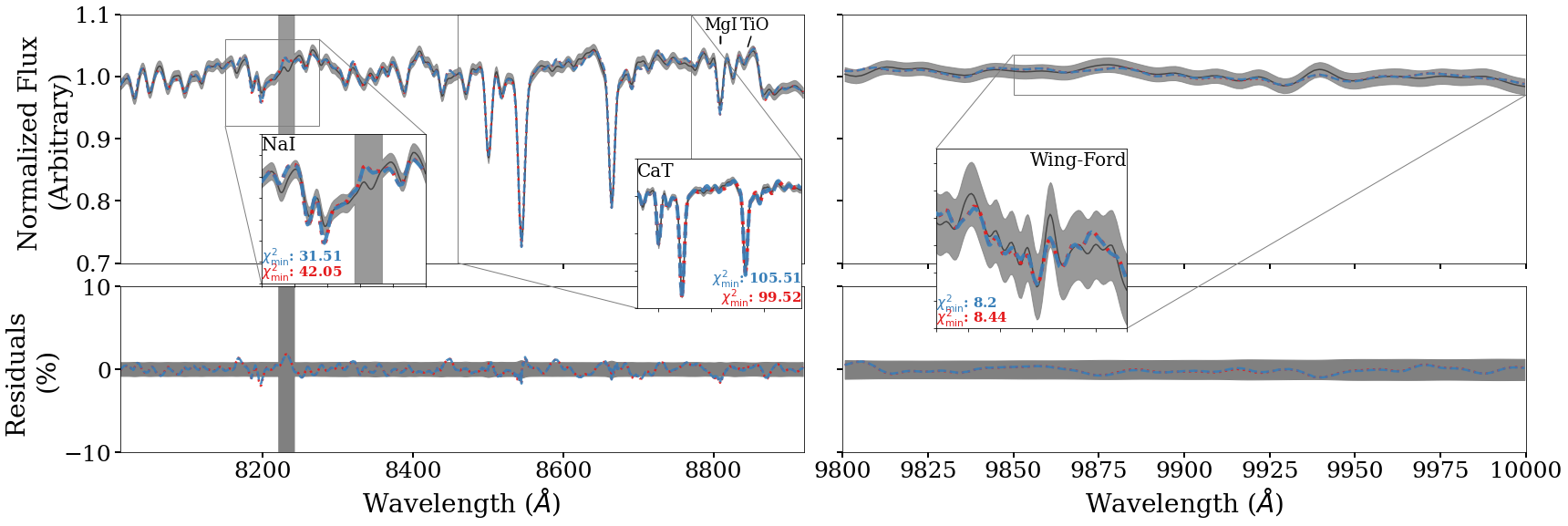}
    \caption{The same as Figure~\ref{fig:paper_fits_vucd7} but for B193 (GC).}
    \label{fig:paper_fits_B193}
\end{figure*}

\begin{figure*}
    \centering
    \renewcommand\thefigure{D9} 
    \includegraphics[width=\textwidth]{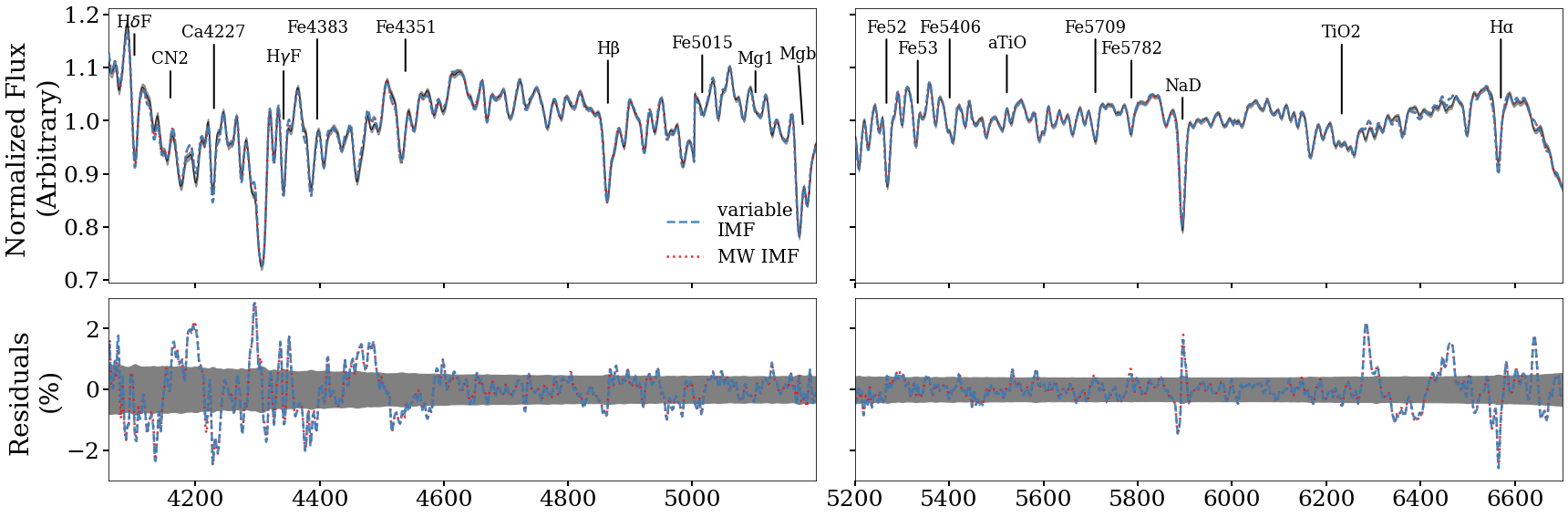}
    \includegraphics[width=\textwidth]{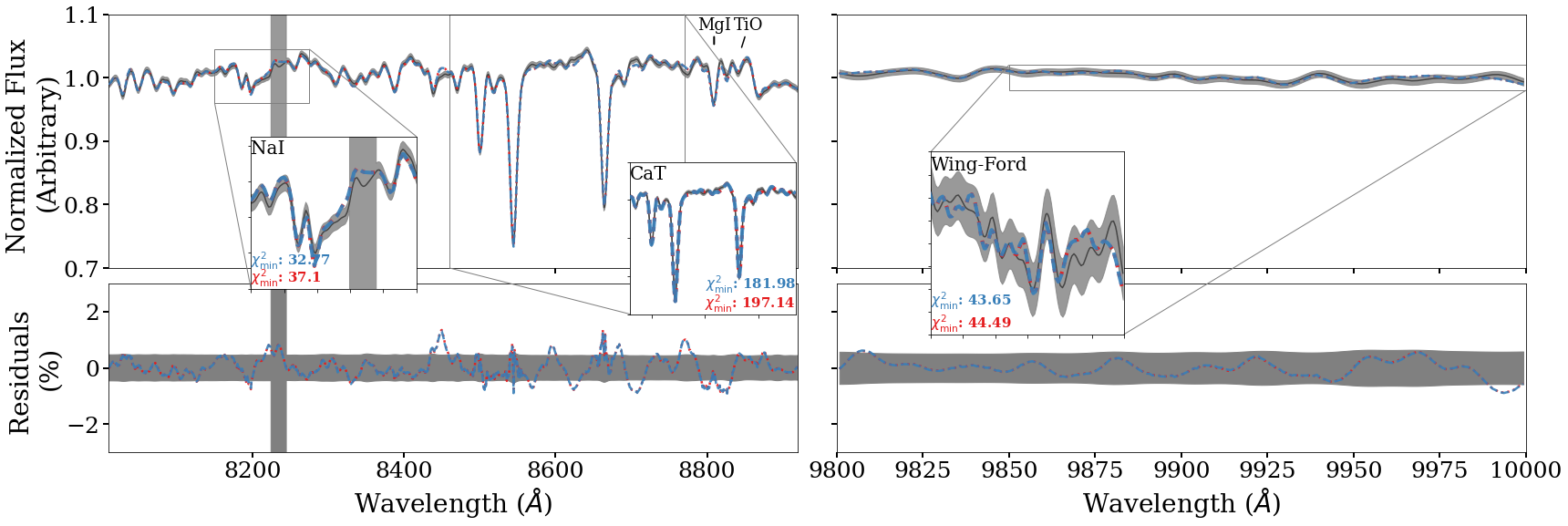}
    \caption{The same as Figure~\ref{fig:paper_fits_vucd7} but for B225 (GC).}
    \label{fig:paper_fits_B225}
\end{figure*}

\begin{figure*}
    \centering
    \renewcommand\thefigure{D10} 
    \includegraphics[width=\textwidth]{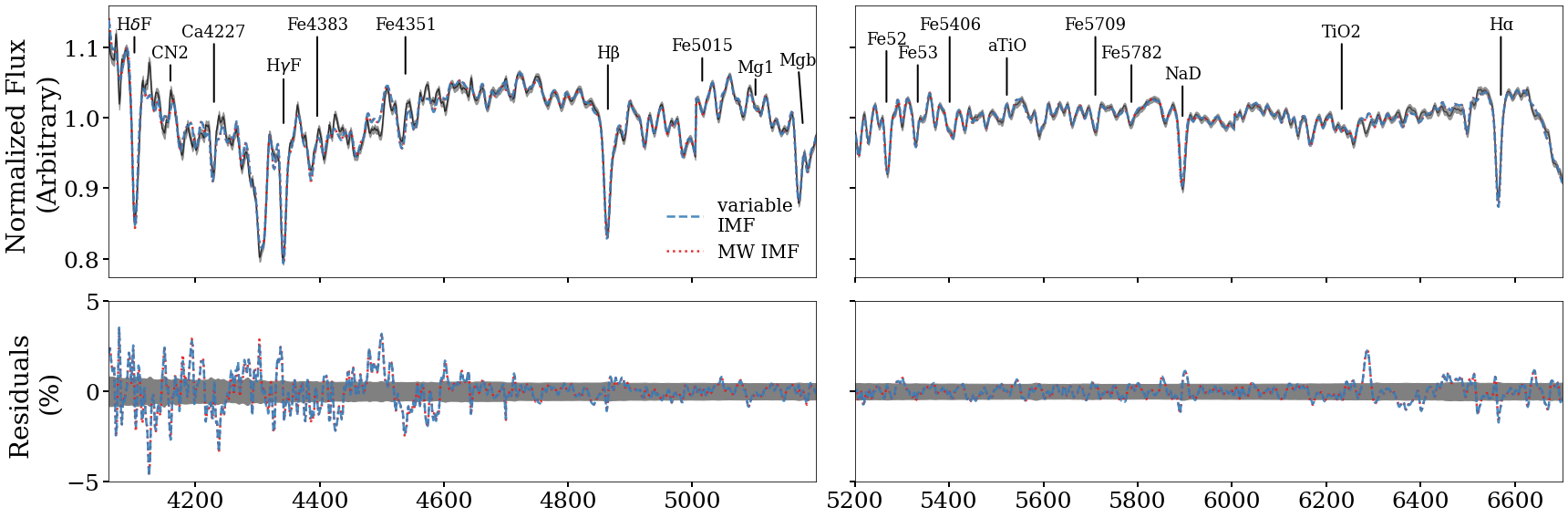}
    \includegraphics[width=\textwidth]{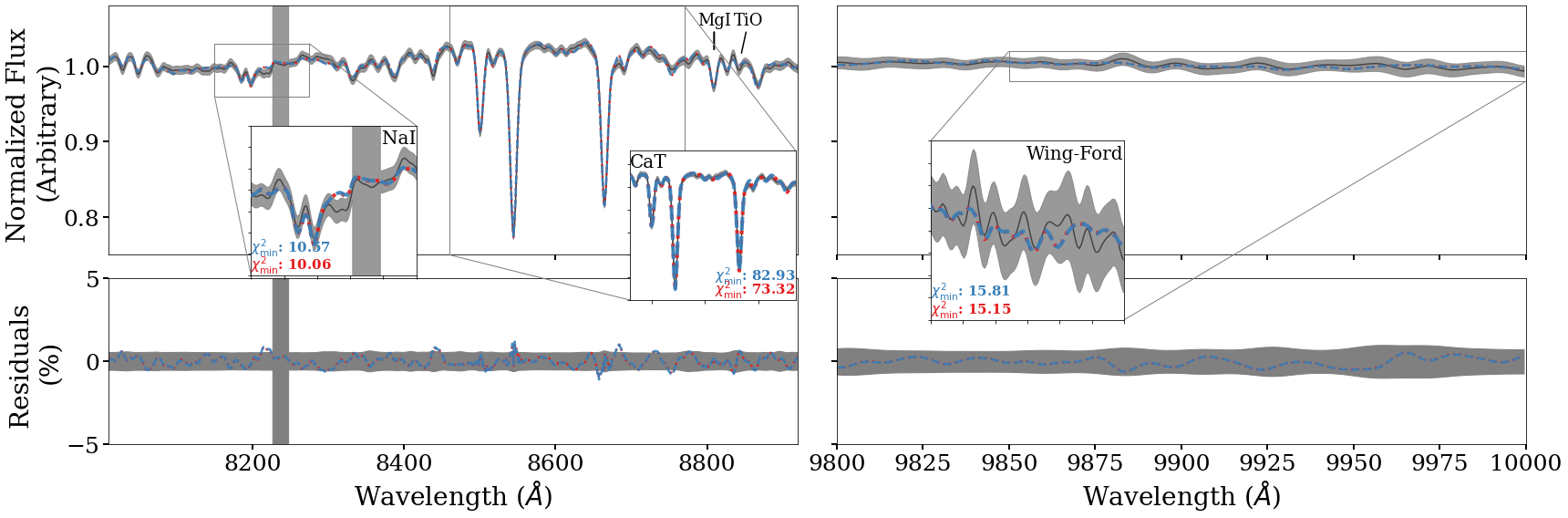}
    \caption{The same as Figure~\ref{fig:paper_fits_vucd7} but for B338 (GC).}
    \label{fig:paper_fits_B338}
\end{figure*}

\begin{figure*}
    \centering
    \renewcommand\thefigure{D11} 
    \includegraphics[width=\textwidth]{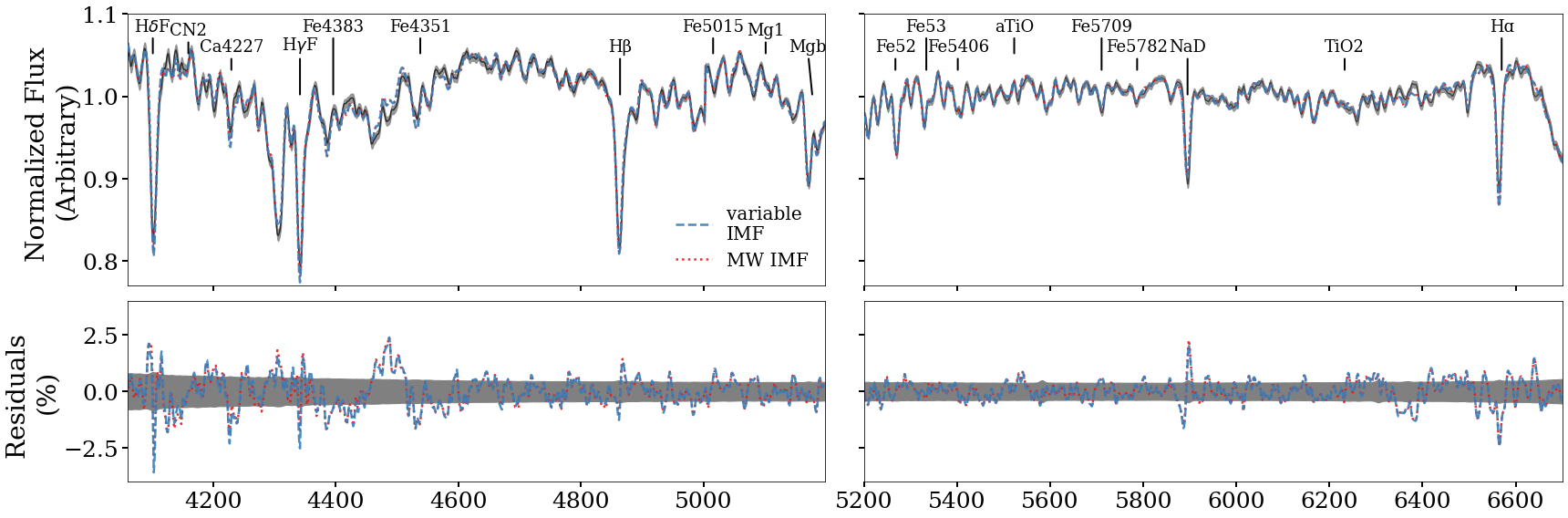}
    \includegraphics[width=\textwidth]{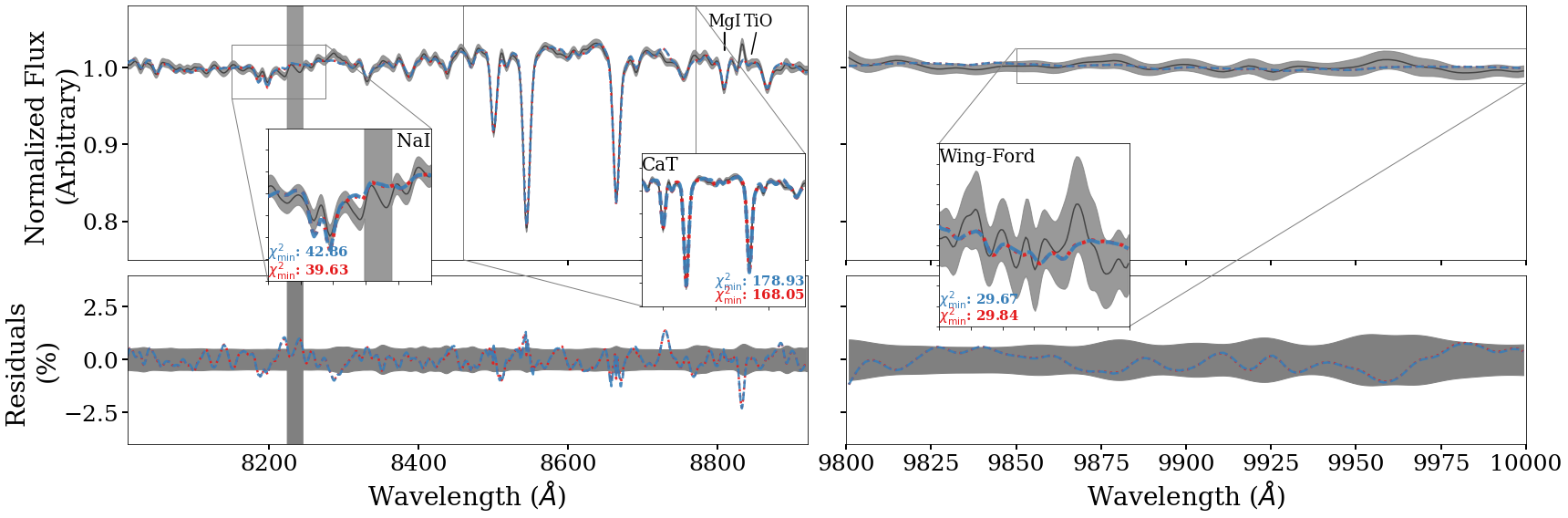}
    \caption{The same as Figure~\ref{fig:paper_fits_vucd7} but for B405 (GC).}
    \label{fig:paper_fits_B405}
\end{figure*}

\begin{figure*}
    \centering
    \renewcommand\thefigure{D12} 
    \includegraphics[width=\textwidth]{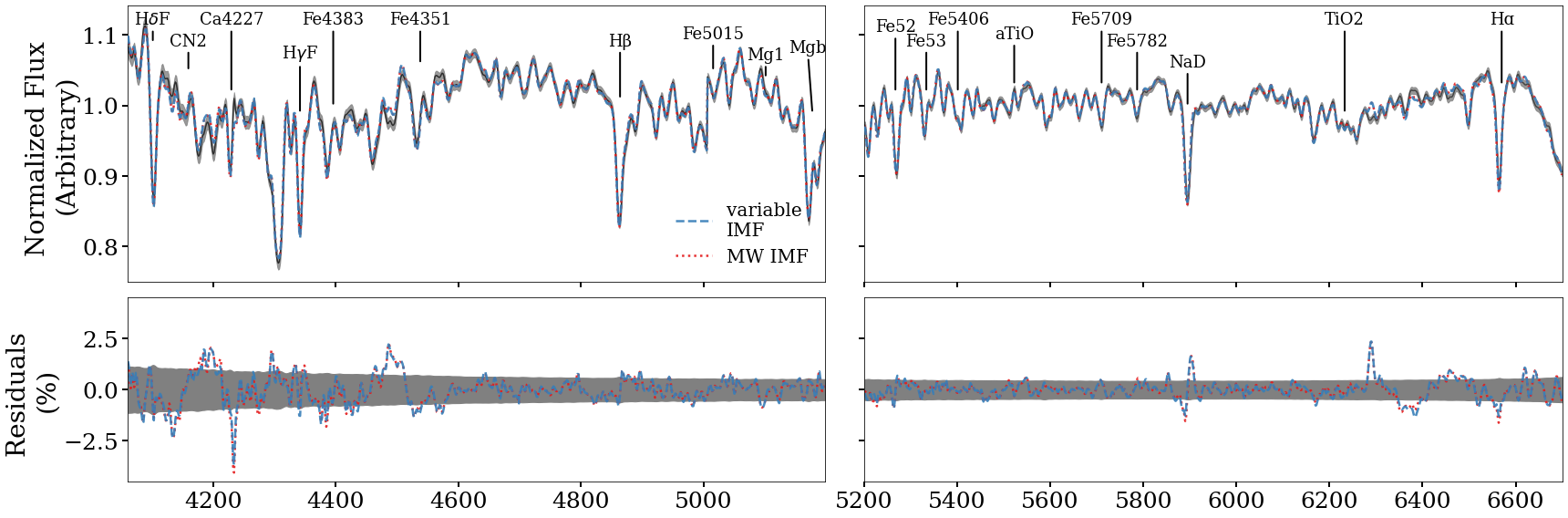}
    \includegraphics[width=\textwidth]{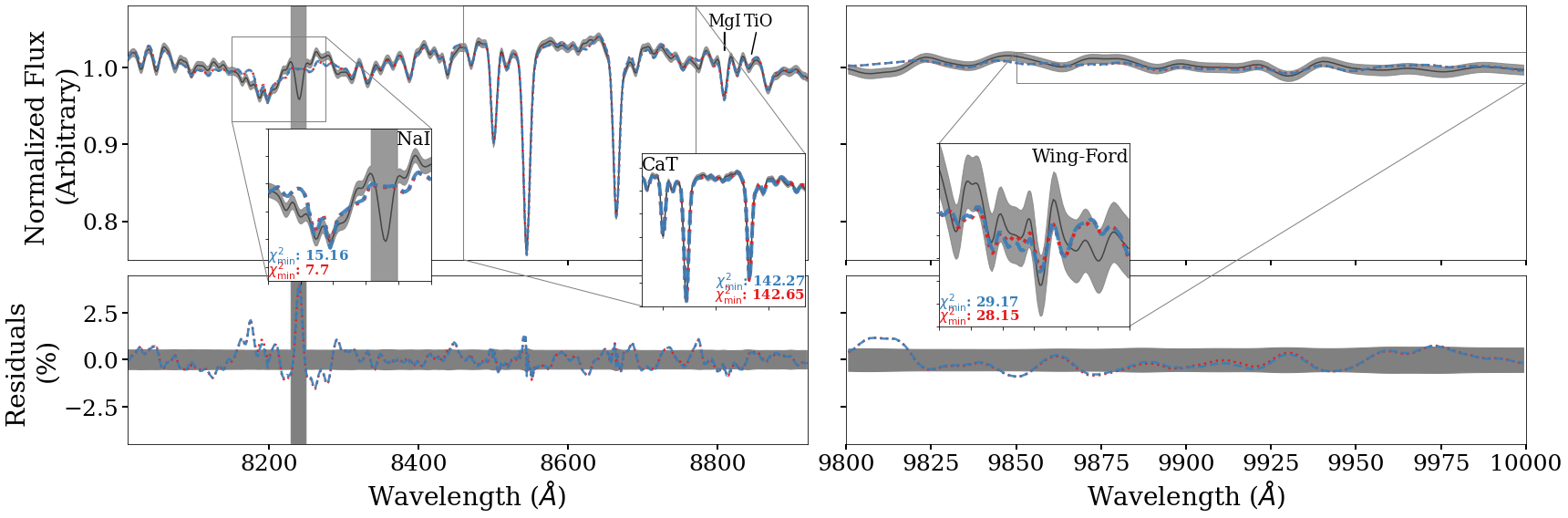}
    \caption{The same as Figure~\ref{fig:paper_fits_vucd7} but for G001 (GC).}
    \label{fig:paper_fits_G001}
\end{figure*}

\begin{figure*}
    \centering
    \renewcommand\thefigure{D13} 
    \includegraphics[width=\textwidth]{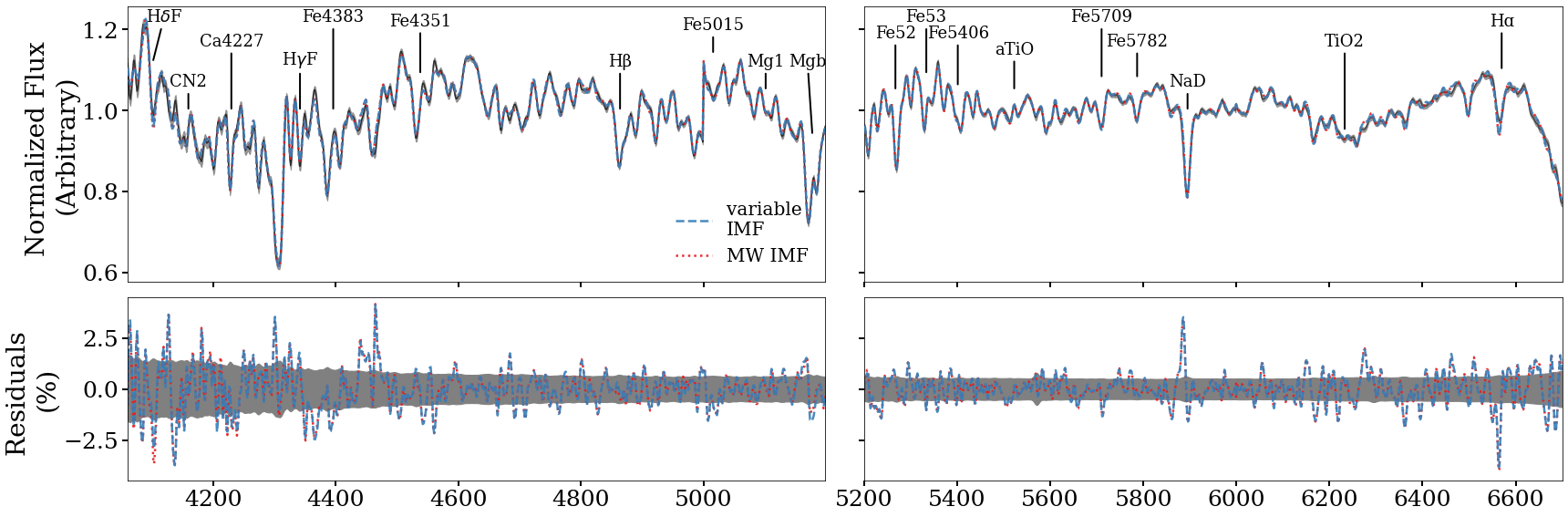}
    \includegraphics[width=\textwidth]{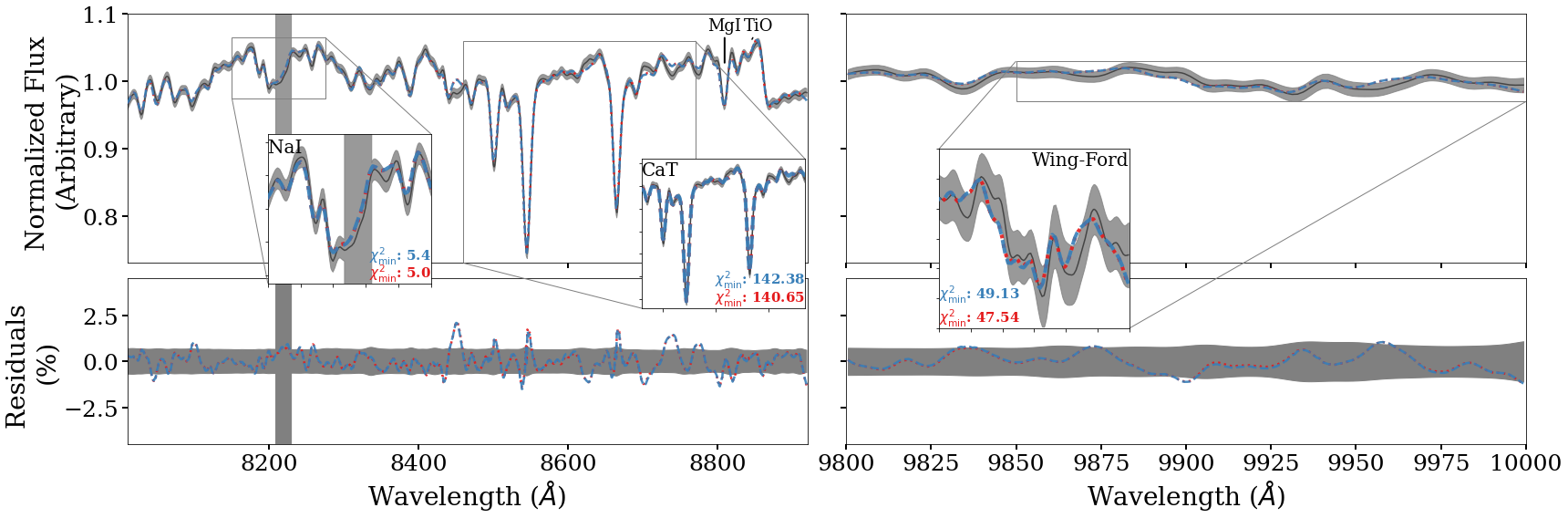}
    \caption{The same as Figure~\ref{fig:paper_fits_vucd7} but for M59-UCD3 (UCD).}
    \label{fig:paper_fits_M59_UCD3}
\end{figure*}

\begin{figure*}
    \centering
    \renewcommand\thefigure{D14} 
    \includegraphics[width=\textwidth]{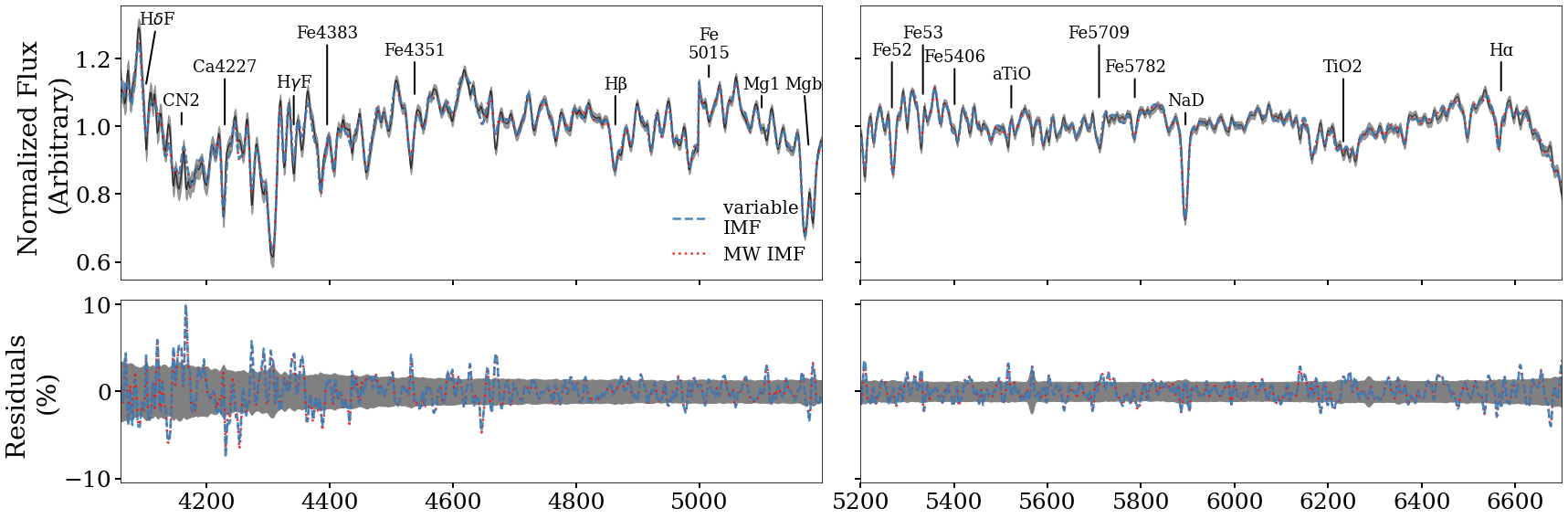}
    \includegraphics[width=\textwidth]{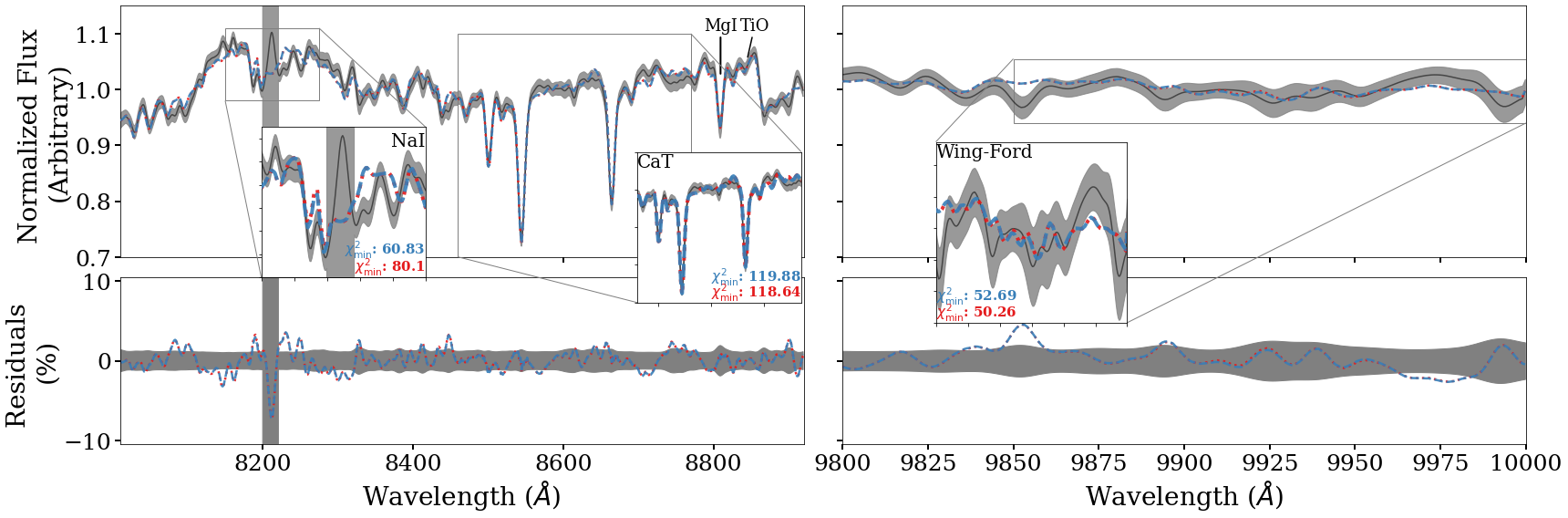}
    \caption{The same as Figure~\ref{fig:paper_fits_vucd7} but for VUCD3 (UCD).}
    \label{fig:paper_fits_VUCD3}
\end{figure*}

\begin{figure*}
    \centering
    \renewcommand\thefigure{D15} 
    \includegraphics[width=\textwidth]{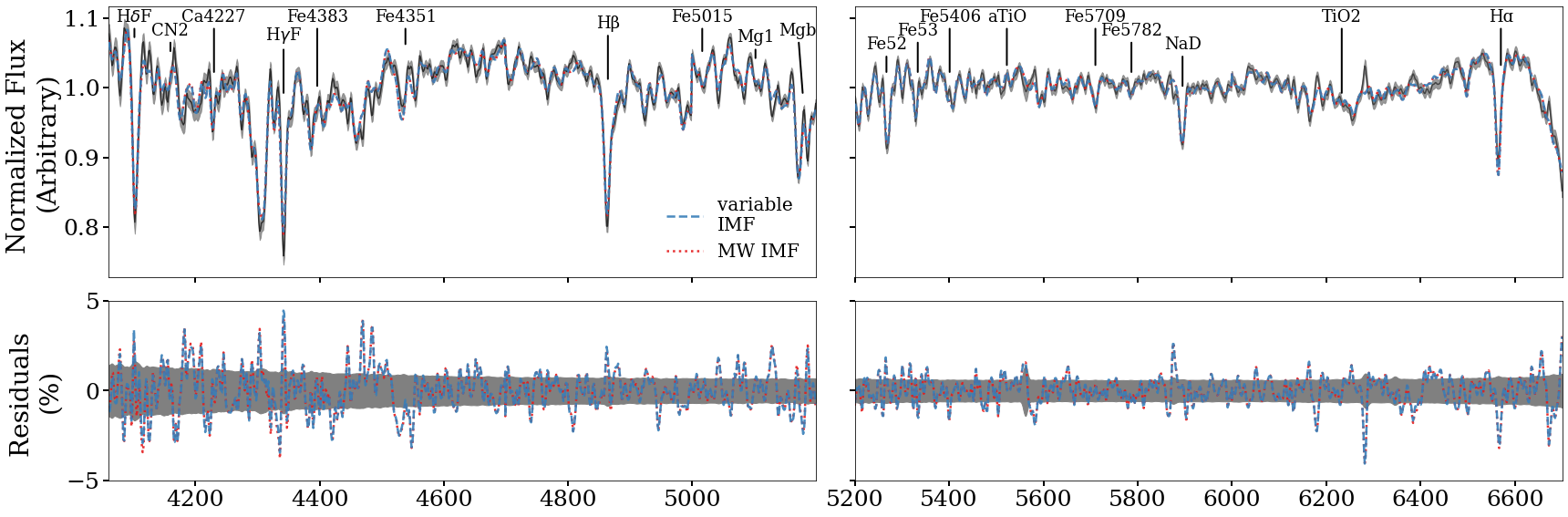}
    \includegraphics[width=\textwidth]{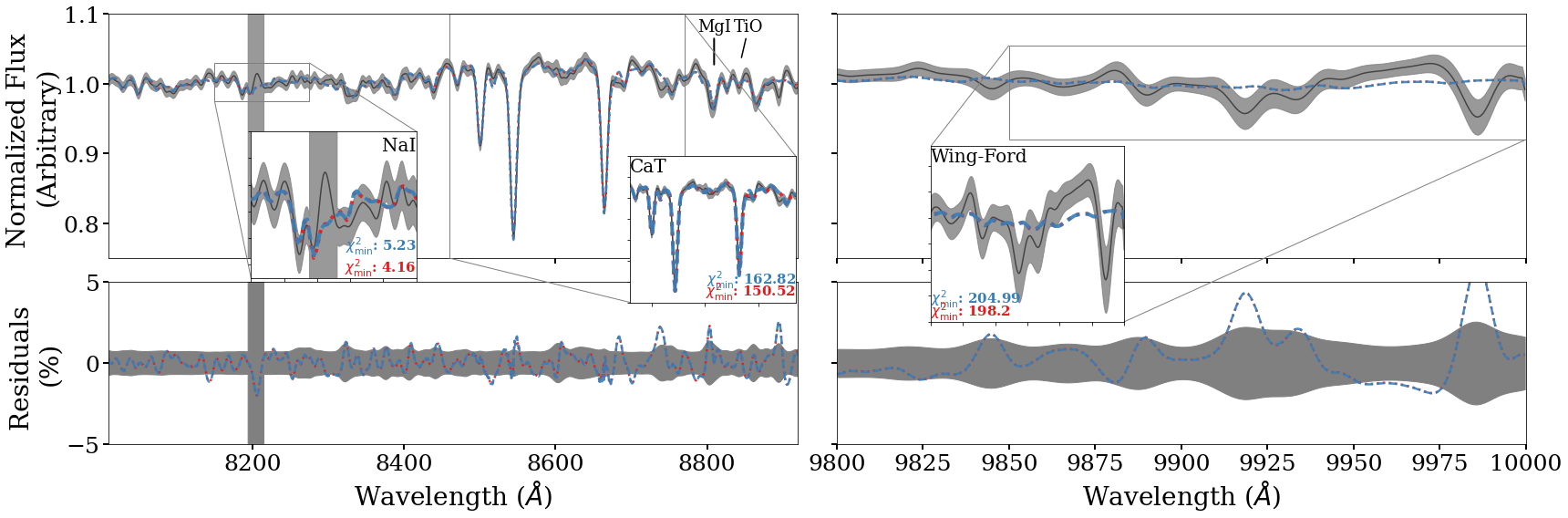}
    \caption{The same as Figure~\ref{fig:paper_fits_vucd7} but for VUCD4 (UCD).}
    \label{fig:paper_fits_VUCD4}
\end{figure*}

\begin{figure*}
    \centering
    \renewcommand\thefigure{D16} 
    \includegraphics[width=\textwidth]{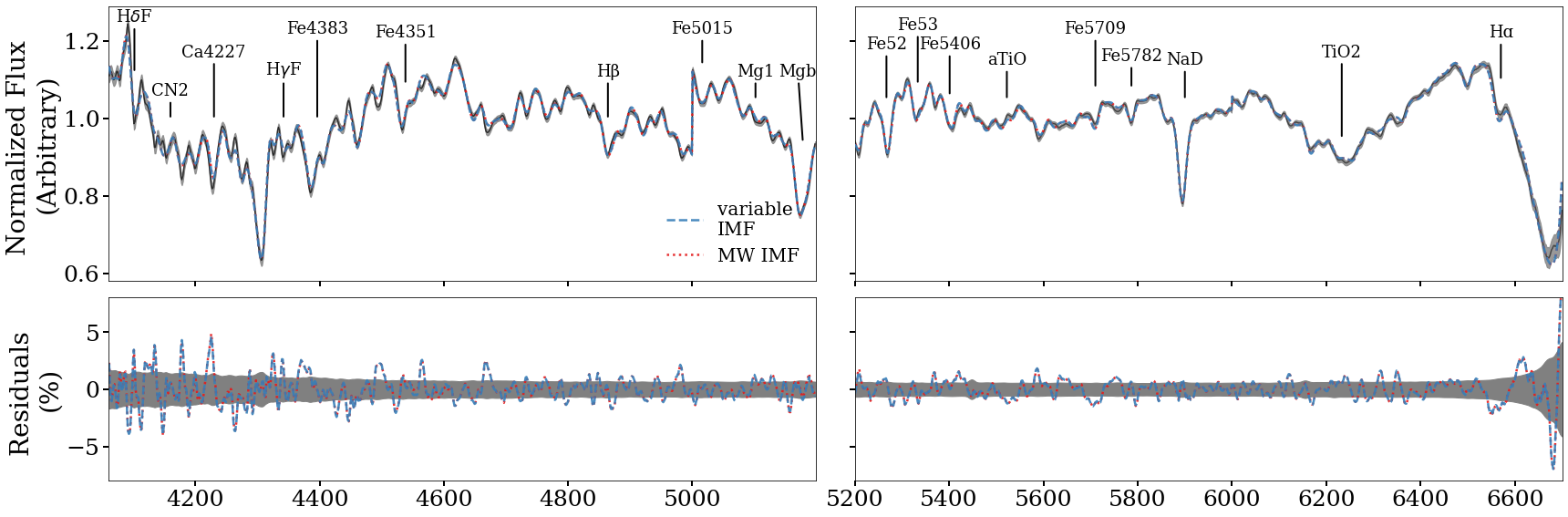}
    \includegraphics[width=\textwidth]{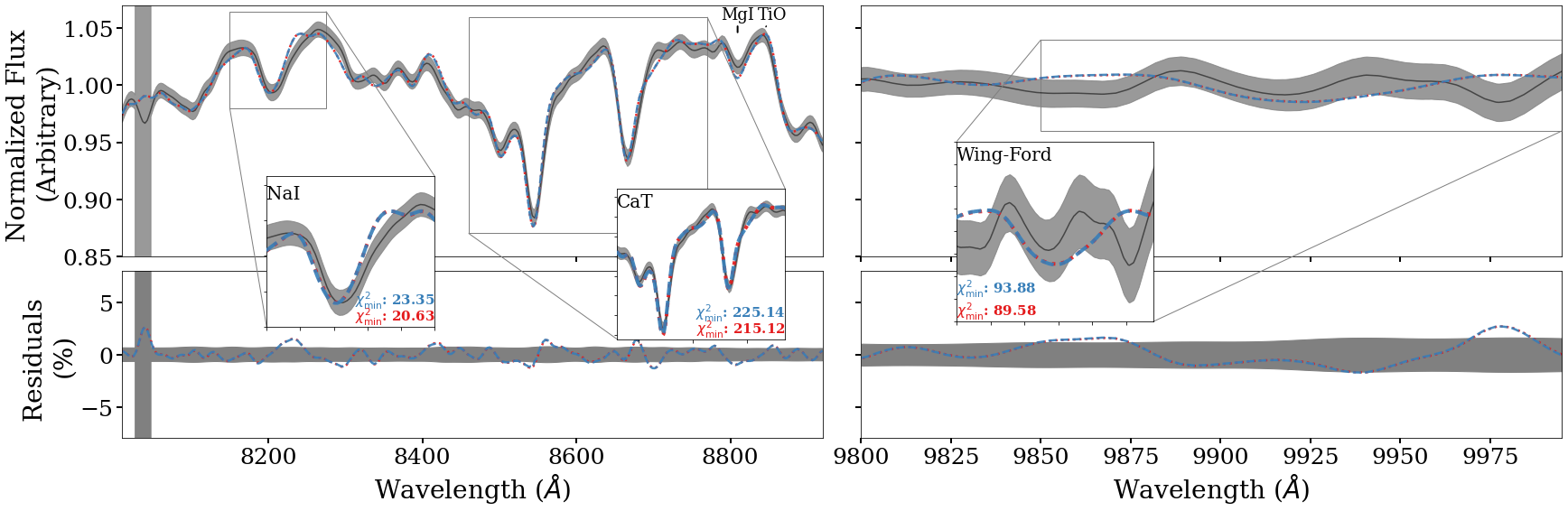}
    \caption{The same as Figure~\ref{fig:paper_fits_vucd7} but for NGC 4874 (BCG).}
    \label{fig:paper_fits_ngc4874}
\end{figure*}

\begin{figure*}
    \centering
    \renewcommand\thefigure{D17} 
    \includegraphics[width=\textwidth]{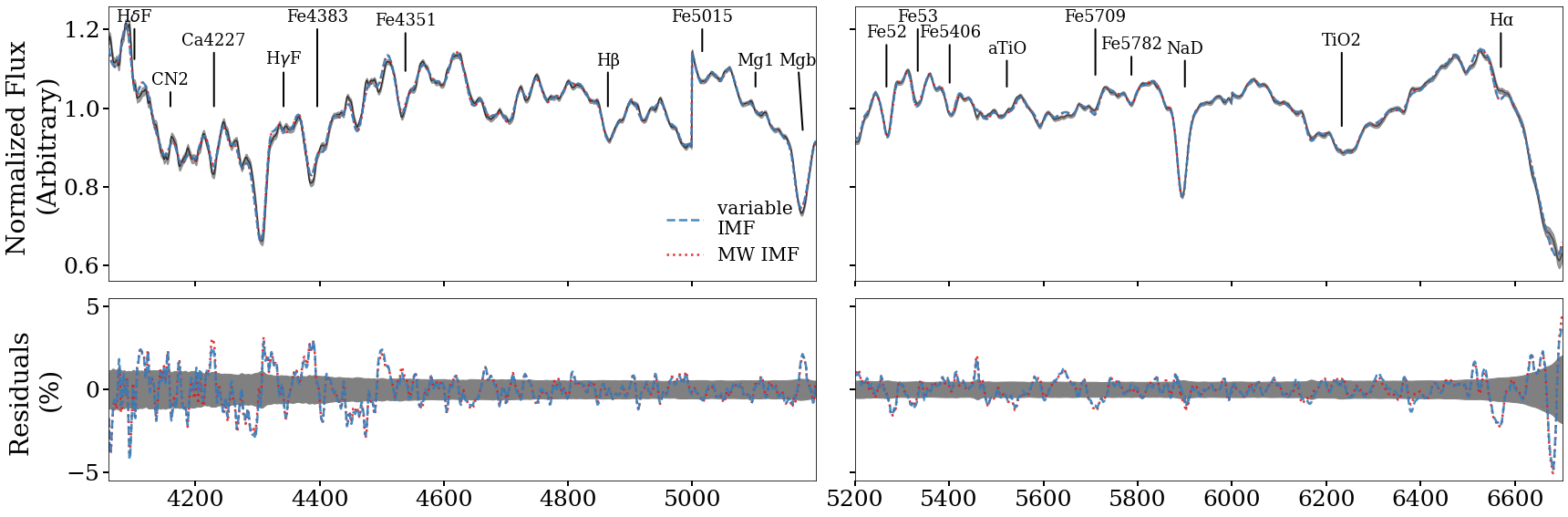}
    \includegraphics[width=\textwidth]{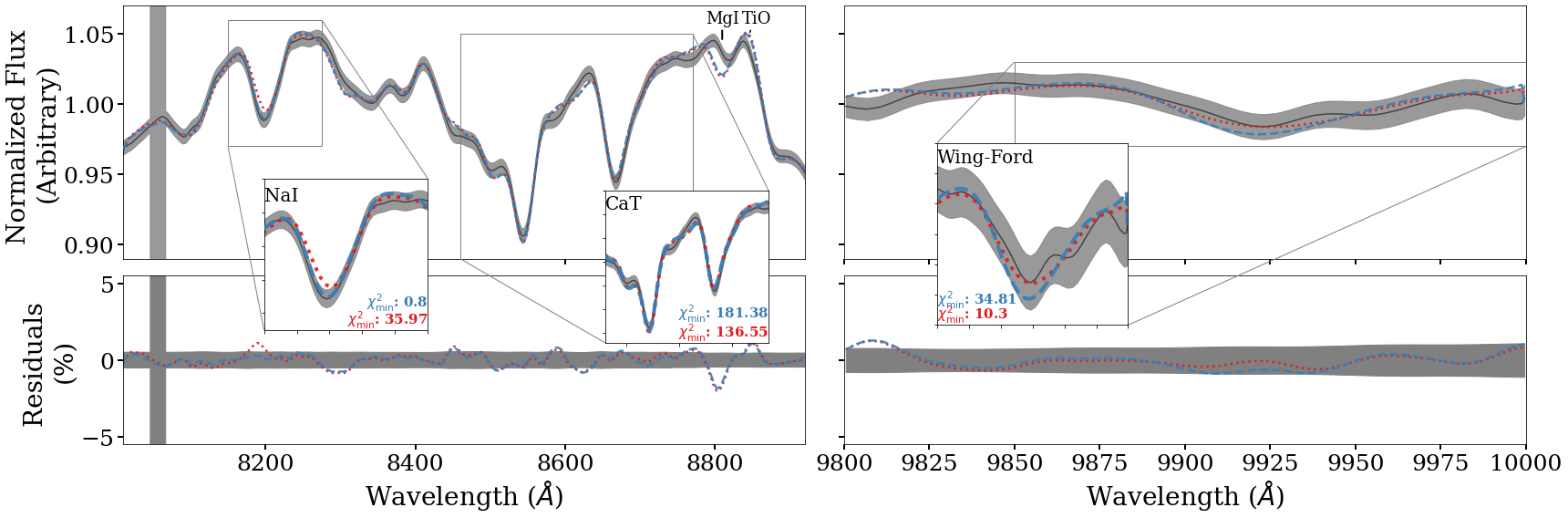}
    \caption{The same as Figure~\ref{fig:paper_fits_vucd7} but for NGC 4889 (BCG).}
    \label{fig:paper_fits_ngc4889}
\end{figure*}

\clearpage
\section{Fitted Stellar Population Parameters}\label{sec:appendix_fitparams}
Here we report the values of all stellar population parameters that we fit with \textsc{alf}, not included in Table~\ref{tab:fitparams}.  These Tables are available online.

\begin{table*}
	\centering
        \renewcommand\thetable{E1} 
	\caption{Values of other fitted parameters from \textsc{alf} for all objects we examine.}
	\label{tab:appendix_fitparams1}
  \begin{threeparttable}    
	\begin{tabular}{ccccc}
		\hline
		ID & $v_r$\tnote{a} (km~s$^{-1}$) & $T_{\mathrm{eff}}$\tnote{b} (K) & $\alpha_1$\tnote{c} & $\alpha_2$\tnote{d}\\
		\hline
		B012	&	$-354.38_{-1.62}^{+1.43}$	&	$-0.39_{-1.16}^{+1.36}$	&	$1.31_{-0.54}^{+0.46}$	&	$0.82_{-0.18}^{+0.19}$		\\
        B058 (2014, 2016)	&	$-212.63_{-0.64}^{+0.66}$, $-216.59_{-0.95}^{+0.96}$	&	$-0.63_{-1.0}^{+1.45}$, $-0.08_{-1.25}^{+1.36}$	&	$1.4_{-0.37}^{+0.26}$, $1.18_{-0.45}^{+0.6}$	&	$0.84_{-0.22}^{+0.27}$, $2.41_{-0.55}^{+0.3}$		\\
        B067	&	$-347.01_{-1.09}^{+0.98}$	&	$-0.48_{-1.06}^{+1.52}$	&	$1.38_{-0.42}^{+0.32}$	&	$0.85_{-0.21}^{+0.24}$		\\
        B074	&	$-436.27_{-1.4}^{+1.27}$	&	$0.01_{-1.34}^{+1.42}$	&	$1.55_{-0.39}^{+0.38}$	&	$0.99_{-0.32}^{+0.39}$		\\
        B107	&	$-323.07_{-1.2}^{+1.13}$	&	$0.49_{-1.67}^{+1.04}$	&	$1.3_{-0.44}^{+0.47}$	&	$2.9_{-0.38}^{+0.32}$		\\
        B163	&	$-163.85_{-0.38}^{+0.29}$	&	$-1.35_{-0.46}^{+1.26}$	&	$1.0_{-0.32}^{+0.44}$	&	$1.18_{-0.39}^{+0.28}$		\\
        B193	&	$-70.52_{-0.83}^{+0.53}$	&	$-0.07_{-1.34}^{+1.31}$	&	$0.96_{-0.31}^{+0.43}$	&	$0.96_{-0.33}^{+0.3}$		\\
        B225	&	$-164.13_{-0.61}^{+0.58}$	&	$-1.0_{-0.74}^{+1.37}$	&	$1.04_{-0.34}^{+0.5}$	&	$1.82_{-0.25}^{+0.19}$		\\
        B338	&	$-264.37_{-0.85}^{+0.9}$	&	$-0.44_{-1.15}^{+1.53}$	&	$1.59_{-0.5}^{+0.44}$	&	$1.26_{-0.48}^{+0.57}$		\\
        B405	&	$-164.0_{-0.57}^{+0.79}$	&	$-0.55_{-1.02}^{+1.43}$	&	$1.78_{-0.35}^{+0.26}$	&	$0.96_{-0.3}^{+0.46}$		\\
        G001	&	$-338.71_{-0.96}^{+1.92}$	&	$-0.39_{-1.25}^{+1.24}$	&	$3.41_{-0.1}^{+0.06}$	&	$0.8_{-0.21}^{+0.19}$		\\
        M59-UCD3	&	$383.13_{-0.36}^{+0.58}$	&	$-0.26_{-1.14}^{+1.58}$	&	$2.37_{-0.51}^{+0.37}$	&	$1.29_{-0.43}^{+0.36}$		\\
        VUCD3	&	$710.47_{-0.57}^{+1.0}$	&	$-0.03_{-1.39}^{+1.34}$	&	$2.68_{-0.6}^{+0.41}$	&	$1.49_{-0.53}^{+0.5}$		\\
        VUCD4	&	$916.95_{-1.59}^{+2.07}$	&	$0.1_{-1.37}^{+1.29}$	&	$1.12_{-0.41}^{+0.41}$	&	$0.98_{-0.29}^{+0.38}$		\\
        VUCD7	&	$986.32_{-0.89}^{+0.75}$	&	$-0.64_{-1.01}^{+1.55}$	&	$2.51_{-0.28}^{+0.17}$	&	$1.18_{-0.38}^{+0.33}$		\\
        NGC 4874	&	$7145.11_{-5.4}^{+11.84}$	&	$-0.03_{-1.42}^{+1.45}$	&	$1.12_{-0.41}^{+0.72}$	&	$2.74_{-0.62}^{+0.5}$		\\
        NGC 4889	&	$6447.56_{-5.49}^{+4.4}$	&	$-0.32_{-1.21}^{+1.51}$	&	$2.43_{-0.23}^{+0.23}$	&	$3.22_{-0.3}^{+0.19}$		\\
		\hline
	\end{tabular}
  \begin{tablenotes}
      \item[a] Recession velocity in km~s$^{-1}$.
      \item[b] Effective temperature in K.
      \item[c] Low-mass IMF slope.
      \item[d] Intermediate-mass IMF slope.
  \end{tablenotes}
  \end{threeparttable}
\end{table*}

\begin{table*}
	\centering
        \renewcommand\thetable{E2} 
	\caption{Values of other fitted elemental abundances from \textsc{alf} for all objects we examine.}
	\label{tab:appendix_fitparams2}
	\begin{tabular}{p{15mm}p{18mm}p{18mm}p{18mm}p{18mm}p{18mm}p{18mm}p{18mm}}
		\hline
		ID & [Ba/Fe] & [C/Fe] & [Co/Fe] & [Cr/Fe] &[Cu/Fe] & [Eu/Fe] & [K/Fe] \\
		\hline
		B012	&	$-0.17_{-0.14}^{+0.18}$	&	$0.14_{-0.04}^{+0.03}$	&	$0.09_{-0.09}^{+0.11}$	&	$0.04_{-0.05}^{+0.04}$	&	$0.36_{-0.23}^{+0.2}$	&	$0.12_{-0.27}^{+0.28}$	&	$0.26_{-0.23}^{+0.29}$	\\
        B058 (2014, 2016)	&	$-0.32_{-0.11}^{+0.12}$, $-0.14_{-0.15}^{+0.12}$	&	$-0.1_{-0.03}^{+0.02}$, $-0.14_{-0.05}^{+0.04}$	&	$-0.16_{-0.05}^{+0.05}$, $0.03_{-0.06}^{+0.06}$	&	$-0.02_{-0.02}^{+0.02}$, $-0.0_{-0.03}^{+0.04}$	&	$-0.14_{-0.11}^{+0.13}$, $0.02_{-0.17}^{+0.17}$	&	$0.37_{-0.08}^{+0.04}$, $0.16_{-0.29}^{+0.23}$	&	$0.34_{-0.14}^{+0.06}$, $0.35_{-0.24}^{+0.12}$	\\
        B067	&	$-0.35_{-0.03}^{+0.07}$	&	$0.05_{-0.04}^{+0.06}$	&	$0.2_{-0.08}^{+0.08}$	&	$0.02_{-0.03}^{+0.03}$	&	$0.24_{-0.16}^{+0.17}$	&	$-0.2_{-0.13}^{+0.19}$	&	$0.22_{-0.22}^{+0.29}$	\\
        B074	&	$-0.27_{-0.12}^{+0.17}$	&	$0.02_{-0.05}^{+0.05}$	&	$0.19_{-0.09}^{+0.11}$	&	$-0.05_{-0.04}^{+0.04}$	&	$0.45_{-0.18}^{+0.14}$	&	$0.11_{-0.3}^{+0.27}$	&	$0.25_{-0.26}^{+0.26}$	\\
        B107	&	$0.12_{-0.21}^{+0.15}$	&	$-0.14_{-0.02}^{+0.03}$	&	$0.28_{-0.06}^{+0.06}$	&	$0.06_{-0.03}^{+0.03}$	&	$0.08_{-0.15}^{+0.19}$	&	$0.31_{-0.3}^{+0.22}$	&	$0.23_{-0.26}^{+0.29}$	\\
        B163	&	$-0.57_{-0.02}^{+0.03}$	&	$0.01_{-0.01}^{+0.01}$	&	$0.04_{-0.02}^{+0.02}$	&	$-0.01_{-0.01}^{+0.01}$	&	$0.25_{-0.12}^{+0.12}$	&	$0.46_{-0.07}^{+0.04}$	&	$0.44_{-0.11}^{+0.05}$	\\
        B193	&	$-0.57_{-0.04}^{+0.08}$	&	$-0.13_{-0.02}^{+0.02}$	&	$0.1_{-0.03}^{+0.03}$	&	$-0.08_{-0.02}^{+0.02}$	&	$0.31_{-0.14}^{+0.1}$	&	$0.43_{-0.07}^{+0.03}$	&	$0.37_{-0.13}^{+0.07}$	\\
        B225	&	$-0.43_{-0.07}^{+0.09}$	&	$-0.03_{-0.02}^{+0.02}$	&	$0.1_{-0.03}^{+0.03}$	&	$0.0_{-0.01}^{+0.01}$	&	$0.15_{-0.12}^{+0.1}$	&	$0.48_{-0.07}^{+0.04}$	&	$0.47_{-0.13}^{+0.05}$	\\
        B338	&	$-0.24_{-0.13}^{+0.12}$	&	$-0.06_{-0.05}^{+0.03}$	&	$-0.13_{-0.07}^{+0.08}$	&	$-0.05_{-0.03}^{+0.04}$	&	$-0.08_{-0.15}^{+0.15}$	&	$-0.54_{-0.06}^{+0.11}$	&	$0.28_{-0.3}^{+0.15}$	\\
        B405	&	$-0.43_{-0.1}^{+0.11}$	&	$-0.06_{-0.05}^{+0.04}$	&	$0.07_{-0.06}^{+0.06}$	&	$-0.02_{-0.03}^{+0.03}$	&	$-0.05_{-0.13}^{+0.15}$	&	$-0.44_{-0.09}^{+0.18}$	&	$0.41_{-0.22}^{+0.09}$	\\
        G001	&	$-0.34_{-0.14}^{+0.14}$	&	$0.01_{-0.02}^{+0.02}$	&	$0.04_{-0.04}^{+0.05}$	&	$-0.01_{-0.03}^{+0.02}$	&	$-0.16_{-0.12}^{+0.14}$	&	$0.06_{-0.22}^{+0.22}$	&	$0.28_{-0.24}^{+0.12}$	\\
        M59-UCD3	&	$-0.48_{-0.03}^{+0.07}$	&	$0.2_{-0.01}^{+0.01}$	&	$0.08_{-0.03}^{+0.03}$	&	$-0.01_{-0.01}^{+0.01}$	&	$-0.04_{-0.11}^{+0.13}$	&	$0.48_{-0.13}^{+0.07}$	&	$0.1_{-0.21}^{+0.25}$	\\
        VUCD3	&	$0.3_{-0.16}^{+0.12}$	&	$0.17_{-0.02}^{+0.02}$	&	$0.06_{-0.05}^{+0.06}$	&	$-0.05_{-0.03}^{+0.03}$	&	$0.25_{-0.21}^{+0.14}$	&	$0.26_{-0.32}^{+0.16}$	&	$0.21_{-0.28}^{+0.19}$	\\
        VUCD4	&	$-0.59_{-0.14}^{+0.19}$	&	$-0.04_{-0.03}^{+0.03}$	&	$-0.23_{-0.09}^{+0.12}$	&	$-0.13_{-0.04}^{+0.04}$	&	$-0.17_{-0.19}^{+0.2}$	&	$0.07_{-0.28}^{+0.17}$	&	$-0.12_{-0.25}^{+0.28}$	\\
        VUCD7	&	$-0.58_{-0.05}^{+0.08}$	&	$-0.02_{-0.01}^{+0.01}$	&	$0.04_{-0.04}^{+0.03}$	&	$-0.14_{-0.02}^{+0.02}$	&	$0.17_{-0.15}^{+0.12}$	&	$0.32_{-0.14}^{+0.09}$	&	$0.33_{-0.19}^{+0.09}$	\\
        NGC 4874	&	$-0.03_{-0.26}^{+0.33}$	&	$0.36_{-0.04}^{+0.03}$	&	$0.2_{-0.11}^{+0.11}$	&	$0.09_{-0.05}^{+0.06}$	&	$0.06_{-0.17}^{+0.27}$	&	$0.18_{-0.42}^{+0.32}$	&	$0.13_{-0.2}^{+0.3}$	\\
        NGC 4889	&	$0.11_{-0.28}^{+0.2}$	&	$0.31_{-0.02}^{+0.02}$	&	$0.33_{-0.07}^{+0.07}$	&	$-0.06_{-0.04}^{+0.03}$	&	$-0.07_{-0.13}^{+0.21}$	&	$0.12_{-0.41}^{+0.29}$	&	$0.1_{-0.24}^{+0.29}$	\\
		\hline
	\end{tabular}
\end{table*}

\begin{table*}
	\centering
        \renewcommand\thetable{E3} 
	\caption{Table~\ref{tab:appendix_fitparams2} continued.}
	\label{tab:appendix_fitparams3}
	\begin{tabular}{p{15mm}p{18mm}p{18mm}p{18mm}p{18mm}p{18mm}p{18mm}p{18mm}}
		\hline
		ID & [Mn/Fe] & [N/Fe] & [Na/Fe] & [Ni/Fe] & [O/Fe] & [Sr/Fe] & [V/Fe] \\
		\hline
		B012	&	$0.28_{-0.06}^{+0.06}$	&	$0.31_{-0.13}^{+0.14}$	&	$1.05_{-0.08}^{+0.09}$	&	$-0.03_{-0.04}^{+0.04}$	&	$1.12_{-0.2}^{+0.11}$	&	$0.01_{-0.06}^{+0.1}$	&	$0.13_{-0.08}^{+0.1}$	\\
            B058 (2014, 2016)	&	$0.21_{-0.03}^{+0.03}$, $0.13_{-0.04}^{+0.04}$	&	$0.44_{-0.04}^{+0.04}$, $0.51_{-0.07}^{+0.12}$	&	$0.5_{-0.02}^{+0.03}$, $0.42_{-0.03}^{+0.05}$	&	$-0.13_{-0.03}^{+0.03}$, $-0.13_{-0.03}^{+0.03}$	&	$0.75_{-0.1}^{+0.07}$, $0.58_{-0.15}^{+0.19}$	&	$-0.09_{-0.07}^{+0.07}$, $-0.12_{-0.12}^{+0.11}$	&	$0.1_{-0.04}^{+0.04}$, $0.01_{-0.05}^{+0.05}$	\\
            B067	&	$0.23_{-0.04}^{+0.04}$	&	$0.35_{-0.07}^{+0.07}$	&	$0.24_{-0.04}^{+0.04}$	&	$0.04_{-0.04}^{+0.04}$	&	$0.63_{-0.15}^{+0.21}$	&	$0.01_{-0.06}^{+0.1}$	&	$-0.07_{-0.02}^{+0.03}$	\\
            B074	&	$0.15_{-0.06}^{+0.06}$	&	$-0.01_{-0.09}^{+0.11}$	&	$0.52_{-0.05}^{+0.04}$	&	$-0.09_{-0.04}^{+0.04}$	&	$0.59_{-0.16}^{+0.22}$	&	$-0.09_{-0.05}^{+0.06}$	&	$-0.1_{-0.04}^{+0.06}$	\\
            B107	&	$-0.01_{-0.06}^{+0.05}$	&	$0.41_{-0.06}^{+0.06}$	&	$1.13_{-0.02}^{+0.02}$	&	$-0.01_{-0.04}^{+0.04}$	&	$0.2_{-0.06}^{+0.08}$	&	$0.09_{-0.14}^{+0.14}$	&	$0.07_{-0.06}^{+0.05}$	\\
            B163	&	$0.08_{-0.02}^{+0.02}$	&	$0.46_{-0.02}^{+0.02}$	&	$0.48_{-0.01}^{+0.01}$	&	$0.01_{-0.02}^{+0.02}$	&	$0.29_{-0.03}^{+0.03}$	&	$0.04_{-0.06}^{+0.06}$	&	$0.02_{-0.02}^{+0.03}$	\\
            B193	&	$0.2_{-0.03}^{+0.03}$	&	$0.02_{-0.03}^{+0.04}$	&	$0.49_{-0.01}^{+0.01}$	&	$0.01_{-0.03}^{+0.03}$	&	$0.17_{-0.04}^{+0.03}$	&	$0.37_{-0.12}^{+0.07}$	&	$0.06_{-0.03}^{+0.03}$	\\
            B225	&	$0.06_{-0.02}^{+0.02}$	&	$0.6_{-0.02}^{+0.02}$	&	$0.44_{-0.02}^{+0.02}$	&	$-0.06_{-0.02}^{+0.02}$	&	$0.41_{-0.04}^{+0.04}$	&	$0.04_{-0.07}^{+0.06}$	&	$0.04_{-0.03}^{+0.02}$	\\
            B338	&	$-0.14_{-0.04}^{+0.04}$	&	$0.54_{-0.05}^{+0.07}$	&	$0.33_{-0.03}^{+0.03}$	&	$-0.14_{-0.04}^{+0.04}$	&	$0.79_{-0.11}^{+0.09}$	&	$-0.16_{-0.09}^{+0.1}$	&	$-0.15_{-0.06}^{+0.05}$	\\
            B405	&	$0.04_{-0.04}^{+0.04}$	&	$0.08_{-0.06}^{+0.06}$	&	$0.53_{-0.03}^{+0.03}$	&	$-0.16_{-0.04}^{+0.03}$	&	$0.69_{-0.18}^{+0.16}$	&	$-0.12_{-0.08}^{+0.08}$	&	$-0.11_{-0.06}^{+0.05}$	\\
            G001	&	$-0.05_{-0.04}^{+0.04}$	&	$0.54_{-0.03}^{+0.04}$	&	$0.4_{-0.02}^{+0.02}$	&	$-0.14_{-0.03}^{+0.03}$	&	$0.7_{-0.06}^{+0.03}$	&	$-0.06_{-0.1}^{+0.1}$	&	$-0.01_{-0.04}^{+0.04}$	\\
            M59-UCD3	&	$0.15_{-0.03}^{+0.02}$	&	$0.12_{-0.03}^{+0.03}$	&	$0.1_{-0.02}^{+0.01}$	&	$0.01_{-0.03}^{+0.03}$	&	$0.29_{-0.03}^{+0.04}$	&	$-0.01_{-0.09}^{+0.09}$	&	$-0.03_{-0.02}^{+0.03}$	\\
            VUCD3	&	$0.07_{-0.04}^{+0.04}$	&	$0.33_{-0.04}^{+0.05}$	&	$0.39_{-0.02}^{+0.02}$	&	$-0.06_{-0.05}^{+0.06}$	&	$0.51_{-0.03}^{+0.02}$	&	$0.26_{-0.17}^{+0.13}$	&	$-0.07_{-0.06}^{+0.06}$	\\
            VUCD4	&	$0.28_{-0.04}^{+0.04}$	&	$0.62_{-0.08}^{+0.07}$	&	$0.01_{-0.05}^{+0.04}$	&	$-0.17_{-0.05}^{+0.05}$	&	$0.74_{-0.14}^{+0.05}$	&	$-0.28_{-0.12}^{+0.17}$	&	$-0.1_{-0.08}^{+0.09}$	\\
            VUCD7	&	$0.16_{-0.03}^{+0.03}$	&	$0.61_{-0.03}^{+0.03}$	&	$0.1_{-0.02}^{+0.02}$	&	$-0.09_{-0.03}^{+0.03}$	&	$0.71_{-0.04}^{+0.03}$	&	$0.01_{-0.07}^{+0.08}$	&	$0.06_{-0.03}^{+0.03}$	\\
            NGC 4874	&	$0.16_{-0.08}^{+0.09}$	&	$0.17_{-0.1}^{+0.08}$	&	$0.38_{-0.06}^{+0.05}$	&	$0.15_{-0.08}^{+0.12}$	&	$0.49_{-0.09}^{+0.08}$	&	$0.25_{-0.24}^{+0.24}$	&	$0.14_{-0.1}^{+0.12}$	\\
            NGC 4889	&	$0.06_{-0.05}^{+0.06}$	&	$0.2_{-0.05}^{+0.06}$	&	$0.44_{-0.03}^{+0.03}$	&	$0.03_{-0.06}^{+0.06}$	&	$0.47_{-0.06}^{+0.05}$	&	$0.17_{-0.2}^{+0.2}$	&	$0.05_{-0.07}^{+0.07}$	\\
		\hline
	\end{tabular}
\end{table*}


\end{document}